\documentclass[preprint,12pt]{elsarticle}
\usepackage{amssymb,amsmath,graphicx,bm,mathrsfs,color,aas_macros,soul}
\biboptions{sort&compress}  
\usepackage[margin=2.5cm]{geometry}  
\newcommand{\be}{\begin{equation}} 
\newcommand{\ee}{\end{equation}}
\newcommand{\bea}{\begin{eqnarray}}
\newcommand{\eea}{\end{eqnarray}}

\newcommand{\Tr}{{\rm Tr}}
\newcommand{\ie}{{\it i.e.}}
\newcommand{\eg}{{\it e.g.}}

\newcommand{\no}{{\nonumber}}
\definecolor{red}{rgb}{0.8,0,0}
\definecolor{violet}{rgb}{0.4,0,0.4}
\definecolor{green}{rgb}{0,0.5,0.0}
\definecolor{navy}{rgb}{0.0,0.0,0.6}
\definecolor{orange}{rgb}{0.8,0.2,0.0}

\usepackage[normalem]{ulem}  

\usepackage[breaklinks,colorlinks,colorlinks=true]{hyperref}
\hypersetup{
  linkcolor=Red,          
  citecolor=Violet,       
  urlcolor=Blue}           

\journal{Annals of Physics}

\begin{document}
\begin{frontmatter}

\title{ Relativistic second-order dissipative hydrodynamics from Zubarev's
non-equilibrium statistical operator}

\author[label1,label2]{Arus Harutyunyan}
\author[label3,label4]{Armen Sedrakian}
\author[label5,label6]{Dirk H.\ Rischke}
\address[label1]{Byurakan Astrophysical Observatory, Byurakan 0213, Armenia}
\address[label2]{Department of Physics, Yerevan State University,
  Yerevan 0025, Armenia}
\address[label3]{Frankfurt Institute for Advanced Studies, 
  Ruth-Moufang-Str.\ 1, D-60438 Frankfurt am Main, Germany} 
\address[label4]{Institute of Theoretical Physics, University of
  Wroc\l{}aw, 50-204 Wroc\l{}aw, Poland}
\address[label5]{Institute for Theoretical Physics,
  Max-von-Laue-Str.\ 1, D-60438 Frankfurt am Main, Germany}
\address[label6]{Helmholtz Research Academy Hesse for FAIR,
  Max-von-Laue-Str.\ 12, D-60438 Frankfurt am Main, Germany}

\begin{abstract}
  We present a new derivation of relativistic second-order 
  dissipative hydrodynamics for quantum systems using Zubarev's
  non-equilibrium statistical-operator formalism. This is achieved
  by a systematic expansion of the energy-momentum tensor 
  and the charge current to second order
  in deviations from equilibrium.  As a concrete example, we obtain the
  relaxation equations for the shear-stress tensor, the bulk-viscous
  pressure, and the charge-diffusion currents required to close the
  set of equations of motion for relativistic second-order dissipative 
  hydrodynamics. We also identify 
  new transport coefficients which describe the relaxation of
  dissipative processes to second-order and express them in terms of
  equilibrium correlation functions, thus establishing new Kubo-type
  formulas for second-order transport coefficients.
\end{abstract}


\begin{keyword}  Hydrodynamics \sep Statistical operator \sep
  Transport coefficients \sep Correlation functions

\end{keyword}
\end{frontmatter}

\section{Introduction}
\label{sec:intro}

Hydrodynamics is a powerful tool to describe
low-frequency and long-wavelength phenomena in statistical systems by
performing an averaging (coarse-graining) over scales that are large
compared to those characteristic for kinetic phenomena, \eg, the
mean free path of a quasiparticle~\cite{Landau1987}. It finds numerous
applications in astrophysics and cosmology~\cite{Weinberg1972},
heavy-ion physics~\cite{Ablyazimov2017EPJA,Friman:2011}, and other
areas. During the last decade, relativistic hydrodynamics has been
successfully applied to describe the behavior of strongly interacting
hot and dense matter created in heavy-ion collision experiments at the
Relativistic Heavy-Ion Collider (RHIC) and the Large Hadron Collider
(LHC).  In these experiments, a new state of matter, the quark--gluon
plasma (QGP), was discovered, which behaves almost like a perfect
fluid (for reviews, see
Refs.~\cite{Busza_2018,Fukushima2017RPPh,Heinz2015IJMPE}). Another
area of applications of relativistic hydrodynamics is the physics of
compact stars. In particular, binary neutron-star mergers and
gravitational waves emitted in these events are modeled by the
coupled dynamics of fluid and space-time in general relativity (for
reviews, see Refs.~\cite{Faber2012:lrr,Baiotti2017RPPh}).

This work aims to obtain the hydrodynamical equations of strongly
correlated matter, such as the non-perturbative QGP in heavy-ion
collisions or hadronic matter in compact stars, by adopting Zubarev's
formalism, known also as the method of the non-equilibrium statistical
operator~\cite{zubarev1974nonequilibrium,zubarev1997statistical,
  Zubarev1979TMP}. This method is based on a generalization of the
Gibbs canonical ensemble to non-equilibrium states, \ie, the
statistical operator is promoted to a non-local functional of the
thermodynamic parameters and their space-time derivatives.  Assuming
that the thermodynamic parameters vary sufficiently smoothly over the
correlation lengths characterizing the system, the statistical
operator is then expanded in a series in gradients of these
parameters, commonly keeping only the first-order terms in the
expansion.  Then, the hydrodynamical equations for the dissipative
currents emerge after statistical averaging of the relevant quantum
operators. At weak coupling, this method is equivalent to hydrodynamics
obtained from moments of the Boltzmann equation for the distribution
function~\cite{Denicol2012PhRvD,Denicol2012PhLB,Molnar2016PhRvD}. An
advantage of Zubarev's formalism is that the transport coefficients of
the system are automatically obtained in the form of Kubo-type
relations, \ie, they are related to certain correlation functions of
the underlying field theory, valid also in the strong-coupling limit.

The application of Zubarev's formalism to quantum fields and
relativistic hydrodynamics is due to the pioneering works of
Zubarev et al.~\cite{Zubarev1979TMP} and Hosoya et
al.~\cite{Hosoya1984AnPhy}, see also Ref.~\cite{Horsley1987NuPhB}. In recent
years, there has been a renewed interest in applications of this
method to relativistic hydrodynamics.  The novel developments include
a formulation of anisotropic magnetohydrodynamics in strong
magnetic fields~\cite{Huang2011AnPhy}, reformulations and reinterpretations
suitable for applications to the QGP~\cite{BecattiniParticles},
second-order expansions of the statistical operator, which lead to
second-order hydrodynamics~\cite{HSR_particles}, and hydrodynamics
with anomalies~\cite{HongoParticles,Hayata2015}.

In this work, we provide a new, alternative derivation of relativistic 
second-order dissipative hydrodynamics within Zubarev's formalism, which 
extends our previous work~\cite{HSR_particles}.  To motivate our strategy, 
we recall that hydrodynamics describes the state of a fluid in terms of 
its energy-momentum tensor and currents of conserved charges, which, in 
the relevant low-frequency and long-wavelength limit, can be expanded 
around their equilibrium values in powers of gradients of thermodynamic 
parameters (so-called thermodynamic forces). The zeroth-order expansion 
corresponds to ideal (non-dissipative) hydrodynamics. At first order, 
the relativistic generalization of Navier--Stokes hydrodynamics emerges 
from a truncation that keeps the first-order terms in 
gradients~\cite{Landau1987,Eckart1940PhRv}. Relativistic second-order 
dissipative theories, which keep the next-to-leading dissipative 
terms in the above expansion were first constructed by Israel and
Stewart~\cite{Israel1976AnPhy,Israel1979AnPhy}. Such theories were
demonstrated to avoid acausalities of the first-order theory and
numerical instabilities associated with it.~\footnote{Recently it 
was shown that these acausalities and instabilities are a consequence 
of the matching procedure to the local-equilibrium reference state. 
Generalizing this matching, causal and stable first-order dissipative 
hydrodynamic theories have been derived \cite{BDN,kovtun}.}  
In the second-order theories, the dissipative currents satisfy 
relaxation equations, which include finite timescales of their 
relaxation towards their (first-order) Navier--Stokes values at 
asymptotically large times~\cite{Baier2008JHEP,Betz2009JPhG,
Romatschke2010CQGra,Tsumura2010PhLB,Betz2011EPJWC,Moore2011PhRvL,
Moore2012JHEP,Jaiswal2013PhRvC,Jaiswal2015PhLB,Florkowski2015PhRvC,
Finazzo2015JHEP,Tinti2017PhRvD,Kovtun2018JHEP}, for reviews see Refs.~\cite{Kovtun2012JPhA,Denicol2014JPhG,Florkowski2018RPPh}. 

The transport coefficients entering dissipative hydrodynamics can be
obtained at weak coupling either in the relaxation-time approximation or,
more systematically, via the method of moments~\cite{Denicol2012PhRvD}, 
or at arbitrary values of the coupling through Kubo formulas. 
Second-order transport coefficients were obtained via an expansion of
the curved-space metric around its flat-space limit through Kubo formulas
in Ref.~\cite{Moore2011PhRvL}.  Here we will show how to obtain these
transport coefficients from the Zubarev formalism. In doing so, we
apply an expansion of the non-equilibrium statistical
operator up to second order in thermodynamic forces, 
which allows us to generate in a systematic way all second-order terms 
in hydrodynamics after statistical averaging the dissipative
currents over the non-equilibrium statistical operator.  A concise
discussion of our approach and some results were reported
earlier~\cite{HSR_particles}, see also Ref.~\cite{ArusPhD}.

This work is structured as follows. To set the stage, in
Section~\ref{sec:Zub_formalism} we briefly review Zubarev's
formalism~\cite{zubarev1974nonequilibrium,zubarev1997statistical}.
Section~\ref{sec:trans_first} recapitulates Navier--Stokes theory and
the Kubo formulas for the first-order transport coefficients. The
second-order transport equations for the dissipative currents and the
relevant transport coefficients are derived in
Section~\ref{sec:trans_second}.  In Section~\ref{sec:hydro_discuss} we
discuss the structure of the transport equations and compare them to
those of earlier studies.  A summary of the results is given in
Section~\ref{sec:hydro_conclusions}.  \ref{app:frames} discusses the
Eckart and Landau frame choices for the fluid velocity.
\ref{app:H-theorem} computes the rate of entropy generation within the
Zubarev formalism. In~\ref{app:Green_func} we derive Kubo formulas for
transport coefficients. Some properties of projection operators are
listed in \ref{app:projectors}.  We work in flat space-time described
by the metric $g_{\mu\nu}={\rm diag}(+,-,-,-)$ and use natural units
throughout.

\section{The non-equilibrium statistical-operator formalism}
\label{sec:Zub_formalism}

In this section, we describe Zubarev's non-equilibrium statistical-operator 
formalism for a generic quantum system which is in the hydrodynamic
regime~\cite{zubarev1974nonequilibrium,zubarev1997statistical,Zubarev1979TMP}.
Our starting point is the operator-valued conservation laws for energy-momentum
and charges
\be\label{eq:cons_laws}
\partial_{\mu} \hat{T}^{\mu\nu} =0\;,\qquad
\partial_{\mu} \hat{N}^{\mu}_a =0\;,
\ee
where $a=1,2,\dots, \ell$ labels the possible conserved charges (\eg, baryonic, 
electric, etc.); $\ell$ is the total number of conserved charges. To obtain the
corresponding conservation laws in hydrodynamics one needs to take the statistical
averages of the operators $\hat{T}^{\mu\nu}$ and $\hat{N}^{\mu}_a$ with respect 
to the full non-equilibrium statistical operator. This operator should be found 
from the quantum Liouville equation with an infinitesimal source
term~\cite{zubarev1974nonequilibrium,zubarev1997statistical,Zubarev1979TMP}
and can then be expanded in a series with respect to the thermodynamic forces up 
to the required order.  Statistical averaging of the dissipative currents leads 
then to constitutive relations for the latter and provides explicit expressions 
for the transport coefficients via certain equilibrium correlation functions of the system.

\subsection{Local-equilibrium statistical operator}
\label{sec:rel_dist}

As is well known, the thermodynamic state of a macroscopic quantum system
is described by the statistical operator $\hat{\rho}(t)$
(\ie, the density matrix). It obeys the Liouville--von Neumann equation, 
which reads in the Schr{\"o}dinger representation~\cite{LandauStat}
\bea\label{eq:Liouville_eq_S}
\frac{\partial \hat{\rho}_S(t)}{\partial t}
+i[\hat{K},\hat{\rho}_S(t)]=0\;,
\eea
where $\hat{K}=\hat{H}-\sum\limits_a\mu_a \hat{\cal N}_a$; here
$\hat{H}$ is the Hamiltonian of the system, $\hat{\cal N}_a$ are the
operators of the conserved charges, $\mu_a$ are the corresponding chemical
potentials, and $[\hat{A} , \, \hat{B}] \equiv \hat{A} \hat{B} - \hat{B} \hat{A}$ 
denotes the commutator of the two operators $\hat{A},\, \hat{B}$. 
In this representation the operators acting on the quantum
states of the system are independent of time, therefore the explicit
time dependence in $\hat{\rho}_S(t)$ arises from the time dependence
of the thermodynamic parameters caused by a perturbation. In the
Heisenberg picture, which we will adopt in this work, the operators
evolve in time, whereas the statistical operator is time-independent,
\bea\label{eq:Liouville_eq_H}
\frac{d\hat{\rho}_H(t)}{dt}=0\;, \quad 
\hat{\rho}_H(t) = \hat\rho_H(0) = {\rm Const.},
\eea
where $\hat\rho_H(0)$ refers to the initial value of
$\hat\rho_H(t)$. The choice of the initial statistical operator is
discussed below in Sec.~\ref{sec:deriv_rho}. In
Eq.~\eqref{eq:Liouville_eq_H}, the differentiation with respect to
time acts simultaneously onto the operators and the thermodynamic
parameters. The two representations are related via the transformation
$\hat{\rho}_S(t)=e^{-i\hat{K}t} \hat{\rho}_H e^{i\hat{K}t}$. From now
on, for the sake of simplicity we will omit the index $H$ for all
quantum operators in the Heisenberg representation.

In addition to the equation of motion \eqref{eq:Liouville_eq_H}, the
statistical operator should also satisfy the normalization condition
$\Tr\,\hat{\rho}=1$. The knowledge of the statistical operator
allows one to compute the thermal expectation value of any quantum
operator $\hat{F}(x)$ via the formula
\bea\label{eq:op_av}
\langle\hat{F}(x)\rangle =\Tr\big[\hat{\rho}\hat{F}(x)\big]\;,
\eea
where $x\equiv (\bm x,t)$ denotes a point in space-time. 

In thermodynamic equilibrium, the statistical operator depends 
only on the integrals of motion. For a system in contact with a
heat bath with temperature $T=\beta^{-1}$ and a charge reservoir
with chemical potentials $\mu_a$, the equilibrium statistical operator 
is given by the grand-canonical (Gibbs) distribution
\bea
\label{eq:Gibbs_dist}
\hat{\rho}_{\rm eq} = e^{\Omega-\beta \hat{K}}\;,\qquad
e^{-\Omega} = \Tr\, e^{-\beta\hat{K}}\;.
\eea
In this case both the Heisenberg and Schr{\"o}dinger 
representations for $\hat{\rho}_{\rm eq}$ coincide. 
Note that the distribution \eqref{eq:Gibbs_dist} is written
in the frame where the system as a whole is at rest. We can generalize
the equilibrium distribution to an arbitrary reference frame via a
Lorentz transformation $\hat{H}\to \hat{\cal P}_\nu U^\nu$, where
$U^\nu$ is the 4-velocity of the system in the chosen frame,
$\hat{\cal P}_\nu$ is the 4-momentum operator and
$\hat{H}\equiv \hat{\cal P}_0$ in the fluid rest frame. In terms of the
energy-momentum tensor and the charge currents, the operators
$\hat{\cal P}_\nu$ and $\hat{\cal N}_a$  are given by
\bea\label{eq:int_motion}
\hat{\cal P}_\nu =\int\! d^3 x\, \hat{T}_{0\nu}(x)\;,\qquad
\hat{\cal N}_a=\int\! d^3 x\, \hat{N}_{a0}(x)\;.
\eea
Substituting Eq.~\eqref{eq:int_motion} into Eq.~\eqref{eq:Gibbs_dist}
we obtain the Lorentz-covariant form of the Gibbs distribution
\footnote{Note that Eqs.~\eqref{eq:Gibbs_inv} and \eqref{eq:Gibbs_inv_Om}
can be generalized to a fully Lorentz-invariant form, where 
the spatial integral is replaced by an integration over an arbitrary space-time 
hypersurface instead of a fixed-time hypersurface~\cite{Zubarev1979TMP}. }
\bea\label{eq:Gibbs_inv}
\hat{\rho}_{\rm eq} &=&
\exp\bigg\{\Omega -\int\! d^3 x\, \beta\Big[U^\nu\hat{T}_{0\nu}(x)
-\sum\limits_a\mu_a \hat{N}_{a0}(x)\Big]\bigg\}\;,\\
\label{eq:Gibbs_inv_Om}
e^{-\Omega} &=& \Tr \exp\bigg\{-\int\! d^3 x\,
\beta\Big[U^\nu\hat{T}_{0\nu}(x)
-\sum\limits_a\mu_a \hat{N}_{a0}(x)\Big]\bigg\}\;.
\eea

Now we consider a system that is out of global thermodynamic
equilibrium, but each small (but still macroscopic) portion of 
the system has reached its local-equilibrium state, \ie, the 
system is in the hydrodynamic regime. Local equilibrium implies 
that each fluid element can be ascribed local values of the hydrodynamic
parameters such as temperature $\beta^{-1}(x)$, chemical potentials 
$\mu_a(x)$, and a macroscopic 4-velocity $u^{\nu}(x)$, which vary 
slowly in space and time. In this case the global-equilibrium distribution 
given by Eqs.~\eqref{eq:Gibbs_inv} and \eqref{eq:Gibbs_inv_Om} is 
replaced by a local-equilibrium statistical operator via
\bea\label{eq:stat_op_eq}
\hat{\rho}_l(t) &=& \exp \bigg\{\Omega_l(t)-\int\! 
d^3x \Big[\beta^\nu(x)\hat{T}_{0\nu}(x)- \sum\limits_a 
\alpha_a(x) \hat{N}_{a0}(x)\Big]\bigg\}\;,\\
\label{eq:Om_eq} 
e^{-\Omega_l(t)} &=& \Tr\exp\bigg\{-\int\! d^3x 
\Big[\beta^\nu(x)\hat{T}_{0\nu}(x)-\sum\limits_a 
\alpha_a(x) \hat{N}_{a0}(x)\Big]\bigg\}\;,
\eea
where 
\be\label{eq:beta_nu_alpha}
\beta^\nu(x) = \beta(x) u^{\nu}(x),\qquad
\alpha_a (x)=\beta(x)\mu_a (x).
\ee
The operator $\hat{\rho}_l$ is determined from the principle of 
maximum entropy for given values of 
$u_\nu\langle\hat{T}^{\mu\nu}\rangle$ and $u_\mu\langle\hat{N}^{\mu}_a\rangle
$~\cite{zubarev1974nonequilibrium,Zubarev1979TMP,zubarev1997statistical}.
The fluid 4-velocity $u^\nu$ can be defined according to either 
Landau or Eckart, see \ref{app:frames}. The local-equilibrium distribution
\eqref{eq:stat_op_eq} is also referred to as {\it relevant statistical
operator}~\cite{zubarev1974nonequilibrium,zubarev1997statistical}.

Next, we define the operators of energy and charge densities in the
comoving frame via $\hat{\epsilon}=u_\mu u_\nu \hat{T}^{\mu\nu}$ and
$\hat{n}_a=u_\mu \hat{N}^\mu_a$.  The local values of the
Lorentz-invariant thermodynamic parameters $\beta$ and $\alpha_a$ are then 
fixed by the given average values of the operators $\hat{\epsilon}$
and $\hat{n}_a$ via the following matching
conditions~\cite{zubarev1974nonequilibrium,zubarev1997statistical,Zubarev1979TMP}
\bea\label{eq:matching}
\langle \hat{\epsilon}(x)\rangle =
\langle\hat{\epsilon}(x)\rangle_l\;,\qquad
\langle\hat{n}_a(x)\rangle
=\langle\hat{n}_a(x)\rangle_l\;,
\eea
where we introduced the notation
\bea\label{eq:stat_av_l}
\langle\hat{F}(x)\rangle_l =
\Tr\big[\hat{\rho}_l(t)\hat{F}(x)\big]\;.
\eea
Note that the conditions \eqref{eq:matching} define the 
temperature and the chemical potentials in general~\cite{Zubarev1972Phy}  as
{\it non-local functionals} of
\begin{equation}
  \label{eq:short-hand1}
  \langle \hat{\epsilon}(x)\rangle
\equiv \epsilon(x), \qquad \langle \hat{n}_a(x)\rangle\equiv
{n}_a(x).
\end{equation}
However, in the hydrodynamic 
description of the fluid one needs to define thermodynamic 
parameters as {\it local} functions of the energy and charge 
densities, as it is the case in global thermodynamic equilibrium. 
This can be done by assuming that all fluid elements where 
local equilibrium is already established are statistically 
independent of each other~\cite{Mori1958PhRv}. In other words, 
the local-equilibrium values $\langle\hat{\epsilon}\rangle_l$ 
and $\langle \hat{n}_a\rangle_l$ in Eq.~\eqref{eq:matching} 
should be evaluated formally at {\it constant values} of
$\beta$ and $\mu_a$, which are then determined by matching
$\langle\hat{\epsilon}\rangle_l$ and $\langle \hat{n}_a\rangle_l$ 
to the real values $\langle \hat{\epsilon} \rangle$ and
$\langle \hat{n}_a \rangle$ of these quantities at the given 
point  $x$ in space-time. This assigns a {\it fictitious 
local-equilibrium state} to any given point, such that it 
reproduces the local values of the energy and charge densities. 
The relevant statistical operator also fixes the local values 
of the 3-momentum or one of the charge currents, when adopting 
the Landau or the Eckart definition of the fluid velocity,
respectively, see~\ref{app:frames}. Using Eq.~\eqref{eq:eps_Landau}
we can write the matching conditions in the Landau frame 
(hereafter L-frame) as follows
\bea\label{eq:matching_L}
u_\mu\langle\hat{T}^{\mu\nu}\rangle =
u_\mu\langle\hat{T}^{\mu\nu}\rangle_{l}\;,\qquad
u_\mu\langle\hat{N}^\mu_a\rangle
= u_\mu\langle\hat{N}^\mu_a\rangle_{l}\;.
\eea
In the Eckart frame (hereafter E-frame) connected to the current
$\hat{N}^\mu_a$ we have instead [see Eq.~\eqref{eq:vel_Eckart_a}]
\bea\label{eq:matching_E}
u_\mu u_\nu\langle\hat{T}^{\mu\nu}\rangle =
u_\mu u_\nu\langle\hat{T}^{\mu\nu}\rangle_{l}\;,\qquad
u_\mu \langle\hat{N}^\mu_b\rangle
=u_\mu \langle\hat{N}^\mu_b\rangle_{l}\;,\qquad
\langle\hat{N}^\mu_a\rangle
=\langle\hat{N}^\mu_a\rangle_{l}\;,
\eea
where in the second relation $b$ runs over all values $1, \ldots, \ell$ except 
$a$. We recall that all 
local-equilibrium averages in Eqs.~\eqref{eq:matching_L} and 
\eqref{eq:matching_E} should be evaluated formally at constant 
values of $\beta$, $\mu_a$, and $u^\mu$, as explained above.

It is useful to also express the relevant distribution in 
terms of the 4-scalars
$\hat{\epsilon}$ and $\hat{n}_a$.
 Going to the local rest frame for each 
fluid element we can write Eqs.~\eqref{eq:stat_op_eq} and 
\eqref{eq:Om_eq} in the following form
\bea\label{eq:stat_op_eq_rest}
\hat{\rho}_l(t) 
&=& \exp \bigg\{\Omega_l(t)-\int\! d^3{\tilde x}\,
\beta(x)\Big[\hat{\epsilon}(x)-
\sum\limits_a \mu_a(x) \hat{n}_a(x)\Big]\bigg\}\;,\\
\label{eq:Om_eq_rest} 
e^{-\Omega_l(t)} &=& \Tr\exp\bigg\{-\int\! d^3\tilde{x} \,
\beta(x)\Big[\hat{\epsilon}(x)-
\sum\limits_a \mu_a(x) \hat{n}_a(x)\Big]\bigg\}\;,
\eea
where $d^3\tilde{x}=u^0(x)d^3 x$ is 
the proper volume of a fluid element.

\subsection{Thermodynamic relations}
\label{sec:therm_rel}

In this section we derive the thermodynamic relations for the local
thermodynamic parameters starting from the relevant distribution
\eqref{eq:stat_op_eq} or \eqref{eq:stat_op_eq_rest}.  Following
Zubarev we first define the entropy operator
as~\cite{zubarev1997statistical,Zubarev1979TMP}
\bea\label{eq:entropy_op}
\hat{S}(t) &=& -\ln\hat{\rho}_l(t) =-\Omega_l(t)+
\int\! d^3x \Big[\beta^\nu(x)\hat{T}_{0\nu}(x)
-\sum\limits_a \alpha_a(x) \hat{N}_{a0}(x)\Big]\nonumber\\
&=& -\Omega_l(t)+ \int\! d^3\tilde{x}\, 
\beta(x)\Big[\hat{\epsilon}(x)
-\sum\limits_a \mu_a(x) \hat{n}_a(x)\Big]\;,
\eea
which allows one to write the relevant statistical operator as
\bea\label{eq:stat_op1}
\hat{\rho}_l(t) = e^{-\hat{S}(t)}\;.
\eea
The thermodynamic entropy in local equilibrium is defined 
as the statistical average of the entropy operator
\bea\label{eq:entropy_eq}
{S}(t) = \langle\hat{S}(t)\rangle_l =-\Omega_l(t)
+\int\! d^3\tilde{x}\, \beta(x)\Big[
\langle \hat{\epsilon}(x)\rangle -\sum\limits_a 
\mu_a(x) \langle\hat{n}_a(x)\rangle\Big]
\equiv \langle\hat{S}(t)\rangle\;,
\eea
where we used the matching conditions \eqref{eq:matching}.

The thermodynamic relations we seek can now be derived using
Eq.~\eqref{eq:Om_eq_rest}. For that purpose we consider small
variations $\delta\epsilon(x)$ and $\delta n_a(x)$ in the local 
energy and charge densities. These variations induce small 
changes in temperature, $\delta\beta(x)$, and chemical potentials,
$\delta\mu_a(x)$, respectively. The corresponding change 
in $\Omega_l(t)$ is
\bea\label{eq:Om_var}
\delta \Omega_l(t)=\int\! d^3\tilde{x} \left[
\frac{\delta \Omega_l(t)}{\delta \beta (x)}
\delta\beta(x)+\sum\limits_a \frac{\delta
\Omega_l(t)}{\delta \mu_a (x)}\delta\mu_a(x)\right]\;,
\eea
where the derivatives in the square brackets are 
Lorentz-invariant functional derivatives of $\Omega_l(t)$. From 
Eqs.~\eqref{eq:stat_op_eq_rest} and \eqref{eq:Om_eq_rest} we obtain
\bea \label{eq:Om_deriv_beta}
\frac{\delta \Omega_l(t)}{\delta \beta (x)}
&=& \epsilon(x)-
\sum\limits_a \mu_a(x)  {n}_a(x)\;,\\
\label{eq:Om_deriv_mu_a}
\frac{\delta \Omega_l(t)}{\delta \mu_a (x)}
&=& -\beta(x) {n}_a(x) \;,
\eea
The infinitesimal change in the entropy can then be 
found from Eqs.~(\ref{eq:entropy_eq})--(\ref{eq:Om_deriv_mu_a}),
\bea\label{eq:entropy_var}
\delta{S}(t) &=& -\delta\Omega_l(t)+ \int\!\! d^3\tilde{x}
\Big[\delta\beta\Big({\epsilon}-
\sum\limits_a \mu_a {n}_a\Big)
-\beta\sum\limits_a {n}_a\delta\mu_a 
+\beta\Big(\delta{\epsilon}-\sum\limits_a \mu_a 
\delta{n}_a\Big) \Big] \nonumber\\
&=& \int\! d^3\tilde{x} \,\beta(x)\Big[ \delta{\epsilon}(x)
-\sum\limits_a \mu_a(x) \delta{n}_a(x)\Big]\;.
\eea
In the next step, we define the invariant entropy 
density ${s}(x)$ such as
\bea\label{eq:entropy_dens}
S(t) = \int\! d^3\tilde{x}\, {s}(x)\;.
\eea
Then Eq.~\eqref{eq:entropy_var} results in
\bea\label{eq:entropy_dens1}
\int\! d^3\tilde{x}\bigg\{ \beta(x)\Big[\delta{\epsilon}(x)-
\sum\limits_a \mu_a(x) \delta{n}_a(x)\Big]-\delta s(x)\bigg\}=0\;.
\eea
Because $\delta\epsilon(x)$ and $\delta n_a(x)$ 
are arbitrary variations and the entropy density 
$s(x)$ is assumed to be a {\it local function} of
$\epsilon(x)$ and $n_a(x)$, [\ie, $s({\epsilon}(x), {n}_a(x)) \equiv s(x)$],
we derive from Eq.~\eqref{eq:entropy_dens1} the relation
\bea\label{eq:thermodyn1}
T(x) \delta s(x)=\delta{\epsilon}(x)
-\sum\limits_a \mu_a (x)\delta{n}_a(x)\;,
\eea
which is the first law of thermodynamics for local variables.

To obtain other thermodynamic relations we recall that the grand
potential in global thermodynamic equilibrium is defined as 
$\Omega_l=-\beta pV$, with $p$ being the pressure and $V$ 
the volume of the system. In local equilibrium, $\Omega_l(t)$ 
given by Eq.~\eqref{eq:Om_eq_rest} is a functional of 
$\epsilon(x)$ and $n_a(x)$, therefore we can define a 
scalar function ${p}(\epsilon(x), 
n_a(x))\equiv p(x)$, such that
\bea\label{eq:omega_pres}
\Omega_l(t)=-\int\! d^3\tilde{x}\, \beta(x) p(x)\;.
\eea
The form of the function $p(\epsilon(x), n_a(x))$ should be
established from Eqs.~\eqref{eq:Om_eq_rest} and \eqref{eq:omega_pres}
and the matching conditions \eqref{eq:matching} (which determine the
temperature and the chemical potentials); it is called the {\it
  equation of state} (EoS).  Using Eq.~\eqref{eq:omega_pres} we can
also write Eq.~\eqref{eq:entropy_eq} as
\bea\label{eq:entropy_eq2} 
{S}(t) = \int\! d^3\tilde{x}\, \beta(x)
\Big[\epsilon(x)+{p}(x)-
\sum\limits_a \mu_a(x) n_a(x)\Big]\;, 
\eea
which in combination with Eq.~\eqref{eq:entropy_dens} 
leads to the well-known thermodynamic relation
\bea\label{eq:enthalpy}
\epsilon(x)+{p}(x)=T(x){s}(x)
+\sum\limits_a \mu_a(x) n_a(x)
\equiv h(x)\;,
\eea
where $h$ is the enthalpy density.
From Eqs.~\eqref{eq:thermodyn1} and \eqref{eq:enthalpy} 
we obtain the Gibbs--Duhem relation
\bea\label{eq:thermodyn2}
\delta p(x)=s(x)\delta T(x)+\sum\limits_a 
n_a(x)\delta\mu_a(x)\;.
\eea
Thus, using the relevant statistical operator, we constructed a full
set of thermodynamic variables (\ie, a fictitious local-equilibrium
state) for given energy-momentum tensor and charge current densities,
as it is required for the hydrodynamic description of the system.

\subsection{Deriving the non-equilibrium statistical operator}
\label{sec:deriv_rho}

As discussed in Sec.~\ref{sec:rel_dist}, the relevant 
statistical operator $\hat{\rho}_l$ defined by 
Eqs.~\eqref{eq:stat_op_eq} and \eqref{eq:Om_eq}
reproduces the local values of the macroscopic observables 
$u_\nu\langle\hat{T}^{\mu\nu}\rangle$ and 
$u_\mu\langle\hat{N}^{\mu}_a\rangle$. However, the 
operator $\hat{\rho}_l$ does not satisfy the Liouville 
equation~\eqref{eq:Liouville_eq_H}, and, therefore, cannot 
describe non-equilibrium thermodynamic processes. 

To proceed further we write the Liouville 
equation~\eqref{eq:Liouville_eq_H} in the form
\bea\label{eq:Liouville_eq_sig}
\frac{d\hat{\sigma}(t)}{dt}=0\;,
\eea
where $\hat{\sigma}(t)=-\ln \hat{\rho}(t)$.
Equation~\eqref{eq:Liouville_eq_sig} simply implies that
$\hat{\sigma}(t)$ is the statistical operator at some initial moment
in time $t'$. Following
Refs.~\cite{zubarev1974nonequilibrium,zubarev1997statistical,
  Robertson1966PhRv,Robertson1967PhRv,Zubarev1970Phy,Onyszkiew1987},
we take the initial condition for Eq.~\eqref{eq:Liouville_eq_sig} in
the form $\hat{\sigma}(t')=\hat{S}(t')$, \ie, the statistical operator
coincides with its local-equilibrium counterpart at the moment of time
$t'$ (note that the matching conditions~\eqref{eq:matching} are
automatically satisfied by this choice of the initial condition).
According to Eq.~\eqref{eq:Liouville_eq_sig}, $\hat{\sigma}$ remains
unchanged for all subsequent times, \ie,
\bea\label{eq:Liouv_sig}
\hat{\sigma}(t)=\hat{S}(t')\;,\quad t\geq t'\;.
\eea

Our next step is to modify the expression \eqref{eq:Liouv_sig} in such
a way that it can incorporate the irreversibility of the thermodynamic
processes. This can be achieved by averaging the right-hand side of
Eq.~\eqref{eq:Liouv_sig} over the initial states in some time interval
$t_0\leq t'\leq t$ around $t'$~\cite{zubarev1974nonequilibrium,
zubarev1997statistical}, \ie,
\bea\label{eq:Liouv_sig1}
\hat{\sigma}^*(t)&\equiv &\frac{1}{t-t_0}
\int_{t_0}^{t}\! dt'\hat{S}(t')=\frac{1}{t-t_0}
\int_{-(t-t_0)}^{0}\!\! dt'\hat{S}(t'+t)\;. 
\eea
Note that information is lost in the averaging process, 
therefore this procedure is irreversible. The time 
interval $t-t_0$ should be sufficiently large for the details of 
the initial state (correlations) to become inessential. Therefore, 
it is natural to take the limit $t_0\to -\infty$. Using Abel's 
theorem~[see Ref.\ \cite{zubarev1997statistical}, Eq.\ (2.3.9)]
\bea\label{eq:Abel}
\lim_{T\to \infty}\frac{1}{T}
\int_{-T}^0\! dt'f(t')=
\lim_{\varepsilon \to 0^+}\varepsilon\!
\int_{-\infty}^0\! dt'e^{\varepsilon t'}f(t')\;,
\eea
which is valid if the function $f(t)$ is sufficiently smooth 
and at least one of the above limits exists\footnote{Eq.~\eqref{eq:Abel} is
  a property of the Laplace transform which is frequently written as
 \bea
 \lim_{\tau \to \infty} \frac{1}{\tau}\int_0^{\tau} f(t)dt
 = \lim_{\varepsilon \to 0} \varepsilon \int_0^{\infty} f(t)
 e^{-\varepsilon t} dt\nonumber
 \eea
 and follows from the general theory of Tauber's
theorems~\cite{Ditkin61,Wiener32,Wiener26}.} we obtain for
Eq.~\eqref{eq:Liouv_sig1}
\bea\label{eq:Liouv_sig2}
\hat{\sigma}^*(t)=\lim_{\varepsilon \to 0^+}\,\varepsilon\!
\int_{-\infty}^0\! dt'e^{\varepsilon t'}\hat{S}(t'+t)
=\lim_{\varepsilon \to 0^+}\,\varepsilon\!
\int_{-\infty}^{t}\! dt'e^{\varepsilon (t'-t)}\hat{S}(t')\;.
\eea
It is easy to show that this operator satisfies the 
Liouville equation \eqref{eq:Liouville_eq_sig}. Let
\bea
\hat{\sigma}_\varepsilon(t) \equiv \varepsilon\!
\int_{-\infty}^{t}\! dt'e^{\varepsilon (t'-t)}\hat{S}(t')\;,
\eea
such that, according to Eq.~\eqref{eq:Liouv_sig2},
$\lim_{\varepsilon \rightarrow 0^+} \hat{\sigma}_\varepsilon(t)
= \hat{\sigma}^*(t)$. Then
\bea\label{eq:Liouv_test}
\frac{d\hat{\sigma}_\varepsilon(t) }{dt} =
\varepsilon\hat{S}(t)-\varepsilon^2\!\!\int_{-\infty}^t\!
dt'e^{\varepsilon (t'-t)} \hat{S}(t')=
-\varepsilon\!\left[\hat{\sigma}_\varepsilon(t)-\hat{S}(t)\right]\;,
\eea
with the right-hand side tending to zero as $\varepsilon\to 0$. 
According to Eq.~\eqref{eq:Liouv_test} the statistical operator
\bea\label{eq:Liouv_sig_pre}
\hat{\rho}_\varepsilon(t) \equiv\exp\left[-\hat{\sigma}_\varepsilon (t)\right]
\eea
satisfies the Liouville
equation up to a source term $\sim \varepsilon$.  The latter breaks the
reversibility of the Liouville equation by choosing its retarded
solution.  Hence we draw the important conclusion that the limit
$\varepsilon\to 0^+$ should be performed only after the thermodynamic
limit is taken. Due to this procedure, one maintains irreversibility
in the evolution of the system until the end of the calculations.
Thus, the statistical average of any operator $\hat{F}(x)$ should be
computed according to the following rule~\cite{zubarev1997statistical}
\bea\label{eq:op_av_full}
\langle\hat{F}(x)\rangle =\lim_{\varepsilon \to 0^+} 
\lim_{{V\to \infty}}
\Tr\big[\hat{\rho}_\varepsilon(t)\hat{F}(x)\big]\;,
\eea
where $V$ is the volume of the system. Furthermore, because the
statistical operator~\eqref{eq:Liouv_sig_pre} incorporates memory
effects, the equations of motion obtained from
Eq.~\eqref{eq:Liouv_sig_pre} are expected to be 
causal~\cite{Zubarev1972Phy,1998Morozov,Koide2007PhRvC,Koide2008PhRvE}.
Thus, we constructed a causal non-equilibrium statistical operator
starting from the relevant statistical operator.

Now we substitute the explicit expression for $\hat{S}(t)$ given by
Eq.~\eqref{eq:entropy_op} into Eq.~\eqref{eq:Liouv_sig_pre} to obtain
(omitting the index $\varepsilon$ for the sake of brevity)
\bea\label{eq:stat_op_non_eq}
\hat{\rho}(t) = Q^{-1}(t) \exp \left\{-\int\! d^3x\, 
\hat{Z}(\bm  x, t) \right\}\;,\quad
Q(t) = \Tr\exp\left\{-\int\! d^3x\, \hat{Z}(\bm  x, t)\right\}\;,
\eea
where 
\bea\label{eq:Z_op}
\hat{Z}(\bm x,t) = \varepsilon\! \int_{-\infty}^t\! dt_1  
e^{\varepsilon (t_1-t)}\Big[\beta^\nu(\bm  x, t_1)
\hat{T}_{0\nu}(\bm  x, t_1)-\sum \limits_a\alpha_a
(\bm  x, t_1) \hat{N}_{a0}(\bm  x, t_1)\Big].
\eea
In the case of one type of conserved charge ($\ell=1$)
Eqs.~\eqref{eq:stat_op_non_eq} and \eqref{eq:Z_op} coincide with
Eqs.~(42) -- (44) of Ref.~\cite{Huang2011AnPhy}, and in the case of
$\ell=0$ we recover Eqs.~(2.4) and (2.8) of Ref.~\cite{Hosoya1984AnPhy}.

We can separate the local-equilibrium contribution from
Eq.~\eqref{eq:Z_op} by an integration by parts:
\bea \label{eq:Z_integrate}
\hat{Z}(\bm  x,t)
 &=& \beta^\nu(\bm  x, t)\hat{T}_{0\nu}(\bm  x, t)
-\sum\limits_a\alpha_a(\bm x,t) \hat{N}_{a0}(\bm x, t)\nonumber\\
&-& \int_{-\infty}^t\! d t_1 e^{\varepsilon (t_1-t)} 
\frac{d}{dt_1}\Big[\beta^\nu(\bm  x, t_1)
\hat{T}_{0\nu}(\bm  x, t_1)-\sum\limits_a\alpha_a
(\bm  x, t_1) \hat{N}_{a0}(\bm  x, t_1)\Big],\no
\eea
where we assumed that the exponential factor $e^{\varepsilon (t_1-t)}$
guarantees the limiting behavior
\bea
\lim_{t_1\to -\infty}e^{\varepsilon (t_1-t)}\hat{F}(t_1)=0,\no
\eea
with the operator $\hat{F}(t_1)$ defined as the term in square brackets in 
Eq.~\eqref{eq:Z_op}.
The conservation laws \eqref{eq:cons_laws} imply the relations
$\partial^{\mu} \hat{T}_{\mu\nu} =
\partial^0 \hat{T}_{0\nu} + \partial^i \hat{T}_{i\nu} = 0$
and $\partial^{\mu} \hat{N}_{a\mu } = \partial^0 \hat{N}_{a0} +
\partial^i \hat{N}_{ai} = 0$, which give 
\bea \label{eq:part_int}
\partial^0\Big(\beta^\nu\hat{T}_{0\nu}-
\sum\limits_a\alpha_a\hat{N}_{a0}\Big)
= \hat{T}_{\mu\nu}\partial^{\mu} \beta^\nu-
\sum\limits_a\hat{N}^\mu_a\partial_\mu\alpha_a
- \partial^i\Big(\beta^\nu\hat{T}_{i\nu}-
\sum\limits_a\alpha_a\hat{N}_{ai}\Big).
\eea
The last term in Eq.~\eqref{eq:part_int} gives a surface term after 
integration over the spatial volume, which vanishes if the surface
is taken at infinity, and we are left with
\bea \label{eq:Z_final}
&&\int\! d^3x\, \hat{Z}(\bm x,t) = \int\! d^3x
\Big[\beta^\nu(\bm x,t)\hat{T}_{0\nu}(\bm x,t)-
\sum\limits_a\alpha_a(\bm x,t) \hat{N}_{a0}(\bm  x, t)\Big]\nonumber\\
&&- \int\! d^3x \! \int_{-\infty}^t \! dt_1 
e^{\varepsilon(t_1-t)}\Big[\hat{T}_{\mu\nu}(\bm  x,t_1)
\partial^{\mu}\beta^\nu(\bm  x,t_1)-
\sum\limits_a\hat{N}^\mu_a(\bm  x,t_1)\partial_\mu\alpha_a
(\bm  x,t_1)\Big],
\eea
where the 4-gradients are implied to act on $(\bm x, t_1)$.
The first term in this expression corresponds to the local-equilibrium
part of the statistical operator. The integrand of the second term is
a thermodynamic ``force" as it involves gradients of temperature,
chemical potentials, and the velocity field. Naturally, the second term in
Eq.~\eqref{eq:Z_final} is then identified with the non-equilibrium
part of the statistical operator.  Using
Eqs.~\eqref{eq:stat_op_non_eq} and \eqref{eq:Z_final} we can write the
full statistical operator as~\cite{Hosoya1984AnPhy,Huang2011AnPhy}
\be\label{eq:stat_op_full}
\hat{\rho}(t) = Q^{-1}e^{-\hat{A}+\hat{B}},
\qquad Q=\Tr e^{-\hat{A}+\hat{B}},
\ee
with
\bea\label{eq:A_op}
\hat{A}(t)&=&\int\! d^3x \Big[\beta^\nu(\bm x,t)
\hat{T}_{0\nu}(\bm x,t)-\sum\limits_a\alpha_a(\bm x,t) 
\hat{N}_{a0}(\bm  x, t)\Big],\\
\label{eq:B_op}
\hat{B}(t)&=& \int\! d^3x\! \int_{-\infty}^t\! 
dt_1 e^{\varepsilon(t_1-t)} \hat{C}(\bm  x,t_1),\\
\label{eq:C_op}
\hat{C}(\bm x,t)&=&\hat{T}_{\mu\nu}(\bm  x,t)
\partial^{\mu}\beta^\nu(\bm  x,t)-\sum\limits_a
\hat{N}^\mu_a(\bm  x,t)\partial_\mu\alpha_a(\bm  x,t).
\eea
The statistical operator given by Eq.~\eqref{eq:stat_op_full} can now
be used to derive the equations of motion for the dissipative
currents. For this purpose, one treats the non-equilibrium part
\eqref{eq:B_op} as a perturbation. Keeping only the first-order terms
in the Taylor expansion of $\hat{\rho}(t)$ with respect to the
operator $\hat{B}(t)$ yields the usual first-order dissipative
hydrodynamic theory~\cite{Hosoya1984AnPhy,Huang2011AnPhy}. In this work, we
will include all second-order terms in the Taylor expansion to
obtain second-order dissipative hydrodynamics within the Zubarev
formalism.

\subsection{Second-order expansion of the statistical operator}

We start from the following formula for two arbitrary 
operators $\hat{A}$ and $\hat{B}$~\cite{Roepke2013nonequilibrium},
\bea\label{eq:op_id_roepke}
e^{-\hat{A}+\hat{B}}e^{\hat{A}} = 1 + \int_0^1\! d\tau
 e^{\tau (-\hat{A}+\hat{B})}\hat{B}e^{\tau \hat{A}},
\eea
which can be obtained by
integrating the identity
\bea\label{eq:op_id_1}
\frac{d}{d\tau}e^{\tau(-\hat{A}+\hat{B})}e^{\tau \hat{A}}
= e^{\tau(-\hat{A}+\hat{B})}\hat{B}e^{\tau \hat{A}}\no
\eea
over the variable $\tau$ from 0 to 1. Keeping only the linear 
term in $\hat{B}$ under the integral we obtain from Eq.~\eqref{eq:op_id_roepke}
\bea\label{eq:op_1st_order}
e^{-\hat{A}+\hat{B}}=e^{-\hat{A}} + \int_0^1\! 
d\lambda e^{-\lambda \hat{A}} \hat{B}e^{\lambda 
\hat{A}}e^{-\hat{A}} +{\cal O}(\hat{B}^2).
\eea
Our next task is to derive the second-order term in $\hat{B}$ in the expansion 
of the non-equilibrium statistical operator given by Eq.~\eqref{eq:stat_op_full}. 
We start with the numerator and expand the
identity \eqref{eq:op_id_roepke} up to the second order. The integrand
in Eq.~\eqref{eq:op_id_roepke} is already of order 
${\cal O}(\hat{B})$, therefore it is sufficient to evaluate the 
operator $e^{\tau (-\hat{A}+\hat{B})}$ using the first-order expansion
\eqref{eq:op_1st_order}. Replacing $\hat{A}\to \tau \hat{A}$, 
$\hat{B}\to \tau \hat{B}$ in Eq.~\eqref{eq:op_1st_order} and 
inserting the result into Eq.~\eqref{eq:op_id_roepke} we find
\bea\label{eq:op_2nd_order3}
e^{-\hat{A}+\hat{B}} =(1 + \hat{\alpha}_1 
+ \hat{\alpha}_2)e^{-\hat{A}},
\eea
with
\bea\label{eq:alpha_12_new}
\hat{\alpha}_1=\int_0^1\! d\tau
\hat{B}_\tau,\qquad
\hat{\alpha}_2 =
\frac{1}{2}\int_0^1\! d\tau\! \int_0^1\!
d\lambda \,\tilde{T}\{\hat{B}_\lambda\hat{B}_\tau\},
\eea 
where we introduced the short-hand notation 
\bea\label{eq:B_tau}
\hat{X}_\tau = e^{-\tau A} \hat{X}e^{\tau A}
\eea
for any operator $\hat{X}$ and $\tilde{T}$ is the 
anti-chronological time-ordering operator with respect 
to the variables $\tau$ and $\lambda$.

In the next step, the trace $Q$ in Eq.~\eqref{eq:stat_op_full} 
is expanded up to second order in a Taylor series, giving
\bea\label{eq:invers_expand}
Q^{-1} = \frac{1}{\Tr e^{-\hat{A}+\hat{B}}}= \frac{1}{\Tr e^{-\hat{A}} }  
\left[1 - \langle \hat{\alpha}_1\rangle_l -\langle \hat{\alpha}_2
\rangle_l + \langle\hat{\alpha}_1\rangle_l^2 \right] ,
\eea
where we used the notation \eqref{eq:stat_av_l}. Substituting 
Eqs.~\eqref{eq:op_2nd_order3}, \eqref{eq:alpha_12_new}, and 
\eqref{eq:invers_expand} into Eq.~\eqref{eq:stat_op_full} 
and dropping higher-order terms we obtain
\bea  \label{eq:stat_full_2nd_order}
\hat{\rho} = \hat{\rho}_l+\hat{\rho}_1+\hat{\rho}_2,
\eea
with
\bea\label{eq:rho_1}
\hat{\rho}_1 &=& \int_0^1\! d\tau \left(\hat{B}_\tau  
- \big\langle \hat{B}_\tau\big\rangle_l\right)
\hat{\rho}_{l},\hspace{4.5cm}\\
 \label{eq:rho_2}
\hat{\rho}_2 &=&
\frac{1}{2}\int_0^1\!  d\tau\! \int_0^1\! d\lambda
\Big[\tilde{T}\{\hat{B}_\lambda\hat{B}_\tau\}-
\big\langle \tilde{T}\{\hat{B}_\lambda\hat{B}_\tau\} \big\rangle_l\no\\
  &&\hspace*{2.3cm} -
\hat{B}_\tau\big\langle \hat{B}_\lambda\big\rangle_l-
\hat{B}_\lambda\big\langle\hat{B}_\tau\big\rangle_l 
+2\big\langle \hat{B}_\tau\big\rangle_l\big\langle
\hat{B}_\lambda\big\rangle_l \Big]\hat{\rho}_{l}.
\eea
Equation \eqref{eq:rho_1} coincides with the result given in
Ref.~\cite{Huang2011AnPhy}.

In the last step, we substitute the explicit expression
for the operator $\hat{B}$ from Eq.~\eqref{eq:B_op},
which gives for the first-order correction
\bea\label{eq:rho_1_final}
\hat{\rho}_1 (t) =  \int\! d^4x_1\! \int_0^1\! d\tau 
\left[\hat{C}_\tau(x_1) - \big\langle \hat{C}_\tau(x_1)
\big\rangle_l\right]\hat{\rho}_{l},
\eea
and for the second-order correction
\bea\label{eq:rho_2_final}
\hat{\rho}_2 (t) 
= \frac{1}{2}\int\! d^4x_1d^4x_2\!
\int_0^1\!  d\tau\! \int_0^1\! d\lambda 
 \Big[\tilde{T} \{\hat{C}_\lambda(x_1)
\hat{C}_\tau(x_2)\} -\big\langle 
\tilde{T}\{ \hat{C}_\lambda(x_1)
\hat{C}_\tau(x_2)\}\big\rangle_l \nonumber\\
-
 \big\langle \hat{C}_\lambda(x_1)\big\rangle_l 
 \hat{C}_\tau(x_2) -\hat{C}_\lambda(x_1)
\big\langle \hat{C}_\tau( x_2)\big\rangle_l  
+ 2 \big\langle \hat{C}_\lambda( x_1)\big\rangle_l 
\big\langle \hat{C}_\tau(x_2)\big\rangle_l\Big]\hat{\rho}_{l},
\eea
where we introduced the abbreviation
\bea\label{eq:int_short}
\int\! d^4x_1 \equiv \int\! d^3x_1\! \int_{-\infty}^t\! 
dt_1 e^{\varepsilon(t_1-t)}.
\eea
Given the generic expansions above, we can now write
down the statistical average of an arbitrary operator 
$\hat{X}(x)$ with the help of
Eqs.~\eqref{eq:op_av}, \eqref{eq:stat_full_2nd_order},
\eqref{eq:rho_1_final}, and \eqref{eq:rho_2_final} as
\bea\label{eq:stat_average}
\langle \hat{X}(x)\rangle 
&=& \langle \hat{X}(x)\rangle_l +
\int\! d^4x_1
\Big(\hat{X}(x),\hat{C}(x_1)\Big) \nonumber\\
&+& \int\! d^4x_1d^4x_2
\Big(\hat{X}(x),\hat{C}(x_1),\hat{C}(x_2)\Big),
\eea
where we defined the two-point correlation function 
\bea\label{eq:2_point_corr}
\Big(\hat{X}(x),\hat{Y}(x_1)\Big) \equiv
\int_0^1\! d\tau\, \Big\langle\hat{X}(x)
\left[\hat{Y}_\tau(x_1)  - \big\langle 
\hat{Y}_\tau(x_1)\big\rangle_l\right]\Big\rangle_l,
\eea
and the three-point correlation function 
\bea\label{eq:3_point_corr}
\Big(\hat{X}(x),\hat{Y}(x_1),\hat{Z}(x_2)\Big) &\equiv&
\frac{1}{2} \int_0^1\!  d\tau\! \int_0^1\! d\lambda \,
\Big\langle \tilde{T} \left\{
\hat{X}(x)\Big[\hat{Y}_\lambda(x_1)\hat{Z}_\tau(x_2) \right. \nonumber\\
&-&
 \big\langle \hat{Y}_\lambda(x_1)\big\rangle_l  
\hat{Z}_\tau(x_2)-   \hat{Y}_\lambda( x_1) 
\big\langle\hat{Z}_\tau(x_2)\big\rangle_l \nonumber\\
&-& \left.
\big\langle\tilde{T}\,\hat{Y}_\lambda(x_1)\hat{Z}_\tau(x_2)\big\rangle_l+
2\big\langle \hat{Y}_\lambda( x_1)\big\rangle_l
\big\langle \hat{Z}_\tau(x_2)\big\rangle_l\Big]\right\}\Big\rangle_l.
\eea
From Eq.~\eqref{eq:3_point_corr} it is straightforward 
to find the symmetry relation
\bea\label{eq:3_point_corr_sym}
\int\! d^4x_1d^4x_2
\Big(\hat{X}(x),\hat{Y}(x_1),\hat{Z}(x_2)\Big) =
\int\! d^4x_1d^4x_2
\Big(\hat{X}(x),\hat{Z}(x_1),\hat{Y}(x_2)\Big),
\eea
which we will utilize below.

\subsection{Hydrodynamic equations}

To separate the dissipative processes related to the viscous and
diffusion currents we need to decompose the energy-momentum tensor and
the charge currents in terms of their equilibrium and dissipative parts. 
The most general decompositions are
\bea \label{eq:T_munu_decomp}
\hat{T}^{\mu\nu} &=& \hat{\epsilon} u^{\mu}u^{\nu}
- \hat{p}\Delta^{\mu\nu} + \hat{q}^{\mu}u^{\nu}
+ \hat{q}^{\nu}u^{\mu} + \hat{\pi}^{\mu\nu},\\
\label{eq:N_a_decomp}
\hat{N}^{\mu}_a &=& \hat{n}_au^\mu +\hat{j}^{\mu}_a,
\eea
where $\Delta^{\mu\nu}=g^{\mu\nu}-u^\mu u^\nu$ is the projection
operator onto the 3-space orthogonal to $u^\mu$. The shear-stress 
tensor $\hat{\pi}^{\mu\nu}$, the heat flux $\hat{q}^\mu$, and the 
diffusion currents $\hat{j}^\mu_a$ are orthogonal to $u_\mu$, and 
$\hat{\pi}^{\mu\nu}$ is traceless:
\bea \label{eq:orthogonality}
\quad
u_{\nu}\hat{q}^{\nu} = 0,\qquad
u_{\nu}\hat{j}^{\nu}_a = 0,\qquad 
u_{\nu}\hat{\pi}^{\mu\nu} = 0,\qquad 
\hat{\pi}_{\mu}^\mu=0.
\eea
Note that here we did not separate the equilibrium part of the
pressure from the bulk-viscous pressure. The statistical average of
the operator $\hat{p}$ gives the actual isotropic (non-equilibrium) 
pressure, which in general differs from the equilibrium pressure 
$p(\langle \hat{\epsilon} \rangle, \langle \hat{n}_a\rangle)$ 
introduced in Sec.~\ref{sec:therm_rel}. The latter is obtained 
by averaging the operator $\hat{p}$ over the local-equilibrium 
distribution (evaluated formally at constant values of the thermodynamic
parameters). The bulk-viscous pressure is defined as the difference
between these two averages, see Sec.~\ref{sec:diss1} for details.

The operators on the right-hand sides of Eqs.~\eqref{eq:T_munu_decomp}
and \eqref{eq:N_a_decomp} are given by projections of
$\hat{T}^{\mu\nu}$ and $\hat{N}_a^\mu$,
\bea\label{eq:proj1_op}
\hat{\epsilon} = u_\mu u_\nu \hat{T}^{\mu\nu},\qquad
\hat{n}_a = u_\mu\hat{N}^{\mu}_a,\qquad
\hat{p}=-\frac{1}{3}\Delta_{\mu\nu}
\hat{T}^{\mu\nu},\\
\label{eq:proj2_op}
\hat{\pi}^{\mu\nu} = \Delta_{\alpha\beta}^{\mu\nu} 
\hat{T}^{\alpha\beta},\qquad
\hat{q}^\mu  = u_\alpha\Delta_{\beta}^{\mu}\hat{T}^{\alpha\beta},\qquad
\hat{j}_a^{\nu}=\Delta_{\mu}^{\nu} \hat{N}^{\mu}_a,
\eea
where the following relations have been used
\bea\label{eq:prop_proj}
u_\mu\Delta^{\mu\nu}=\Delta^{\mu\nu}u_\nu=0,\qquad
\Delta^{\mu\nu}\Delta_{\nu\lambda}=\Delta^{\mu}_\lambda,
\qquad \Delta^{\mu}_{\mu}=3.
\eea
In Eq.~\eqref{eq:proj2_op} we also introduced the 
rank-4 traceless projector orthogonal to $u^\mu$ via
\bea\label{eq:projector_delta4}
\Delta_{\mu\nu\rho\sigma}= \frac{1}{2}
\left(\Delta_{\mu\rho}\Delta_{\nu\sigma}
+\Delta_{\mu\sigma}\Delta_{\nu\rho}\right)
-\frac{1}{3}\Delta_{\mu\nu}\Delta_{\rho\sigma},
\eea
which has the properties 
\bea\label{eq:prop_projector4_1} 
\Delta_{\mu\nu\rho\sigma}=\Delta_{\nu\mu\rho\sigma}
=\Delta_{\rho\sigma\mu\nu},\quad
u^\mu \Delta_{\mu\nu\rho\sigma}=0,\quad
\Delta_{\alpha}^{\mu} \Delta_{\mu\nu\rho\sigma}=
\Delta_{\alpha\nu\rho\sigma},\\
\label{eq:prop_projector4_2}
\Delta_{\mu~\rho\sigma}^{~\mu}=0,\quad
\Delta_{\nu\mu~\sigma}^{\quad\mu}=
\frac{5}{3}\Delta_{\nu\sigma},\quad
\Delta_{\mu\nu}^{\quad\mu\nu}=5,\quad
\Delta_{\mu\nu\rho\sigma}\Delta^{\rho\sigma}_{\alpha\beta}
=\Delta_{\mu\nu\alpha\beta}.
\eea
In the local rest frame, one finds from
Eqs.~\eqref{eq:proj1_op} and \eqref{eq:proj2_op}
\bea \label{eq:currents_rest1_op}
\hat{\epsilon} = \hat{T}^{00},\qquad
\hat{n}_a = \hat{N}^{0}_a,\qquad
\hat{p} =-\frac{1}{3}\hat{T}^k_k,\hspace{1.5cm}\\
\label{eq:currents_rest2_op}
\hat{\pi}_{kl} = \left(\delta_{ki}\delta_{lj}-
\frac{1}{3}\delta_{kl}\delta_{ij}
\right) \hat{T}_{ij},\qquad
\hat{q}^i  = \hat{T}^{0i},\qquad
\hat{j}_a^{i} = \hat{N}^{i}_a.
\eea

Averaging Eqs.~\eqref{eq:T_munu_decomp} and \eqref{eq:N_a_decomp} 
over the non-equilibrium statistical operator and substituting them 
into Eq.~\eqref{eq:cons_laws} leads to the equations of 
{\it dissipative hydrodynamics},
\bea \label{eq:hydro1}
Dn_a +n_a\theta +\partial_\mu j^{\mu}_a &=& 0,\\
\label{eq:hydro2}
D\epsilon + (h+\Pi)\theta +\partial_\mu q^{\mu}-
q^{\mu}D u_\mu -\pi^{\mu\nu}\sigma_{\mu\nu} &=& 0,\\
\label{eq:hydro3}
(h+\Pi) D u_{\alpha}- \nabla_\alpha (p+\Pi)
+ \Delta_{\alpha\mu}D q^{\mu}+
q^{\mu}\partial_\mu u_{\alpha} +q_{\alpha}\theta 
+\Delta_{\alpha\nu}\partial_\mu \pi^{\mu\nu} &=&0,
\eea
where $\epsilon \equiv \langle \hat{\epsilon} \rangle$, 
$n_a \equiv \langle \hat{n}_a \rangle$, $q^\mu \equiv \langle \hat{q}^\mu \rangle$, 
$\pi^{\mu\nu}\equiv \langle \hat{\pi}^{\mu \nu} \rangle$, and $j_a^{\mu}\equiv
\langle \hat{j}_a^\mu \rangle$ 
are the statistical averages of the corresponding operators;
$p\equiv p(\epsilon,n_a)$ is the pressure in local equilibrium, 
\ie, the pressure given by the EoS, whereas $\Pi$ is
the non-equilibrium part of the pressure (see Sec.~\ref{sec:diss1}
for details); $D\equiv u^\mu\partial_\mu$ is the comoving derivative (equal to the
time derivative in the local rest frame),
$\nabla_\alpha\equiv \Delta_{\alpha\beta}\partial^\beta$ is the covariant
spatial derivative, $\sigma_{\mu\nu}\equiv \Delta_{\mu\nu}^{\alpha\beta}
\partial_\alpha u_\beta$ is the shear tensor, and 
$\theta \equiv \partial_\mu u^\mu$ is the expansion scalar.
The latter quantity is a measure for the rate of the fluid expansion 
(for $\theta > 0$) or contraction (for $\theta < 0$) and is therefore also
called the {\it fluid expansion rate}. Equations~\eqref{eq:hydro2} 
and \eqref{eq:hydro3} are obtained by contracting the 
first equation~\eqref{eq:cons_laws} by $u_\nu$ and $\Delta_{\nu\alpha}$, respectively.
The system of Eqs.~\eqref{eq:hydro1} -- \eqref{eq:hydro3} contains 
$\ell+4$ equations, whereas the number of independent variables is $4\ell+10$. 
In order to close the system we need additional equations of motion. In the L-frame,
where $q^\mu =0$, these are 
$3\ell$ equations for the independent components of the
diffusion currents, 5 equations for
the independent components of the shear-stress tensor,
and one equation for the bulk-viscous pressure (recall that the equilibrium pressure 
is fixed by the EoS). In the E-frame, one of the diffusion currents can 
be eliminated, but then the determination of the independent components of the 
heat flux requires 3 additional equations of motion.

The averages of the dissipative operators over the local-equilibrium 
distribution vanish~\cite{Zubarev1979TMP}:
\bea\label{eq:diss_currents_av_eq}
\langle \hat{q}^{\mu}\rangle_l =0,\qquad
\langle \hat{j}^{\mu}_a\rangle_l =0,\qquad
\langle \hat{\pi}^{\mu\nu}\rangle_l =0.
\eea
Indeed, the relevant distribution given by
Eqs.~\eqref{eq:stat_op_eq_rest} and \eqref{eq:Om_eq_rest} depends only
on the scalar operators $\hat{\epsilon}$ and $\hat{n}_a$, which are
not correlated with vector and tensor quantities due to Curie's
theorem~\cite{Zubarev1979TMP,Hosoya1984AnPhy,Groot1963PhT,Groot1963AmJPh}.
As a result, averaging Eqs.~\eqref{eq:T_munu_decomp} and
\eqref{eq:N_a_decomp} over the relevant distribution and substitution
into Eqs.~\eqref{eq:cons_laws} leads to the equations of {\it ideal
  hydrodynamics}, namely
\bea \label{eq:ideal_hydro2}
Dn_a+ n_a\theta =0,\qquad
D\epsilon+ h \theta =0,\qquad
h D u_\alpha = \nabla_\alpha p.
\eea
The first two equation in Eq.~\eqref{eq:ideal_hydro2} are
the covariant forms of the charge- and the energy-conservation laws,
respectively. The third equation is the relativistic Euler
equation. We see that the rest-mass density is replaced here by the
enthalpy density $h$, which therefore is the appropriate measure of
inertia for relativistic fluids.  To include dissipative phenomena,
one needs to take into account the deviation of the statistical
operator from its local-equilibrium form.

\section{First-order dissipative hydrodynamics}
\label{sec:trans_first}

The expansion of the non-equilibrium statistical operator obtained in
the previous section allows us to derive hydrodynamics of a
dissipative fluid order by order. We have seen that at zeroth order
this expansion leads to ideal hydrodynamics given by
Eq.~\eqref{eq:ideal_hydro2}. In this section, we review the
derivation of the relativistic Navier--Stokes equations by exploiting the
first-order expansion of the statistical operator in the thermodynamic
forces~\cite{Zubarev1979TMP,Hosoya1984AnPhy,Huang2011AnPhy,Hayata2015}.
This procedure will allow us to obtain Kubo formulas for the transport
coefficients entering first-order dissipative hydrodynamics.

\subsection{Decomposition into different dissipative processes}
\label{sec:C_decomp1}

For our further computations it is convenient to decompose the
operator $\hat{C}$ given by Eq.~\eqref{eq:C_op} into the different
dissipative quantities entering Eqs.~\eqref{eq:T_munu_decomp} and
\eqref{eq:N_a_decomp}.  Similar decompositions were performed in
Refs.~\cite{Zubarev1979TMP,Hosoya1984AnPhy,Huang2011AnPhy}.
Using Eqs. \eqref{eq:orthogonality} and \eqref{eq:prop_proj},
as well as $\beta^\nu \equiv \beta u^\nu$, we obtain
\bea\label{eq:op_C_decompose1}
\hat{C}
&=&\hat{\epsilon} D\beta - \hat{p}
\beta\theta -\sum\limits_a\hat{n}_aD\alpha_a
+ \hat{q}^{\sigma}(\beta Du_{\sigma}+\partial_{\sigma}\beta)-
\sum\limits_a\hat{j}^{\sigma}_a \partial_\sigma\alpha_a
+ \beta\hat{\pi}_{\rho\sigma}\partial^{\rho}u^{\sigma}.
\eea
The first three terms correspond to the scalar, the next two terms
to the vector, and the last term to the tensor dissipative
processes. To lowest order in dissipative currents and gradients,
the terms $D\beta$, $D\alpha_a$, and $Du^\sigma$ can be eliminated 
using the equations of ideal hydrodynamics~\eqref{eq:ideal_hydro2}.  
(Higher-order corrections from the fully dissipative hydrodynamic 
equations \eqref{eq:hydro1} -- \eqref{eq:hydro3} are necessary, 
and will be taken into account, in the derivation of second-order 
dissipative hydrodynamics, see~Sec.~\ref{sec:C_decomp2}.) 
Choosing $\epsilon$ and $n_a$ as independent thermodynamic variables 
and using the first two equations \eqref{eq:ideal_hydro2} we find
\bea\label{eq:D_beta}
D\beta &=& \left.\frac{\partial\beta}
{\partial\epsilon}\right|_{n_a}D\epsilon +
\sum\limits_a
\left.\frac{\partial\beta}{\partial n_a}
\right|_{\epsilon, n_b\neq n_a}\!\!\!Dn_a= 
-\theta \left( h \left. \frac{\partial\beta}
{\partial\epsilon}\right|_{n_a} +\sum\limits_a
n_a \left.\frac{\partial\beta}{\partial n_a}
\right|_{\epsilon, n_b\neq n_a}\right),\\
\label{eq:D_alpha}
D\alpha_c &=& \left.\frac{\partial\alpha_c}
{\partial\epsilon}\right|_{n_a}D\epsilon +
\sum\limits_a
\left.\frac{\partial\alpha_c}{\partial n_a}
\right|_{\epsilon, n_b\neq n_a}\!\!\!Dn_a=
 -\theta \left( h \left.\frac{\partial\alpha_c}
{\partial\epsilon}\right|_{n_a} +\sum\limits_a
n_a \left.\frac{\partial\alpha_c}{\partial n_a}
\right|_{\epsilon, n_b\neq n_a}\right).
\eea
For the sake of convenience we write the thermodynamic relations 
\eqref{eq:thermodyn1}, \eqref{eq:enthalpy}, and 
\eqref{eq:thermodyn2} in the following form
\bea\label{eq:thermodyn3}
ds=\beta d\epsilon -\sum\limits_a\alpha_a dn_a,
\qquad \beta dp  =-hd\beta +\sum\limits_a n_ad\alpha_a.
\eea
We obtain from the first equation the set of Maxwell relations
\bea\label{eq:rel_alpha_beta}
\left.\frac{\partial \beta}{\partial n_a}
\right|_{\epsilon,n_b\neq n_a}=-
\left.\frac{\partial \alpha_a}{\partial\epsilon}
\right|_{n_b},\qquad
\left.\frac{\partial \alpha_c}{\partial n_a}
\right|_{\epsilon, n_b\neq n_a}=
\left.\frac{\partial \alpha_a}{\partial n_c}
\right|_{\epsilon, n_b\neq n_c},
\eea
and from the second equation we immediately read off
\bea\label{eq:h_n_a}
h=-\beta\left.\frac{\partial p}{\partial\beta}
\right|_{\alpha_a},\qquad
n_a=\beta\left.\frac{\partial p}{\partial\alpha_a}
\right|_{\beta,\alpha_b\neq \alpha_a}.
\eea
Substituting Eqs.~\eqref{eq:rel_alpha_beta} and \eqref{eq:h_n_a} into
Eqs.~\eqref{eq:D_beta} and \eqref{eq:D_alpha} we obtain
\bea\label{eq:D_beta1}
D\beta 
&=& \beta\theta \left( \left.\frac{\partial p}{\partial\beta}
\right|_{\alpha_a} \left.\frac{\partial\beta}{\partial\epsilon}\right|_{n_a} +
\sum\limits_a
\left.\frac{\partial p}{\partial\alpha_a}
\right|_{\beta,\alpha_b\neq\alpha_a}
\left.\frac{\partial \alpha_a}{\partial\epsilon}\right|_{n_b} \right) \equiv
\beta\theta \gamma,\\
\label{eq:D_alpha1}
D\alpha_c 
&=& -\beta\theta \left( \left.\frac{\partial p}{\partial\beta}
\right|_{\alpha_a} \left.\frac{\partial \beta}{\partial n_c}
\right|_{\epsilon,n_b\neq n_c}+\sum\limits_a
\left.\frac{\partial p}{\partial\alpha_a}
\right|_{\beta,\alpha_b\neq \alpha_a}
\left.\frac{\partial \alpha_a}{\partial n_c}
\right|_{\epsilon, n_b\neq n_c}\right) \equiv -\beta\theta\delta_c,
\eea
where
\bea\label{eq:gamma_delta_a}
\gamma \equiv \left.\frac{\partial p}{\partial\epsilon}
\right|_{n_a},\qquad
\delta_a \equiv \left.\frac{\partial p}
{\partial n_a} \right|_{\epsilon, n_b\neq n_a}.
\eea
The first three terms in Eq.~\eqref{eq:op_C_decompose1}
can then be combined as follows,
\bea\label{eq:scalar_part}
\hat{\epsilon} D\beta- \hat{p}
\beta\theta -\sum\limits_a\hat{n}_a D\alpha_a =
- \beta\theta \hat{p}^*,
\eea
where
\bea\label{eq:p_star}
\hat{p}^* 
= \hat{p} - \gamma\hat{\epsilon}
-\sum\limits_a\delta_a \hat{n}_a.
\eea
Note that the coefficient $\gamma$ coincides with the square of 
the speed of sound $c_s$ in the case where conserved charges are
absent. However, in general $\gamma\neq c_s^2$. (For a single conserved 
charge $n$, we have $c_s^2 = \partial p/\partial \epsilon|_{s/n}$.)

Using the second relation~\eqref{eq:thermodyn3} and
the third equation~\eqref{eq:ideal_hydro2} we obtain
\bea\label{eq:q_mod}
T\nabla_\sigma\beta +Du_\sigma =
T\sum\limits_a\frac{n_a}{h}\nabla_\sigma \alpha_a.
\eea
Using Eq.~\eqref{eq:q_mod}, as well as
$\hat{q}^{\sigma}\nabla_\sigma =\hat{q}^{\sigma}\partial_\sigma$ and
$\hat{j}_a^{\sigma}\nabla_\sigma =\hat{j}_a^{\sigma}\partial_\sigma$, which 
is a consequence of Eq.~\eqref{eq:orthogonality}, we modify the vector
terms in Eq.~\eqref{eq:op_C_decompose1} as follows,
\bea\label{eq:vector_part}
\hat{q}^{\sigma}(\beta Du_{\sigma}+
\partial_{\sigma}\beta)=
\sum\limits_a\frac{n_a}{h}
\hat{q}^{\sigma}\nabla_\sigma \alpha_a,\qquad
\hat{j}^{\sigma}_a\partial_\sigma\alpha_a
=\hat{j}^{\sigma}_a\nabla_\sigma\alpha_a.
\eea
Finally, using the properties of the shear-stress tensor
$\hat{\pi}_{\rho\sigma}$ we can replace $\partial^{\rho}u^{\sigma}\to
\sigma^{\rho\sigma}=\Delta^{\rho\sigma}_{\mu\nu}\partial^{\mu}u^{\nu}$
in the last term in Eq.~\eqref{eq:op_C_decompose1}.  
Now combining Eqs.~\eqref{eq:op_C_decompose1},
\eqref{eq:scalar_part}, and \eqref{eq:vector_part}
we obtain the final form of the operator
$\hat{C}$ to first order in gradients
\bea \label{eq:C_decompose}
\hat{C} = - \beta \theta \hat{p}^*+\beta
\hat{\pi}_{\rho\sigma}\sigma^{\rho\sigma}-
\sum\limits_a\mathscr{\hat{J}}^\sigma_a\nabla_\sigma\alpha_a,
\eea
where 
\bea\label{eq:diff_currents}
\mathscr{\hat{J}}_a^\sigma =\hat{j}^{\sigma}_a
-\frac{n_a}{h}\hat{q}^{\sigma}
\eea
are the charge-diffusion currents with respect to the
energy current, as shown in \ref{app:frames}.

As seen from Eq.~\eqref{eq:C_decompose}, the operator $\hat{C}$
depends linearly on the thermodynamic forces $\theta$,
$\sigma^{\rho\sigma}$, and $\nabla_\sigma\alpha_a$ which correspond to
the bulk-viscous, the shear-viscous, and the flavor-diffusion effects,
respectively.  We will proceed below to obtain the linear
Navier--Stokes relations between these thermodynamic forces and the
dissipative currents.

\subsection{Computing the dissipative quantities}
\label{sec:diss1}

According to Curie's theorem, in an isotropic medium the correlations
between operators of different rank
vanish~\cite{Groot1963PhT,Groot1963AmJPh}. Using this fact,
we obtain from Eqs.~\eqref{eq:stat_average} and
\eqref{eq:C_decompose} for the shear-stress tensor to leading order
\bea \label{eq:stress_av1}
\langle \hat{\pi}_{\mu\nu}(x)\rangle_1
=\int\! d^4x_1
\Big(\hat{\pi}_{\mu\nu}(x),\hat{\pi}_{\rho\sigma}
(x_1)\Big) \beta(x_1) \sigma^{\rho\sigma}(x_1).
\eea
The main contribution to the integrand in Eq.~\eqref{eq:stress_av1}
comes from the range $|x_1-x|\lesssim \lambda$, where $\lambda$ is a
typical {\it microscopic} length scale over which the shear-stress
correlation function decays. This correlation length characterizes the
range of the interaction and is thus proportional to the mean free
path between particle scatterings. On the other hand, in the
hydrodynamic regime, the thermodynamic parameters and the fluid
velocity vary over a {\it macroscopic} length scale $L \gg \lambda$.
The ratio Kn$= \lambda/L \ll 1$ is commonly called the Knudsen number.
Corrections to ideal hydrodynamics can be formally arranged in a power
series in Kn. First-order dissipative hydrodynamics takes into account
terms of linear order in Kn.

Since the thermodynamic force $\beta \sigma^{\rho \sigma}$ varies 
over the scale $L$, while the integrand in Eq.~\eqref{eq:stress_av1} 
contributes only over a range $\sim \lambda$, we may take 
$\beta \sigma^{\rho\sigma}$ as a constant and factor it out from
the integral. In doing so, we take $\beta \sigma^{\rho \sigma}$ at 
the space-time point $x$~\cite{Zubarev1979TMP,Hosoya1984AnPhy,Huang2011AnPhy}, 
in accordance with the mean-value theorem of integral calculus.
This leads to a local, linear relation between the shear-stress 
tensor and the shear tensor
\bea \label{eq:stress_av2}
\pi_{\mu \nu} (x) \equiv \langle \hat{\pi}_{\mu\nu}(x)\rangle_1
=\beta (x) \sigma^{\rho\sigma}(x) \int\! d^4x_1
\Big(\hat{\pi}_{\mu\nu}(x),\hat{\pi}_{\rho\sigma}
(x_1)\Big).
\eea
Since the effective range of integration in Eq.~\eqref{eq:stress_av1} 
is of order $\lambda$ and since $|\sigma^{\rho\sigma}|\simeq |u^{\rho}|/L$,
we observe that the right-hand side of Eq.~\eqref{eq:stress_av2} is
of order $\lambda/L =$ Kn, \ie, of order one in the Knudsen number.
However, as we will show in the next section, the non-locality 
of the thermodynamic forces is crucial for the derivation
of causal equations of motion for the dissipative 
currents~\cite{Zubarev1972Phy,1998Morozov,Koide2007PhRvC,Koide2008PhRvE}.
Therefore, we need to go beyond this approximation to maintain the
causality of the theory.

Next we turn to the bulk-viscous pressure $\Pi$, which is defined 
as the deviation of the actual isotropic pressure $\langle\hat{p}
\rangle=\langle\hat{p}\rangle_l +\langle\hat{p}\rangle_1$ from its 
equilibrium value $p(\epsilon,n_a)$ given by the EoS as a result 
of fluid expansion or compression
\bea\label{eq:bulk}
\Pi = \langle\hat{p}\rangle-{p}(\epsilon,n_a)=
\langle\hat{p}\rangle_l +\langle\hat{p}\rangle_1
-{p}(\epsilon,n_a).
\eea
We have also $\epsilon=\langle\hat{\epsilon}\rangle_l 
+\langle\hat{\epsilon}\rangle_1$, $n_a=\langle\hat{n}_a
\rangle_l +\langle\hat{n}_a \rangle_1$, therefore at 
first order in gradients we have 
\bea\label{eq:pressure_expand}
\langle\hat{p}\rangle_l \equiv
{p}\big(\langle\hat{\epsilon}\rangle_l,
\langle\hat{n}_a\rangle_l\big)=
{p}\big({\epsilon}-\langle\hat{\epsilon}\rangle_1,{n}_a-
\langle\hat{n}_a \rangle_1\big)= p({\epsilon},{n}_a)
-\gamma\langle\hat{\epsilon}\rangle_1-
\sum\limits_a\delta_a\langle\hat{n}_a\rangle_1,
\eea
where the coefficients $\gamma$, $\delta_a$ are defined in
Eq.~\eqref{eq:gamma_delta_a}. 
Note that the corrections $\langle\hat{\epsilon}\rangle_1$ 
and $\langle\hat{n}_a\rangle_1$ vanish if 
the matching conditions~\eqref{eq:matching} are imposed. We prefer to keep 
them for the sake of generality, so that the final 
expressions will be independent of the choice of the matching conditions.
 Substituting Eq.~\eqref{eq:pressure_expand} in 
 Eq.~\eqref{eq:bulk} for the bulk-viscous pressure we obtain
\bea\label{eq:bulk_1}
\Pi = \langle\hat{p}-\gamma \hat{\epsilon}-\sum\limits_a
\delta_a \hat{n}_a\rangle_1 =\langle\hat{p}^*\rangle_1,
\eea
where we used the definition \eqref{eq:p_star} of $\hat{p}^*$. 
From Eqs.~\eqref{eq:stat_average} and \eqref{eq:C_decompose} 
we now obtain the first-order correction to the bulk viscous
pressure (applying similar arguments as in the derivation of 
Eq.~\eqref{eq:stress_av2})
\bea\label{eq:pressure_av} 
\Pi (x) = \langle \hat{p}^*(x)\rangle_1 = -\beta(x) \theta(x) \!
\int\! d^4x_1 \Big(\hat{p}^* (x),\hat{p}^*(x_1)\Big).  
\eea
Thus, the bulk-viscous pressure is expressed in terms of a
symmetric correlator between two $\hat{p}^*$ operators.

Finally, again using Curie's theorem, 
we find for the charge-diffusion currents
\bea\label{eq:charge_currents}
\mathscr{J}_a^\mu (x) \equiv  \langle\mathscr{\hat{J}}_a^{\mu}(x)\rangle_1 =
 - \sum\limits_b \left[ \nabla_\sigma\alpha_b(x)\right] 
\int\! d^4 x_1
\left(\mathscr{\hat{J}}_a^{\mu}(x),
\mathscr{\hat{J}}_b^{\sigma}(x_1)\right),
\eea
which are expressed in terms of symmetric correlators as well.

\subsection{Transport coefficients for an isotropic medium}
\label{sec:trans}

The isotropy of the medium together with the conditions
\eqref{eq:orthogonality} further implies~\cite{Hosoya1984AnPhy}
\bea \label{eq:corr1_current}
\Big(\mathscr{\hat{J}}_a^{\mu}( x),\mathscr{\hat{J}}_b^{\nu}(x_1)\Big)& = & 
\frac{1}{3}\Delta^{\mu\nu}(x)\Big(\mathscr{\hat{J}}_a^{\lambda}(x),
\mathscr{\hat{J}}_{b\lambda}(x_1)\Big),\\
\label{eq:corr1_stress}
\Big(\hat{\pi}_{\mu\nu}(x),
\hat{\pi}_{\rho\sigma}( x_1)\Big)
 & =& \frac{1}{5} \Delta_{\mu\nu\rho\sigma}(x)
 \Big(\hat{\pi}^{\lambda\eta}(x),
\hat{\pi}_{\lambda\eta}( x_1)\Big).
\eea
Defining the shear and the bulk viscosities as 
\bea \label{eq:shear_def}
\eta(x) &\equiv & \frac{\beta(x)}{10}\!\int\! d^4x_1
\Big(\hat{\pi}_{\mu\nu}(x),\hat{\pi}^{\mu\nu}(x_1)\Big),\\
 \label{eq:bulk_def}
\zeta (x)&\equiv& \beta(x)\!\int\! d^4x_1
\Big(\hat{p}^*(x), \hat{p}^*(x_1)\Big),
\eea
we obtain from Eqs.~\eqref{eq:stress_av2}, 
\eqref{eq:pressure_av}, and \eqref{eq:corr1_stress} 
\bea \label{eq:shear_bulk_1} 
{\pi}_{\mu\nu} = 2\eta
\sigma_{\mu\nu},
\qquad \Pi =-\zeta\theta.
\eea
To write down $\eta(x)$ in the local rest frame, we use the first
relation in Eq.~\eqref{eq:proj2_op} in Eq.~\eqref{eq:shear_def} and
make the transition (indicated by the arrow)
\bea\label{eq:eta_rest}
\eta(x) = \frac{\beta(x)}{10} \Delta_{\alpha\beta\gamma\delta}(x)\!
\int\! d^4x_1\Big(\hat{T}^{\alpha\beta}(x),\hat{T}^{\gamma\delta}(x_1)\Big)
 \longrightarrow \frac{\beta(x)}{10}\! \int\! d^4x_1
\Big(\hat{\pi}_{ij}(x),\hat{\pi}^{ij}(x_1)\Big),
\eea
where the $\Delta(x_1)$ projector is taken out of the integral at the
value $x$ (which is correct up to the required accuracy) and 
\bea\label{eq:delta4_rest}
\Delta_{\alpha\beta\gamma\delta}(x)\longrightarrow
\Delta_{ijkl}=\frac{\delta_{ik}\delta_{jl}
+\delta_{il}\delta_{jk}}{2}-\frac{1}{3}
\delta_{ij}\delta_{kl}\no
\eea
in the local rest frame.

The two-point correlators in Eqs.~\eqref{eq:shear_def} and
\eqref{eq:bulk_def} should be evaluated at constant values of
the thermodynamic parameters, which, nevertheless, can change 
from space-time point to space-time point.

Using Eqs.~\eqref{eq:charge_currents} and 
\eqref{eq:corr1_current} we obtain for the diffusion currents
\bea\label{eq:charge_currents1}
\mathscr{J}_a^\mu
= \sum\limits_b\chi_{ab}\nabla^\mu \alpha_b,\quad
\eea
where we defined the matrix of diffusion coefficients
\bea \label{eq:chi_ab_def}
\chi_{ab} (x) = -\frac{1}{3}\int\! d^4x_1\!
\left(\mathscr{\hat{J}}_a^{\lambda}
(x),\mathscr{\hat{J}}_{b\lambda}(x_1)\right).
\eea 
In the case of a single conserved charge species we define the 
heat-flux operator as (see \ref{app:frames} for details)
\bea\label{eq:op_heat_flux}
\hat{h}^\mu= \hat{q}^{\mu}
-\frac{h}{n}\hat{j}^{\mu}=-\frac{h}{n}\mathscr{\hat{J}}^{\mu},
\eea
the averaging of which gives
\bea\label{eq:heat_current1}
h^\mu = -\kappa\frac{nT^2}{h}\nabla^\mu \alpha,
\eea
where the thermal conductivity is defined as
\bea\label{eq:kappa_def}
\kappa(x) = -\frac{\beta^2(x)}{3}\!\int\! d^4x_1\!
\left(\hat{h}^{\lambda}
(x),\hat{h}_{\lambda}(x_1)\right).
\eea

Equations~\eqref{eq:shear_bulk_1}, \eqref{eq:charge_currents1}, and
\eqref{eq:heat_current1} establish the required linear relations
between the dissipative currents and the thermodynamic forces. 
We thus conclude that the non-equilibrium statistical operator correctly 
reproduces the Navier--Stokes limit of relativistic dissipative hydrodynamics.

The shear viscosity, the bulk viscosity, and the thermal conductivity
given by Eqs.~\eqref{eq:shear_def}, \eqref{eq:bulk_def}, and
\eqref{eq:kappa_def}, respectively, are positive quantities, and the
matrix $\chi_{ab}$ is positive semidefinite in flavor space due to
the fact that all diffusion currents are spatial. These properties
guarantee the entropy increase in dissipative fluids, see
\ref{app:H-theorem}.

As shown in ~\ref{app:Green_func}, the transport coefficients
defined in Eqs.~\eqref{eq:shear_def}, \eqref{eq:bulk_def},
\eqref{eq:chi_ab_def}, and  \eqref{eq:kappa_def} can be expressed via
two-point retarded Green's functions as follows
\bea \label{eq:shear_bulk_mod}
\eta = -\frac{1}{10}\frac{d}{d\omega} {\rm Im}
G^R_{\hat{\pi}_{\mu\nu}\hat{\pi}^{\mu\nu}}(\omega)
\bigg\vert_{\omega=0},\qquad
\zeta = -\frac{d}{d\omega} {\rm Im}G^R_{\hat{p}^{*}
\hat{p}^*}(\omega)\bigg\vert_{\omega=0},\\
\label{eq:kappa_mod}
\chi_{ab} = \frac{T}{3}\frac{d}{d\omega}
{\rm Im}G^R_{\hat{\mathscr{J}}_a^{\lambda}
\hat{\mathscr{J}}_{b\lambda}}(\omega)
\bigg\vert_{\omega=0},\qquad
\kappa = \frac{1}{3T}\frac{d}{d\omega} {\rm Im}
G^R_{\hat{h}^{\lambda}\hat{h}_{\lambda}}(\omega)
\bigg\vert_{\omega=0},
\eea
where 
\bea\label{eq:green_func}
G^R_{\hat{X}\hat{Y}}(\omega) 
= -i\!\int_{0}^{\infty}\!\! dt e^{i\omega t}\!\!
\int\! d^3x\,\big\langle\big[\hat{X}(\bm x, t),
\hat{Y}(\bm 0,0)\big]\big\rangle_l
\eea
is the Fourier transform of the retarded two-point correlator taken in
the zero-wavenumber limit and the square brackets denote the
commutator. Some of these relations were obtained within the Zubarev
formalism in
Refs.~\cite{Hosoya1984AnPhy,Huang2011AnPhy,Horsley1987NuPhB}.
The relations \eqref{eq:shear_bulk_mod} and \eqref{eq:kappa_mod} are known as
the Kubo formulas for the transport
coefficients~\cite{Kubo1957,Kubo1957_2}.

\section{Second-order dissipative hydrodynamics}
\label{sec:trans_second}

In this section, we systematically compute all second-order corrections
to the dissipative currents.  Along with the thermodynamic
forces these currents are regarded as {\it first-order} quantities in deviations from
equilibrium. The second-order terms are those which involve 
either space-time derivatives of the dissipative currents or products of 
the dissipative currents with the thermodynamic forces. It is
easy to see that such second-order contributions arise not only from the
three-point correlators in Eq.~\eqref{eq:stat_average}, which are quadratic
in the operator $\hat{C}$, but also from the two-point
correlators, where one should take into
account the second-order corrections to the operator $\hat{C}$, which
were neglected in the derivation of Eq.~\eqref{eq:C_decompose}. Apart
from that, additional second-order gradient terms arise from the
non-locality of the thermodynamic forces in the two-point
correlators, \ie, when the difference between the
space-time arguments $x$ and $x_1$ in the dissipative
currents and the thermodynamic forces are taken into account. As we will
show below in Secs.~\ref{sec:2nd_shear} -- \ref{sec:2nd_diff}, these
non-local effects generate relaxation terms in the transport equations
which are required to maintain causality.

\subsection{Decomposing the thermodynamic forces up to second order}
\label{sec:C_decomp2}

In this subsection, we repeat the decomposition of the operator
$\hat{C}$, keeping all second-order corrections that were neglected in
Sec.~\ref{sec:C_decomp1}.  Our starting point is
Eq.~\eqref{eq:op_C_decompose1}, which on account of Eq.~\eqref{eq:orthogonality}
we can write in the form
\bea\label{eq:C_decomp_2nd}
\hat{C} =\hat{\epsilon} D\beta - \hat{p}
\beta\theta -\sum\limits_a\hat{n}_aD\alpha_a 
+ \hat{q}^{\sigma}(\beta Du_{\sigma}+\nabla_{\sigma}\beta)
-\sum\limits_a\hat{j}^{\sigma}_a \nabla_\sigma\alpha_a
+ \beta\hat{\pi}_{\rho\sigma}\sigma^{\rho\sigma}.
\eea
We now use the dissipative hydrodynamical equations of motion
\eqref{eq:hydro1}, \eqref{eq:hydro2} 
to eliminate the terms $D\beta$, $D\alpha_a$ in
Eq.~\eqref{eq:C_decomp_2nd}. Instead of Eqs.~\eqref{eq:D_beta},
\eqref{eq:D_alpha}, \eqref{eq:D_beta1}, and \eqref{eq:D_alpha1} we now
have
\bea\label{eq:D_beta_2nd}
D\beta 
&=&\beta\theta \gamma- (\Pi\theta +\partial_\mu q^{\mu}-
q^{\mu}Du_\mu 
-\pi^{\mu\nu}\sigma_{\mu\nu})
\left.\frac{\partial\beta}{\partial\epsilon}\right|_{n_a} -
\sum\limits_a\partial_\mu j^{\mu}_a
\left.\frac{\partial\beta}{\partial n_a}\right|_{\epsilon, n_b\neq n_a},\\
\label{eq:D_alpha_2nd}
D\alpha_c &=& -\beta\theta\delta_c
-(\Pi\theta +\partial_\mu q^{\mu}-q^{\mu}Du_\mu 
-\pi^{\mu\nu}\sigma_{\mu\nu})
\left.\frac{\partial\alpha_c}{\partial\epsilon}\right|_{n_a} -
\sum\limits_a\partial_\mu j^{\mu}_a \left.\frac{\partial
\alpha_c}{\partial n_a}\right|_{\epsilon, n_b\neq n_a}\!.
\eea
The first three terms in Eq.~\eqref{eq:C_decomp_2nd}
can be combined as follows 
\bea\label{eq:scalar_part_2nd}
\hat{\epsilon} D\beta- \hat{p}
\beta\theta -\sum\limits_a \hat{n}_a D\alpha_a  = 
- \beta\theta \hat{p}^*- \hat{\beta}^*(\Pi\theta +\partial_\mu q^{\mu}-
q^{\mu}Du_\mu -\pi^{\mu\nu}\sigma_{\mu\nu})
+\sum\limits_a \hat{\alpha}_a^*\partial_\mu j^{\mu}_a ,
\eea
where we used Eqs.~\eqref{eq:rel_alpha_beta} and 
\eqref{eq:p_star} and defined new operators
\bea\label{eq:beta_star}
\hat{\beta}^* &=& \hat\epsilon\left.\frac{\partial\beta}
{\partial\epsilon}\right|_{n_a}
+\sum\limits_a\hat{n}_a\left.\frac{\partial \beta}
{\partial n_a} \right|_{\epsilon,n_b\neq n_a},\\
\label{eq:alpha_star}
\hat{\alpha}^*_a &=& \hat\epsilon\left.\frac{\partial \alpha_a}
{\partial\epsilon}\right|_{n_b}
+\sum\limits_c\hat{n}_c\left.\frac{\partial\alpha_a}
{\partial n_c}\right|_{\epsilon, n_b\neq n_c}.
\eea
Next we use Eq.~\eqref{eq:hydro3} in the form
[the gradient of pressure is modified according to 
the second relation in Eq.~\eqref{eq:thermodyn3}]
\bea\label{eq:Du_modify_IS2}
h D u_{\sigma}
&=&-hT\nabla_\sigma \beta +
T\sum\limits_a n_a\nabla_\sigma \alpha_a+
\nabla_\sigma\Pi
\no\\
&-&\Pi D u_\sigma - \Delta_{\sigma\mu}D q^{\mu}
-q^{\mu}\partial_\mu u_{\sigma}-q_{\sigma}\theta 
-\Delta_{\sigma\nu}\partial_\mu \pi^{\mu\nu}, 
\eea
to modify the vector term involving 
$\hat{q}^\sigma$ in Eq.~\eqref{eq:C_decomp_2nd},
\bea\label{eq:vector_part_2nd}
\hat{q}^{\sigma}(\beta Du_{\sigma}+
\nabla_{\sigma}\beta)& =& \sum\limits_a\frac{n_a}{h}
\hat{q}^{\sigma}\nabla_\sigma \alpha_a \nonumber \\
& - &
 \hat{q}^{\sigma}\beta h^{-1}
  (-\nabla_\sigma \Pi
+\Pi Du_\sigma + D q_{\sigma}
 + q^{\mu}\partial_\mu u_{\sigma} +q_{\sigma}\theta 
 +\partial_\mu \pi^{\mu}_{\sigma}).
\eea
Combining Eqs.~\eqref{eq:C_decomp_2nd}, \eqref{eq:scalar_part_2nd},
and \eqref{eq:vector_part_2nd} we obtain 
\bea \label{eq:C_decompose_2nd}
\hat{C}(x)=\hat{C}_1(x)+\hat{C}_2(x),
\eea
where $\hat{C}_1$ and $\hat{C}_2$ are the first- 
and the second-order contributions, respectively:
\bea\label{eq:op_C1} 
\hat{C}_1(x)&=&-\beta \theta \hat{p}^* +
\beta\hat{\pi}_{\rho\sigma}\sigma^{\rho\sigma}-
\sum\limits_a\hat{\mathscr J}^{\sigma}_a\nabla_\sigma\alpha_a,\\
\label{eq:op_C2}
\hat{C}_2(x)&=&-
\hat{\beta}^*\big(\Pi\theta +\partial_\mu q^{\mu}-
q^{\mu}Du_\mu -\pi^{\mu\nu}\sigma_{\mu\nu}\big)
+\sum\limits_a \hat{\alpha}_a^* \partial_\mu j^{\mu}_a \nonumber\\
&-& \hat{q}^{\sigma}\beta h^{-1}\big (-\nabla_\sigma \Pi
+\Pi Du_\sigma + D q_{\sigma} + q^{\mu}\partial_\mu u_{\sigma}
 +q_{\sigma}\theta +\partial_\mu \pi^{\mu}_{\sigma}\big).
\eea
We observe that the operator $\hat{C}_2$ contains only scalar and
vector terms, and, therefore, contributes only to the bulk-viscous
pressure and the diffusion currents.  The reason for this is that
$\hat{C}_2$ originates from the dissipative terms of the
hydrodynamic equations~\eqref{eq:hydro1} -- \eqref{eq:hydro3},
which have either scalar or vector structure. It is natural to denote the expressions
contained in parentheses in Eq.~\eqref{eq:op_C2} as
{\it generalized} or {\it extended thermodynamic forces}. They involve either
space-time derivatives of the dissipative currents or their products
with the ``ordinary" thermodynamic forces.  However, in these generalized
thermodynamic forces the comoving derivatives (\ie, 
time derivatives in the local rest frame)
\bea\label{eq:dot_def} 
\dot{\Pi}\equiv D\Pi,\qquad
\dot{\pi}_{\mu\nu}\equiv\Delta_{\mu\nu\rho\sigma}
D{\pi}^{\rho\sigma},\qquad
\mathscr{\dot{J}}_{a\mu}\equiv \Delta_{\mu\nu} 
D\!\!\mathscr{J}^\nu_a,
\eea
do not appear, but should be present in a causal theory. As we
will show below, these terms arise from the {\it non-locality} of the
thermodynamic forces involved in Eq.~\eqref{eq:op_C1}.  [Note that 
in Eq.~\eqref{eq:op_C2} we have a comoving derivative for the energy flow,
\ie, $Dq_\sigma$. This term, however, is absent in the
L-frame. Therefore, we actually have only space-like gradients of the
dissipative currents in Eq.~\eqref{eq:op_C2}.]
  
Now using Eq.~\eqref{eq:stat_average} for the statistical 
average of an {\it arbitrary operator} $\hat{X}(x)$ we can 
write up to second order 
\bea\label{eq:stat_average_2nd}
\langle \hat{X}(x)\rangle 
= \langle \hat{X}(x)\rangle_l +
\langle \hat{X}(x)\rangle_1 +
\langle \hat{X}(x)\rangle_2.
\eea
The first-order correction is given by
\bea
\label{eq:stat_average_C1}
\langle \hat{X}(x)\rangle_1 =
\int\! d^4x_1 \Big(\hat{X}(x),\hat{C}_1(x_1)
\Big)\Big\vert_{\rm loc},
\eea
where the index ``${\rm loc}$'' indicates that the thermodynamic 
forces in the integrand are approximated by their local 
values at the point $x$, \ie, the non-local effects are 
neglected, as explained in Sec.~\ref{sec:diss1}.

The second-order correction $\langle \hat{X}(x)\rangle_2$ can be
decomposed into three terms,
\bea\label{eq:stat_average_2_new}
\langle \hat{X}(x)\rangle_2 =
\langle \hat{X}(x)\rangle_2^1
+\langle \hat{X}(x)\rangle_2^2
+\langle \hat{X}(x)\rangle_2^3,
\eea
with
\bea\label{eq:stat_average_21}
\langle \hat{X}(x)\rangle_2^1 &=&
\int\! d^4x_1 \Big(\hat{X}(x),\hat{C}_1(x_1)\Big)
-\langle \hat{X}(x)\rangle_1,\\
\label{eq:stat_average_22}
\langle \hat{X}(x)\rangle_2^2 &=&
\int\! d^4x_1 \Big(\hat{X}(x),\hat{C}_2(x_1)\Big),\\
\label{eq:stat_average_23}
\langle \hat{X}(x)\rangle_2^3 &=&
\int\! d^4x_1d^4x_2
\Big(\hat{X}(x),\hat{C}_1(x_1),\hat{C}_1(x_2)\Big).
\eea
The first term in Eq.~\eqref{eq:stat_average_2_new} collects those
corrections which arise from the non-locality of the thermodynamic
forces involved in the operator $\hat{C}(x_1)$. These corrections
are of second order, because they involve the differences of a
thermodynamic force, \eg, $\sigma^{\mu\nu}$, at points $x_1$ and
$x$, as seen from Eqs.~\eqref{eq:op_C1}, \eqref{eq:stat_average_C1},
and \eqref{eq:stat_average_21}.  Therefore, we can approximate
$\sigma^{\mu\nu}(x_1)-\sigma^{\mu\nu}(x)\simeq
(x_1-x)^\alpha \partial_\alpha\sigma^{\mu\nu}(x)\sim
\mathrm{Kn}\, \sigma^{\mu\nu}(x)$, 
because $x_1-x\sim \lambda$ and $\partial\sim L^{-1}$,
as mentioned in Sec.~\ref{sec:diss1}. Thus, the corrections of the
type \eqref{eq:stat_average_21} contain an additional power of the
Knudsen number  as compared to the first-order expression
\eqref{eq:stat_average_C1}, and, therefore, are at least of second
order in the hydrodynamic expansion.

The second term in Eq.~\eqref{eq:stat_average_2_new} includes the
corrections from the generalized thermodynamic forces. Finally, the
third term stands for the corrections which are nonlinear (quadratic)
in the three thermodynamic forces $\theta$, $\sigma_{\rho\sigma}$, and
$\nabla_\sigma\alpha_a$, which appear in the Navier--Stokes limit.

In order to properly derive the non-local corrections
\eqref{eq:stat_average_21} to the dissipative currents, we should first
generalize the expressions for the two-point correlators given by
Eqs.~\eqref{eq:corr1_current} and \eqref{eq:corr1_stress}, which were
initially written to provide first-order accuracy only.

\subsection{Non-local generalization of two-point correlators}

The two-point correlators \eqref{eq:corr1_current} and
\eqref{eq:corr1_stress} can be generalized in a straightforward manner
to incorporate the non-locality of the spatial projectors
\bea\label{eq:corr1_curr_2nd}
\Big(\mathscr{\hat{J}}_a^{\mu}(x),
\mathscr{\hat{J}}_b^{\nu}(x_1)\Big)& = & 
\frac{1}{3}\Delta^{\mu\nu}(x,x_1)
\Big(\mathscr{\hat{J}}_a^{\lambda}(x),
\mathscr{\hat{J}}_{b\lambda}(x_1)\Big),\\
\label{eq:corr1_stress_2nd}
\Big(\hat{\pi}_{\mu\nu}(x),
\hat{\pi}_{\rho\sigma}(x_1)\Big)
 & =& \frac{1}{5} \Delta_{\mu\nu\rho\sigma}(x,x_1)
 \Big(\hat{\pi}^{\lambda\eta}(x),\hat{\pi}_{\lambda\eta}(x_1)\Big).
\eea
Here the new projectors
\bea\label{eq:projector_2nd_1}
\Delta_{\mu\nu}(x,x_1) &=& \Delta_{\mu\lambda}(x)
\Delta^\lambda_{\nu}(x_1),\\
\label{eq:projector_2nd_2}
\Delta_{\mu\nu\rho\sigma}(x,x_1)&=& 
\Delta_{\mu\nu\alpha\beta}(x)
\Delta^{\alpha\beta}_{\rho\sigma}(x_1)
\eea
are the natural non-local generalizations of the second-rank
$\Delta_{\mu\nu}$  and fourth-rank $\Delta_{\mu\nu\rho\sigma}$                                                             
projectors, respectively. The normalization of the right-hand sides 
of Eqs.~\eqref{eq:corr1_curr_2nd} and \eqref{eq:corr1_stress_2nd} is
performed at the leading order in velocity gradients, see
\ref{app:projectors} for details.  The non-local form of the
projectors~\eqref{eq:projector_2nd_1} and \eqref{eq:projector_2nd_2}
guarantees that the orthogonality conditions~\eqref{eq:orthogonality}
are satisfied for the correlation functions given by
Eqs.~\eqref{eq:corr1_curr_2nd} and \eqref{eq:corr1_stress_2nd} at both
points $x$ and $x_1$.

For our calculations, it is sufficient to keep only the linear terms in the
difference $x_1-x$ of the expansion of the non-local projectors around
$x_1=x$. As shown in \ref{app:projectors}, the first derivative of
the fourth-rank projector is
\bea\label{eq:delta_deriv2}
\frac{\partial}{\partial x_1^\alpha}
\Delta_{\mu\nu\rho\sigma}(x,x_1)\bigg\vert_{x_1=x}=
-(\Delta_{\mu\nu\rho\beta}u_{\sigma}+
\Delta_{\mu\nu\sigma\beta}u_{\rho})
\partial_\alpha u^\beta,
\eea
which we will utilize below. In addition, we will assume that 
Curie's theorem holds also in this approximation, \ie,
 the two-point correlations between tensors of different rank vanish.

\subsection{Second-order corrections to the shear-stress tensor}
\label{sec:2nd_shear}

According to
Eqs.~\eqref{eq:stat_average_2_new} -- \eqref{eq:stat_average_23}, the
second-order corrections to the dissipative currents arise from three
different sources. In this subsection we
compute all these corrections separately for the shear-stress tensor. 
In the following two subsections we do this for
the bulk-viscous pressure and the diffusion currents.

\subsubsection{Non-local corrections from the two-point correlation function}
\label{sec:non_loc_shear}

Substituting the decomposition \eqref{eq:op_C1} into
Eq.~\eqref{eq:stat_average_21} and recalling Curie's theorem we obtain
\bea\label{eq:shear_av_21}
\langle \hat{\pi}_{\mu\nu}(x)\rangle_2^1 
= \int\! d^4x_1\Big(\hat{\pi}_{\mu\nu}(x),
\hat{\pi}_{\rho\sigma}(x_1)\Big) \beta(x_1) 
\sigma^{\rho\sigma}(x_1)-2\eta(x) \sigma_{\mu\nu}(x),
\eea
where we used the first-order relation for
$\langle \hat{\pi}_{\mu\nu}(x)\rangle_1$
given by Eq.~\eqref{eq:shear_bulk_1}. The
thermodynamic force $\beta(x_1) \sigma^{\rho\sigma}(x_1)$ cannot be
factored out from the integral with its value at $x$, but should be
expanded around that value to first order in the
difference $x_1-x$.  Here we note that, in order to obtain all
second-order corrections which have non-local origin, we should take
into account also the non-locality of the fluid velocity $u^\lambda$
in the expression
$\hat{\pi}^{\lambda\eta}(x_1)=\Delta^{\lambda\eta}
_{\gamma\delta}(x_1)\hat{T}^{\gamma\delta}(x_1)$ [see Eq.~\eqref{eq:proj2_op}].  
In contrast to $\Delta^{\lambda\eta}_{\gamma\delta}(x_1)$, which is 
a {\it hydrodynamic} quantity, the energy-momentum tensor 
$\hat{T}^{\gamma\delta}(x_1)$ is a {\it microscopic} quantity, and, 
therefore, does not require an expansion.

Now we substitute the two-point correlation function given by
Eqs.~\eqref{eq:corr1_stress_2nd} and \eqref{eq:projector_2nd_2} into
Eq.~\eqref{eq:shear_av_21} and use the definition of the shear-stress
tensor given in Eq.~\eqref{eq:proj2_op} as well as
$\Delta^{\alpha \beta}_{\rho \sigma} \sigma^{\rho \sigma} \equiv \sigma^{\alpha \beta}$
to obtain,
\bea\label{eq:shear_av_21_2}
\langle \hat{\pi}_{\mu\nu}(x)\rangle_2^1 
&=& \frac{1}{5} \Delta_{\mu\nu\rho\sigma}(x)\!
\int\! d^4x_1\Big(\hat{\pi}^{\lambda\eta}(x),
\hat{\pi}_{\lambda\eta}(x_1)\Big)
\beta (x_1) \sigma^{\rho\sigma}(x_1) -2\eta(x) \sigma_{\mu\nu}(x)\nonumber\\
&=&\frac{1}{5} \Delta_{\mu\nu\rho\sigma}(x)\!
\int\! d^4x_1\Big(\hat{T}^{\alpha\beta}(x),
\hat{T}^{\gamma\delta}(x_1)\Big)
\Delta_{\alpha\beta\gamma\delta}(x,x_1)
\beta(x_1) \sigma^{\rho\sigma}(x_1) -2\eta(x) \sigma_{\mu\nu}(x).\no
\eea
In the following step we expand
$\Delta_{\alpha\beta\gamma\delta}(x,x_1) \beta(x_1)
\sigma^{\rho\sigma}(x_1)$ around $x$ up to first order in $x_1-x$.
Using Eq.~\eqref{eq:delta_deriv2}, as well as
$\Delta_{\alpha\beta\gamma\delta}(x,x) =
\Delta_{\alpha\beta\rho\sigma}(x)\Delta^{\rho
  \sigma}_{\gamma\delta}(x) = \Delta_{\alpha\beta\gamma\delta}(x)$, we
obtain
\bea\label{eq:shear_av_21_3}
\langle \hat{\pi}_{\mu\nu}(x)\rangle_2^1 
&=& \frac{1}{5} \Delta_{\mu\nu\rho\sigma}(x)\!
\int\! d^4x_1\Big(\hat{T}^{\alpha\beta}(x),
\hat{T}^{\gamma\delta}(x_1)\Big)
\bigg\{\Delta_{\alpha\beta\gamma\delta}(x)\beta(x) \sigma^{\rho\sigma}(x)
\nonumber\\
&+ &(x_1-x)^\tau 
 \Big[\Delta_{\alpha\beta\gamma\delta}
\partial_\tau(\beta \sigma^{\rho\sigma})-\beta \sigma^{\rho\sigma}
(\Delta_{\alpha\beta\gamma\lambda}u_{\delta}+
\Delta_{\alpha\beta\delta\lambda}u_{\gamma})
\partial_\tau u^\lambda\Big]_x\bigg\}-2\eta (x)\sigma_{\mu\nu}(x)\nonumber\\
&=& 2a^\tau(x)\beta^{-1}(x)\Delta_{\mu\nu\rho\sigma}(x)
\partial_\tau\left[\beta(x) \sigma^{\rho\sigma}(x) \right]
-4b^\tau_\lambda(x) \sigma_{\mu\nu}(x) \partial_\tau u^\lambda(x).
\eea
Here we used the fact that the first term in curly brackets 
cancels with the last term $-2 \eta \sigma_{\mu \nu}$, on 
account of Eqs.~\eqref{eq:proj2_op} and \eqref{eq:shear_def}. 
We also defined
\bea\label{eq:tensors_ab}
a^\tau(x)  \equiv  \Delta_{\alpha\beta\gamma\delta}(x)
I^{\alpha\beta\gamma\delta,\tau}(x),\qquad
b^\tau_\lambda (x) \equiv 
\Delta_{\alpha\beta\gamma\lambda}(x) u_{\delta}(x)
I^{\alpha\beta\gamma\delta,\tau}(x),
\eea
with 
\bea\label{eq:integral}
I^{\alpha\beta\gamma\delta,\tau}(x)=
\frac{\beta(x)}{10}\! \int\! d^4x_1
\Big(\hat{T}^{\alpha\beta}(x),
\hat{T}^{\gamma\delta}(x_1)\Big)(x_1-x)^\tau.
\eea

Recalling the relations~\eqref{eq:proj1_op} and \eqref{eq:proj2_op} we
can write the expressions \eqref{eq:tensors_ab} and
\eqref{eq:integral} in the following form
\bea\label{eq:tensor_a}
a^\tau (x) &=&
\frac{\beta(x)}{10}\! \int\! d^4x_1
\Big(\hat{\pi}_{\alpha\beta}(x),
\hat{\pi}^{\alpha\beta}(x_1)\Big)(x_1-x)^\tau,\\
\label{eq:tensor_b}
b^\tau_\lambda(x) &=&
\frac{\beta(x)}{10}\! \int\! d^4x_1
\Big(\hat{\pi}_{\alpha\lambda}(x),
\hat{q}^{\alpha}(x_1)\Big)(x_1-x)^\tau =0,
\eea
where we have approximated $u^\mu(x)\simeq u^\mu(x_1)$, because the
quantities \eqref{eq:tensor_a} and \eqref{eq:tensor_b} are already multiplied
with second-order terms in Eq.~\eqref{eq:shear_av_21_3}. The tensor
$b^\tau_\lambda $ vanishes on account of Curie's theorem, and the vector
$a^\tau$ can be written in the following form [see
~\ref{app:Green_func},
Eqs.~\eqref{eq:vector_I0_rest} -- \eqref{eq:vec_gen}]
\bea \label{eq:vec_a}
a^\tau =-\eta\tau_\pi u^\tau,
\eea
where we defined
\bea\label{eq:eta_relax}
\eta\tau_\pi = -i\frac{d}{d\omega}\eta(\omega)\bigg\vert_{\omega=0}=
\frac{1}{20}\frac{d^2}{d\omega^2} {\rm Re}G^R_{\hat{\pi}_{ij}
\hat{\pi}^{ij}}(\omega)\bigg\vert_{\omega=0}.
\eea
Here the retarded Green's function is given by
Eq.~\eqref{eq:green_func}, and the frequency-dependent shear viscosity
$\eta(\omega)$ is defined by analogy with Eq.~\eqref{eq:I_XY}.  As
seen from Eq.~\eqref{eq:eta_relax}, the new coefficient $\tau_\pi$ has
the dimension of time and can be regarded as a relaxation time for the
shear-stress tensor.

Combining Eqs.~\eqref{eq:shear_av_21_3},
\eqref{eq:tensor_b}, and \eqref{eq:vec_a} we obtain 
\bea\label{eq:shear_av_21_final}
\langle \hat{\pi}_{\mu\nu}\rangle_2^1 =
-2\eta\tau_\pi\beta^{-1}\Delta_{\mu\nu\rho\sigma}
D(\beta \sigma^{\rho\sigma})=
-2\eta\tau_\pi (\Delta_{\mu\nu\rho\sigma}D
\sigma^{\rho\sigma}+\gamma\theta \sigma_{\mu\nu}),
\eea
where we used Eq.~\eqref{eq:D_beta1} in the second step and for the sake of
brevity omitted the argument $x$ at all quantities.

\subsubsection{Corrections from extended thermodynamic forces}
\label{sec:C2_shear}

Since the operator $\hat{C}_2$ given by Eq.~\eqref{eq:op_C2} does not
have a tensor part, the correction of $\hat{\pi}_{\mu\nu}$ from this
term vanishes due to Curie's theorem
\bea\label{eq:shear_av_22}
\langle \hat{\pi}_{\mu\nu}(x)\rangle_2^2 =
\int d^4x_1
\Big(\hat{\pi}_{\mu\nu}(x),\hat{C}_2(x_1)\Big)=0.
\eea

\subsubsection{Corrections from the three-point correlation function}
\label{sec:ninlin_shear}

From Eqs.~\eqref{eq:op_C1} and \eqref{eq:stat_average_23} we have
\bea\label{eq:shear_av_23}
\langle \hat{\pi}_{\mu\nu}(x)\rangle_2^3 
& =& \int\! d^4x_1 d^4 x_2 \Big(\hat{\pi}_{\mu\nu}(x),
\Big[-\beta \theta \hat{p}^* +
 \beta\hat{\pi}_{\rho\sigma}\sigma^{\rho\sigma}
 -\sum\limits_a\hat{\mathscr
   J}^{\sigma}_a\nabla_\sigma\alpha_a\Big]_{x_1},\no\\
& & \hspace*{3.5cm} \Big[-\beta \theta \hat{p}^*+
\beta\hat{\pi}_{\alpha\beta}\sigma^{\alpha\beta}
-\sum\limits_b\hat{\mathscr J}^{\alpha}_b
\nabla_\alpha\alpha_b \Big]_{x_2}\Big).
\eea
In contrast to the two-point correlators, where only operators of the
same rank are coupled, in the three-point correlators, one may have a
mixing between operators of different ranks.  The mixed three-point
correlators in Eq.~\eqref{eq:shear_av_23} which in general do not
vanish are
\bea\label{eq:cor_pi_mixed1}
\Big(\hat{\pi}_{\mu\nu}(x),\hat{p}^*(x_1),\hat{\pi}_{\alpha\beta}(x_2)\Big)
&=& \frac{1}{5}\Delta_{\mu\nu\alpha\beta}(x)
\Big(\hat{\pi}_{\gamma\delta}(x),\hat{p}^*(x_1),\hat{\pi}^{\gamma\delta}(x_2)\Big),\\
\label{eq:cor_pi_mixed2}
\Big(\hat{\pi}_{\mu\nu}(x),\hat{\mathscr J}_{a\sigma}(x_1),\hat{\mathscr
  J}_{b\alpha}(x_2)\Big)&=& \frac{1}{5} \Delta_{\mu\nu\sigma\alpha}(x)
\Big(\hat{\pi}_{\gamma\delta}(x),\hat{\mathscr J}^{\gamma}_a(x_1),
\hat{\mathscr J}^{\delta}_b(x_2)\Big),
\eea
where we exploited the symmetries of the relevant operators 
and of the three-point correlator, cf.~Eq.~\eqref{eq:3_point_corr_sym}.
The odd-rank mixed three-point correlator  
$\Big(\hat{\pi}_{\mu\nu}(x),\hat{p}^*(x_1),
\hat{\mathscr J}_{a\sigma}(x_2)\Big)$
vanishes on account of the space-time symmetry of the
operators involved. For similar reasons, the correlator 
$\Big(\hat{\pi}_{\mu\nu}(x),\hat{p}^*(x_1),
\hat{p}^*(x_2)\Big)$ can only be proportional to $\Delta_{\mu \nu}(x)$, 
but this is not traceless, so cannot contribute to the constitutive 
relation for $\langle \hat{\pi}_{\mu \nu}(x) \rangle$.
Because the correlators \eqref{eq:cor_pi_mixed1} and
\eqref{eq:cor_pi_mixed2} are accompanied by second-order terms 
in the thermodynamic forces, it is sufficient to evaluate all
$\Delta$ projectors on the right-hand sides of these relations at the point $x$.

The most general expression for the correlator of three 
shear-stress tensors which satisfies the orthogonality 
condition $u^\mu \hat{\pi}_{\mu\nu}=0$ is
\bea\label{eq:cor_pi_pi_pi}
\lefteqn{\hspace*{-1.5cm} \Big(\hat{\pi}_{\mu\nu}(x),\hat{\pi}_{\rho\sigma}(x_1),
\hat{\pi}_{\alpha\beta}(x_2)\Big) = a\big(\Delta_{\mu\nu}
\Delta_{\rho\sigma\alpha\beta}+\Delta_{\rho\sigma}
\Delta_{\mu\nu\alpha\beta}+\Delta_{\alpha\beta}
\Delta_{\mu\nu\rho\sigma}\big)
+b\Delta_{\mu\nu}\Delta_{\rho\sigma}\Delta_{\alpha\beta}}
\nonumber\\
&+&c\big(\Delta_{\mu\rho}\Delta_{\nu\alpha}\Delta_{\sigma\beta}
+\Delta_{\mu\rho}\Delta_{\nu\beta}\Delta_{\sigma\alpha}+
\Delta_{\mu\sigma}\Delta_{\nu\alpha}\Delta_{\rho\beta}+
\Delta_{\mu\sigma}\Delta_{\nu\beta}\Delta_{\rho\alpha}
\nonumber\\
&+&
\Delta_{\mu\alpha}\Delta_{\nu\rho}\Delta_{\sigma\beta}
+\Delta_{\mu\alpha}\Delta_{\nu\sigma}\Delta_{\rho\beta}
+\Delta_{\mu\beta}\Delta_{\nu\rho}\Delta_{\sigma\alpha}
+\Delta_{\mu\beta}\Delta_{\nu\sigma}\Delta_{\rho\alpha}\big),\quad
\eea
where the coefficients $a,b,c$ are functions
of $x,x_1,$ and $x_2$, while the $\Delta$ projectors can all
be taken at point $x$, since the three-point correlator is accompanied
by second-order terms. We now determine the coefficients $a,b,c$.
Since $\hat{\pi}_\alpha^\alpha=0$, we obtain by contracting the indices
$\alpha$ and $\beta$ in Eq.~\eqref{eq:cor_pi_pi_pi} and using the
properties \eqref{eq:prop_proj}, \eqref{eq:prop_projector4_1}, 
and \eqref{eq:prop_projector4_2},
\bea
\Big(\hat{\pi}_{\mu\nu}(x),\hat{\pi}_{\rho\sigma}(x_1),
\hat{\pi}_{\alpha}^{\alpha}(x_2)\Big) &=& 3a\Delta_{\mu\nu\rho\sigma}
+3b\Delta_{\mu\nu}\Delta_{\rho\sigma}
+4c(\Delta_{\mu\sigma}\Delta_{\nu\rho}+
\Delta_{\mu\rho}\Delta_{\nu\sigma})
\nonumber\\
&=& (3a+8c)\Delta_{\mu\nu\rho\sigma}
+\left(3b+\frac{8c}{3}\right)\Delta_{\mu\nu}
\Delta_{\rho\sigma}=0,\no
\eea
which implies that $b=a/3$, $c=-3a/8$.  In order to 
determine $a$ we compute the mixed contraction
\bea\label{eq:coeff_a}
\Big(\hat{\pi}_{\mu\nu}(x),\hat{\pi}^{\nu\alpha}(x_1),
\hat{\pi}_{\alpha\beta}(x_2)\Big)
&=& (5a + b + 22c) \Delta_{\mu\beta}
= -\frac{35}{12}a\Delta_{\mu\beta}\nonumber\\
 \Longrightarrow \qquad
a& =& -\frac{4}{35}\Big(\hat{\pi}_{\mu}^{\nu}(x),
\hat{\pi}_{\nu}^{\alpha}(x_1),\hat{\pi}_{\alpha}^{\mu}(x_2)\Big).\no
\eea
Then Eq.~\eqref{eq:cor_pi_pi_pi} can be cast into the compact form
\bea\label{eq:cor_pi_pi_pi_2}
\Big(\hat{\pi}_{\mu\nu}(x),\hat{\pi}_{\rho\sigma}(x_1),
\hat{\pi}_{\alpha\beta}(x_2)\Big) 
&=& \frac{1}{35}\Big[3(\Delta_{\rho\alpha}\Delta_{\mu\nu\sigma\beta}+
\Delta_{\rho\beta}\Delta_{\mu\nu\sigma\alpha}+
\Delta_{\sigma\alpha}\Delta_{\mu\nu\rho\beta}
+\Delta_{\sigma\beta}\Delta_{\mu\nu\rho\alpha})\nonumber\\
&&
-4(\Delta_{\rho\sigma}\Delta_{\mu\nu\alpha\beta}+
\Delta_{\alpha\beta}\Delta_{\mu\nu\rho\sigma})\Big]
\Big(\hat{\pi}_{\gamma}^{\delta}(x),\hat{\pi}_{\delta}^{\lambda}(x_1),
\hat{\pi}_{\lambda}^{\gamma}(x_2)\Big).\quad
\eea
 
Inserting the correlation functions given by
Eqs.~\eqref{eq:cor_pi_mixed1}, \eqref{eq:cor_pi_mixed2}, and 
\eqref{eq:cor_pi_pi_pi_2} into Eq.~\eqref{eq:shear_av_23}, 
factoring out the thermodynamic forces from the integral with 
their values at $x$, taking into account the symmetry 
property~\eqref{eq:3_point_corr_sym}, and defining the set of
transport coefficients
\bea\label{eq:lambda_pi}
\lambda_\pi &=& \frac{12}{35}\beta^2\!\int\! d^4x_1d^4x_2
\Big(\hat{\pi}_{\gamma}^{\delta}(x),\hat{\pi}_{\delta}^{\lambda}
(x_1),\hat{\pi}_{\lambda}^{\gamma}(x_2)\Big),\\
\label{eq:lambda_2}
\lambda_{\pi\Pi} &=& -\frac{\beta^2 }{5}\!\int\! d^4x_1d^4x_2
\Big(\hat{\pi}_{\gamma\delta}(x),\hat{\pi}^{\gamma\delta}(x_1),
\hat{p}^*(x_2)\Big),\\
\label{eq:lambda1_ab}
\lambda_{\pi\!{\mathscr J}}^{ab} &=& \frac{1}{5}\!\int\! 
d^4x_1d^4x_2 \Big(\hat{\pi}_{\gamma\delta}(x),
\hat{\mathscr J}^{\gamma}_a(x_1),
\hat{\mathscr J}^{\delta}_b(x_2)\Big),
\eea
we finally obtain
\bea\label{eq:shear_av_23_2}
\langle \hat{\pi}_{\mu\nu}\rangle_2^3 
=2\lambda_{\pi\Pi}\theta \sigma_{\mu\nu}
+\lambda_\pi \sigma_{\alpha<\mu}\sigma_{\nu>}^{\alpha}
+\sum\limits_{ab}\lambda_{\pi\!{\mathscr J}}^{ab}
\nabla_{<\mu}\alpha_a\nabla_{\nu>}\alpha_b,
\eea
where we again suppressed the $x$-dependence for the sake of brevity,
employed the notation 
\bea\label{eq:A_mu_nu}
A_{<\mu\nu>}\equiv \Delta_{\mu\nu}^{\alpha\beta}A_{\alpha\beta},
\eea
and applied the identities $\sigma^{<\alpha\beta>}=\sigma^{\alpha\beta }$,
$\sigma^{\alpha\beta}\Delta_{\beta\lambda}=\sigma^{\alpha}_{\lambda}$, 
and $\sigma^{\alpha}_{\alpha}=0$.

\subsubsection{Final result for the shear-stress tensor}
\label{sec:final_shear}

Combining all contributions from Eqs.~\eqref{eq:shear_bulk_1},
\eqref{eq:shear_av_21_final}, \eqref{eq:shear_av_22}, and
\eqref{eq:shear_av_23_2} and using
Eqs.~\eqref{eq:diss_currents_av_eq}, \eqref{eq:stat_average_2nd}, and
\eqref{eq:stat_average_2_new} we obtain the complete second-order
expression for the shear-stress tensor
\bea\label{eq:shear_total_final}
{\pi}_{\mu\nu}
&=&  2\eta \sigma_{\mu\nu}
-2\eta\tau_\pi (\Delta_{\mu\nu\rho\sigma}
D \sigma^{\rho\sigma}+ \gamma \theta \sigma_{\mu\nu})\nonumber\\
 & +&   2\lambda_{\pi\Pi}\theta \sigma_{\mu\nu}
 + \lambda_\pi \sigma_{\alpha<\mu}\sigma_{\nu>}^{\alpha}
 +\sum\limits_{ab}\lambda_{\pi\! {\mathscr J}}^{ab}
 \nabla_{<\mu}\alpha_a\nabla_{\nu>}\alpha_b.
\eea
Here, the second-order terms in the first line represent the non-local
corrections, whereas the second line collects the nonlinear
corrections from the three-point correlations. The physical
interpretation of the various terms in Eq.~\eqref{eq:shear_total_final} is
discussed in Sec.~\ref{sec:shear_compare}.

In order to derive a relaxation-type equation for ${\pi}_{\mu\nu}$ from 
expression~\eqref{eq:shear_total_final}, 
we follow Refs.~\cite{Baier2008JHEP,Jaiswal2013PhRvC,Finazzo2015JHEP}
and use the first-order Navier--Stokes relation \eqref{eq:shear_bulk_1} to replace $2\sigma^{\rho\sigma} \to \eta^{-1}{\pi}^{\rho\sigma}$
in the second term of the right-hand-side of
Eq.~\eqref{eq:shear_total_final}.
This substitution is justified because that term is already of second
order in space-time gradients, so any correction to
the Navier--Stokes result would be at least of third order. We then have
\bea\label{eq:shear_relax}
-2\eta\tau_\pi \Delta_{\mu\nu\rho\sigma}D \sigma^{\rho\sigma}
\simeq  -\tau_\pi \dot{\pi}_{\mu\nu}+ \tau_\pi\beta\eta^{-1}
\bigg(\gamma \frac{\partial\eta}{\partial \beta}
-\sum\limits_a\delta_a \frac{\partial\eta}
{\partial \alpha_a}\bigg)\theta {\pi}_{\mu\nu}, 
\eea
where we used Eqs.~\eqref{eq:D_beta1}, \eqref{eq:D_alpha1}, and \eqref{eq:dot_def}.
Combining Eqs.~\eqref{eq:shear_total_final} and
\eqref{eq:shear_relax} and introducing the coefficients
\bea\label{eq:lambda}
\lambda &=& 2(\lambda_{\pi\Pi}-\gamma\eta\tau_\pi),\\
\label{eq:tilde_lambda_pi}
\tilde{\lambda}_\pi &=& \tau_\pi \beta \eta^{-1}
\bigg(\gamma \frac{\partial\eta}{\partial \beta}
-\sum\limits_a\delta_a \frac{\partial\eta}{\partial \alpha_a}\bigg),
\eea
we finally obtain the following relaxation equation for the shear-stress tensor,
\bea\label{eq:shear_final_relax}
\tau_\pi \dot{\pi}_{\mu\nu}+{\pi}_{\mu\nu}
= 2\eta \sigma_{\mu\nu}
+\tilde{\lambda}_\pi\theta\pi_{\mu\nu}
 + \lambda \theta {\sigma}_{\mu\nu} + \lambda_\pi
 \sigma_{\alpha<\mu}\sigma_{\nu>}^{\alpha}
+\sum\limits_{ab}\lambda_{\pi\!{\mathscr J}}^{ab}
\nabla_{<\mu}\alpha_a\nabla_{\nu>}\alpha_b.
\eea

\subsection{Second-order corrections to the bulk-viscous pressure}
\label{sec:2nd_bulk}

In order to evaluate the bulk-viscous pressure to second 
order, we should include also the second-order corrections to
Eqs.~\eqref{eq:pressure_expand} and \eqref{eq:bulk_1}. 
Denoting $\Delta\epsilon =\langle\hat{\epsilon}\rangle_1
+\langle\hat{\epsilon}\rangle_2$ and $\Delta n_a =
\langle\hat{n}_a\rangle_1+\langle\hat{n}_a\rangle_2$ we obtain
\bea\label{eq:pressure_expand1}
\langle\hat{p}\rangle_l = {p}\big({\epsilon}-
\Delta{\epsilon},{n}_a-\Delta{n}_a \big) =
p({\epsilon},{n}_a)-\gamma\Delta\epsilon -
\sum\limits_a\delta_a\Delta n_a \nonumber\\
+\psi_{\epsilon\epsilon}\Delta\epsilon^2 + 
2\sum\limits_a\psi_{\epsilon a} \Delta\epsilon\Delta n_a+
\sum\limits_{ab}\psi_{ab} \Delta n_a\Delta n_b,
\eea
where we defined 
\bea\label{eq:xi_ab}
\psi_{\epsilon\epsilon} =\frac{1}{2}
\frac{\partial^2 p}{\partial\epsilon^2},\qquad
\psi_{\epsilon a} =\frac{1}{2} \frac{\partial^2 p}
{\partial\epsilon\partial n_a},\qquad
\psi_{ab} =\frac{1}{2} \frac{\partial^2 p}
{\partial n_a\partial n_b}.
\eea
Then we obtain for the bulk-viscous pressure
\bea
\Pi & = & 
\langle\hat{p}\rangle_l +\langle\hat{p}\rangle_1+
\langle\hat{p}\rangle_2-{p}(\epsilon,n_a) 
= \langle\hat{p}\rangle_1+\langle\hat{p}\rangle_2 \no \\
&&  
-\gamma\Delta\epsilon-\sum\limits_a\delta_a\Delta n_a 
 +\psi_{\epsilon\epsilon} \Delta\epsilon^2+
2\sum\limits_a\psi_{\epsilon a} \Delta\epsilon\Delta n_a +
\sum\limits_{ab}\psi_{ab} \Delta n_a\Delta n_b.
\eea
Substituting $\Delta\epsilon$ and $\Delta n_a$ 
and dropping higher-order terms we obtain
\bea\label{eq:bulk_2}
\Pi =
\langle\hat{p}^*\rangle_1+\langle\hat{p}^*\rangle_2
+\psi_{\epsilon\epsilon}\langle\hat{\epsilon}\rangle_1^2+
2\sum\limits_a\psi_{\epsilon a} \langle\hat{\epsilon}\rangle_1
\langle\hat{n}_a\rangle_1 +\sum\limits_{ab}\psi_{ab} 
\langle\hat{n}_a\rangle_1 \langle\hat{n}_b\rangle_1,
\eea
where we used the definition \eqref{eq:p_star} of $\hat{p}^*$.

Upon introducing the coefficients [see Eq.~\eqref{eq:I_XY5} of
\ref{app:Green_func}]
\bea\label{eq:zeta_ep}
\zeta_\epsilon &=& \beta\! \int\! d^4x_1
\Big(\hat{\epsilon}(x),\hat{p}^*(x_1)\Big)
=-\frac{d}{d\omega} {\rm Im}G^R_{\hat{\epsilon}
\hat{p}^*}(\omega)\bigg\vert_{\omega=0},\\
\label{eq:zeta_n_a}
\zeta_{a} &=& \beta\! \int\! d^4x_1
\Big(\hat{n}_a(x),\hat{p}^*(x_1)\Big)=
-\frac{d}{d\omega} {\rm Im}G^R_{\hat{n}_a
\hat{p}^*}(\omega)\bigg\vert_{\omega=0},
\eea
according to Eqs.~\eqref{eq:op_C1} and
\eqref{eq:stat_average_C1} the averages $\langle\hat{\epsilon}\rangle_1$ and
$\langle\hat{n}_a\rangle_1$ can be written as
\bea\label{eq:ep_n_1st_order}
\langle \hat{\epsilon}\rangle_1
= -\zeta_\epsilon\theta,\qquad
\langle \hat{n}_a\rangle_1
= -\zeta_{a}\theta,
\eea
where we have used Curie's theorem, \ie, 
$\Big(\hat{\epsilon}(x),\hat{\pi}_{\rho \sigma}(x_1)\Big)
= \Big(\hat{\epsilon}(x),\hat{\mathscr{J}}^\sigma_a(x_1)\Big) = 0$.
Then we have from Eqs.~\eqref{eq:shear_bulk_1}, \eqref{eq:bulk_2}, and
\eqref{eq:ep_n_1st_order}
\bea\label{eq:bulk_sum}
\Pi =-\zeta\theta 
+\Big(\psi_{\epsilon\epsilon}\zeta_\epsilon^2+
2\zeta_\epsilon \sum\limits_a\psi_{\epsilon a}  \zeta_{a} +
 \sum\limits_{ab}\psi_{ab} \zeta_{a} \zeta_{b} \Big)\theta^2
+\langle\hat{p}^*\rangle_2.
\eea

\subsubsection{Non-local corrections from the two-point correlation function}
\label{sec:non_loc_bulk}

From Eqs.~\eqref{eq:shear_bulk_1}, \eqref{eq:op_C1}, and \eqref{eq:stat_average_21}
we have
\bea\label{eq:bulk_av_21}
\langle \hat{p}^*(x)\rangle_2^1 = -
\int\! d^4x_1\Big(\hat{p}^*(x),\hat{p}^*(x_1)\Big)
\beta(x_1) \theta(x_1)+\zeta(x)\theta(x).
\eea
Now we need to expand all {\it hydrodynamic} quantities which are
evaluated at the point $x_1$ around the point $x$, as explained in
Sec.~\ref{sec:2nd_shear}. For this purpose we use
Eqs.~\eqref{eq:proj1_op} and \eqref{eq:p_star}  to express the operator
$\hat{p}^*$ in terms of the operators $\hat{T}^{\mu\nu}$ and
$\hat{N}^{\mu}_a$
\bea\label{eq:p_star2}
\hat{p}^*(x_1)  = -\frac{1}{3}\Delta_{\mu\nu}(x_1)
\hat{T}^{\mu\nu}(x_1) - \gamma(x_1) u_\mu (x_1)u_\nu(x_1)
\hat{T}^{\mu\nu}(x_1) - \sum\limits_a\delta_a(x_1)u_\mu(x_1)\hat{N}^{\mu}_a(x_1).
\eea
Expanding the hydrodynamic quantities, \ie, $u_\mu (x_1)$,
$\gamma(x_1)$, and $\delta_a(x_1)$ in Eq.~\eqref{eq:p_star2} in terms of
Taylor series around $x_1=x$ and keeping only the linear terms we obtain
\bea\label{eq:p_star_expand}
\hat{p}^*(x_1)  = \hat{p}^*(x_1)_x+
 (x_1-x)^\tau \partial_\tau\hat{p}^*(x),
\eea
where $\hat{p}^*(x_1)_x$ is obtained from $\hat{p}^*(x_1)$, Eq.~\eqref{eq:p_star2},
by replacing the arguments $x_1$ of all hydrodynamic 
quantities (but not of the microscopic operators $\hat{T}^{\mu\nu}$ 
and $\hat{N}^{\mu}_a$!) with $x$, and
\bea\label{eq:delta_p_star}
\partial_\tau\hat{p}^* &=&  2\left(\frac{1}{3}-\gamma\right)
\hat{q}^\mu \partial_\tau u_\mu - 
2\hat{\epsilon}\Big(\psi_{\epsilon\epsilon} \partial_\tau\epsilon
+\sum\limits_a \psi_{\epsilon a} \partial_\tau n_a\Big) \nonumber\\
&  - & 2\sum\limits_a \hat{n}_a\Big(\psi_{\epsilon a} \partial_\tau \epsilon
+\sum\limits_{b} \psi_{ab} \partial_\tau n_{b}\Big)
-\sum\limits_a\hat{j}^{\mu}_a\delta_a \partial_\tau u_\mu,
\eea
where we have used Eqs.~\eqref{eq:gamma_delta_a} and \eqref{eq:xi_ab},
as well as Eqs.~\eqref{eq:proj1_op} and \eqref{eq:proj2_op}, at the same time
approximating $u_\mu(x)\approx u_\mu(x_1)$, since the difference is of higher order.
All operators in Eq.~\eqref{eq:delta_p_star} are evaluated at $x_1$, 
while all hydrodynamic quantities are evaluated at $x$. 

Substituting Eq.~\eqref{eq:p_star_expand} into
Eq.~\eqref{eq:bulk_av_21} and expanding also the thermodynamic force
$\beta\theta$ around $x_1=x$ we obtain up to the second order in
gradients 
\bea\label{eq:bulk_av_21_1}
\langle \hat{p}^*(x)\rangle_2^1 \!\!&=&\!\! -
\int\! d^4x_1\Big(\hat{p}^*(x),\hat{p}^*(x_1)_x
+\partial_\tau\hat{p}^*(x) (x_1-x)^\tau\Big) 
\big[\beta \theta +\partial_\tau(\beta \theta)
(x_1-x)^\tau\big]_x+\zeta(x) \theta(x) \nonumber\\
&=&  -\partial_\tau(\beta \theta)_x\!\int\! d^4x_1\Big
(\hat{p}^*(x),\hat{p}^*(x_1)_x\Big)(x_1-x)^\tau \nonumber\\
&&- \beta(x)\theta(x) \int\! d^4x_1\Big (\hat{p}^*(x),
\partial_\tau\hat{p}^*(x)\Big)(x_1-x)^\tau,
\eea
where we canceled the zeroth-order term of the expansion with
$\zeta\theta$ noting that the definition \eqref{eq:bulk_def} in the present
context  is rewritten as 
\bea\label{eq:def_zeta_1st}
\zeta &=&\beta\!\int\!
d^4x_1\Big(\hat{p}^*(x),\hat{p}^*(x_1)_x\Big), 
\eea
where the index $x$ means that the slowly varying thermodynamic
quantities (but not the operators) are taken at the position $x$.

Now inserting Eq.~\eqref{eq:delta_p_star} into
Eq.~\eqref{eq:bulk_av_21_1}, taking into account Curie's theorem, and
using Eqs.~\eqref{eq:vector_I0_rest} -- \eqref{eq:vec_gen} in
\ref{app:Green_func} we obtain
\bea\label{eq:bulk_av_21_2}
\langle \hat{p}^*(x)\rangle_2^1 
&=& - \partial_\tau(\beta \theta)_x\int\! d^4x_1
\Big(\hat{p}^*(x),\hat{p}^*(x_1)\Big)(x_1-x)^\tau \no \\
& + & 2\theta\Big(\psi_{\epsilon\epsilon} \partial_\tau\epsilon 
 +\sum\limits_a \psi_{\epsilon a} \partial_\tau n_a\Big) 
  \beta\!\int\! d^4x_1\Big(\hat{p}^*(x), \hat{\epsilon}(x_1)\Big)(x_1-x)^\tau \no \\
  &+ & 2\theta 
\sum\limits_a\Big(\psi_{\epsilon a} \partial_\tau \epsilon 
+\sum\limits_{b} \psi_{ab} \partial_\tau n_{b}\Big)
\beta\! \int\! d^4x_1\Big (\hat{p}^*(x), 
\hat{n}_a(x_1)\Big)(x_1-x)^\tau \no \\
& = &
 D(\beta \theta)\beta^{-1}\zeta\tau_\Pi-2\theta\Big(\psi_{\epsilon\epsilon} D\epsilon 
+\sum\limits_a \psi_{\epsilon a} D n_a\Big)\zeta_\epsilon\tau_\epsilon \no \\
&- &
2\theta \sum\limits_a\Big (\psi_{\epsilon a} D\epsilon
+ \sum\limits_{b} \psi_{ab} D n_{b}\Big)\zeta_{a}\tau_a, 
\eea
where we neglected the difference $\hat{p}^*(x_1)_x-\hat{p}^*(x_1)$ in
the first term because this term is already multiplied by a
second-order gradient $\partial_\tau(\beta \theta)$. The new
coefficients $\tau_\Pi$, $\tau_\epsilon$, and $\tau_a$ introduced in
Eq.~\eqref{eq:bulk_av_21_2} are given by
\bea\label{eq:zeta_relax}
\zeta \tau_\Pi &=& -i\frac{d}{d\omega}
\zeta(\omega) \bigg\vert_{\omega=0}=
\frac{1}{2}\frac{d^2}{d\omega^2} {\rm Re}G^R_{\hat{p}^*
\hat{p}^*}(\omega) \bigg\vert_{\omega=0},\\
\label{eq:zeta_tilde_ep}
\zeta_\epsilon \tau_\epsilon &=&
-i\frac{d}{d\omega}
\zeta_\epsilon(\omega) \bigg\vert_{\omega=0}= 
\frac{1}{2}\frac{d^2}{d\omega^2} {\rm Re}G^R_{\hat{p}^*
\hat{\epsilon}}(\omega) \bigg\vert_{\omega=0},\\
\label{eq:zeta_tilde_a}
\zeta_{a} \tau_a &=& -i\frac{d}{d\omega}
\zeta_{a}(\omega) \bigg\vert_{\omega=0}=
\frac{1}{2}\frac{d^2}{d\omega^2} {\rm Re}G^R_{\hat{p}^*
\hat{n}_a}(\omega) \bigg\vert_{\omega=0},
\eea
where $\zeta_\epsilon$ and $\zeta_a$ in the limit
$\omega\to 0$ are defined in Eqs.~\eqref{eq:zeta_ep} 
and \eqref{eq:zeta_n_a}. In the case of $\omega\neq 0$ 
the formula \eqref{eq:I_XY} should be used with the
relevant choices of the operators $\hat{X}$ and
$\hat{Y}$. [Note that there is no summation on the 
left-hand side of Eq.~\eqref{eq:zeta_tilde_a}.]

The derivatives $D\beta$, $D\epsilon$, and $Dn_a$ can be eliminated
from Eq.~\eqref{eq:bulk_av_21_2} by employing
Eqs.~\eqref{eq:ideal_hydro2}, \eqref{eq:D_beta1}, and
\eqref{eq:D_alpha1}. Denoting
\bea\label{eq:zeta_star0}
{\zeta}^* 
= \gamma\zeta\tau_\Pi +2\zeta_\epsilon
\tau_\epsilon\Big(\psi_{\epsilon\epsilon} h +\sum\limits_a n_a\psi_{\epsilon a}\Big) +
2\sum\limits_a\zeta_{a}\tau_a\Big (\psi_{\epsilon a} h 
+\sum\limits_{b} \psi_{ab}  n_{b}\Big),
\eea
we obtain from Eqs.~\eqref{eq:bulk_av_21_2} -- \eqref{eq:zeta_star0}
\bea\label{eq:bulk_av_21_3}
\langle \hat{p}^*\rangle_2^1 
= \zeta \tau_\Pi D\theta +\zeta^*\theta^2.
\eea

\subsubsection{Corrections from extended thermodynamic forces}
\label{sec:C2_bulk}

Inserting Eq.~\eqref{eq:op_C2} into Eq.~\eqref{eq:stat_average_22} 
and taking into account Curie's theorem we obtain
\bea\label{eq:bulk_av_22}
\langle \hat{p}^*(x)\rangle_2^2 =
-\zeta_\beta \big(\Pi\theta +\partial_\mu q^{\mu}-
q^{\mu}Du_\mu -\pi^{\mu\nu}\sigma_{\mu\nu}\big)+
\sum\limits_a \zeta_{\alpha_a}
\big(\partial_\mu j^{\mu}_a\big),
\eea
where we factored out all thermodynamic forces from the integral 
with their values at $x$ and introduced short-hand notations
\bea\label{eq:zeta_beta}
\zeta_\beta =\int\! d^4x_1
\Big(\hat{p}^*(x),\hat{\beta}^*(x_1)\Big)
=T \frac{\partial\beta} {\partial\epsilon} \,
\zeta_\epsilon +\sum\limits_c  T
\frac{\partial \beta}{\partial n_c}\, \zeta_c,\\
\label{eq:zeta_alpha_a}
\zeta_{\alpha_a} =\int\! d^4x_1 
\Big(\hat{p}^*(x),\hat{\alpha}_a^*(x_1)\Big)
=T \frac{\partial \alpha_a}{\partial\epsilon}\,
\zeta_\epsilon  +\sum\limits_c  T 
\frac{\partial\alpha_a} {\partial n_c}\,\zeta_c\,.
\eea
In the second step we used the definitions~\eqref{eq:beta_star}, 
\eqref{eq:alpha_star}, \eqref{eq:zeta_ep} and \eqref{eq:zeta_n_a}.

It is convenient to modify the second term in
Eq.~\eqref{eq:bulk_av_22} using Eq.~\eqref{eq:diff_curr1}
\bea\label{eq:j_mu_mod}
\partial_\mu j^{\mu}_a=
\partial_\mu {\mathscr J}^{\mu}_a+
n_ah^{-1}\partial_\mu q^\mu +
q^\mu \partial_\mu (n_ah^{-1}).\no
\eea
Using this and defining 
\bea\label{eq:tilde_zeta_beta}
\tilde{\zeta}_\beta ={\zeta}_\beta-h^{-1}
\sum\limits_a n_a\zeta_{\alpha_a},
\eea
 we obtain for Eq.~\eqref{eq:bulk_av_22}
\bea\label{eq:bulk_av_22_1}
\langle \hat{p}^*\rangle_2^2 &=&\!\!
\sum\limits_a \zeta_{\alpha_a}
\partial_\mu {\mathscr J}^{\mu}_a 
-\zeta_\beta (\Pi\theta -\pi^{\mu\nu}\sigma_{\mu\nu})-
\tilde{\zeta}_\beta \partial_\mu q^{\mu}+
q^\mu \Big[\zeta_\beta Du_\mu +\sum\limits_a \zeta_{\alpha_a} 
\nabla_\mu (n_ah^{-1})\Big].\qquad
\eea

\subsubsection{Corrections from the three-point correlation function}
\label{sec:nonlin_bulk}

From Eqs.~\eqref{eq:op_C1} and \eqref{eq:stat_average_23} we have
\bea\label{eq:bulk_av_23}
\langle \hat{p}^*(x)\rangle_2^3 =
\int\! d^4x_1 d^4x_2 \Big(\hat{p}^*(x),
\Big [-\beta \theta \hat{p}^* +
\beta\hat{\pi}_{\rho\sigma}\sigma^{\rho\sigma}-
\sum\limits_a\hat{\mathscr J}^{\sigma}_a
\nabla_\sigma\alpha_a\Big]_{x_1},\no\\
\Big[-\beta \theta \hat{p}^* +
\beta\hat{\pi}_{\alpha\beta}\sigma^{\alpha\beta}-
\sum\limits_a\hat{\mathscr J}^{\alpha}_b
\nabla_\alpha\alpha_b\Big]_{x_2}\Big).
\eea
The non-vanishing correlators in Eq.~\eqref{eq:bulk_av_23} 
are $\Big(\hat{p}^*(x),\hat{p}^*(x_1),\hat{p}^*(x_2)\Big)$ and
\bea\label{eq:cor_p_star1}
\Big(\hat{p}^*(x),\hat{\mathscr J}_{a\sigma}(x_1),
\hat{\mathscr J}_{b\alpha}(x_2)\Big)=
\frac{1}{3} \Delta_{\sigma\alpha}(x)
\Big(\hat{p}^*(x),\hat{\mathscr J}_{a\gamma}(x_1),
\hat{\mathscr J}^{\gamma}_b(x_2)\Big),\\
\label{eq:cor_p_star2}
\Big(\hat{p}^*(x),\hat{\pi}_{\rho\sigma}(x_1),
\hat{\pi}_{\alpha\beta}(x_2)\Big) = \frac{1}{5} 
\Delta_{\rho\sigma\alpha\beta}(x)
\Big(\hat{p}^*(x),\hat{\pi}_{\gamma\delta}(x_1),
\hat{\pi}^{\gamma\delta}(x_2)\Big).
\eea
Inserting these expressions into Eq.~\eqref{eq:bulk_av_23} we obtain
\bea\label{eq:bulk_av_23_2}
\langle \hat{p}^*\rangle_2^3 =
\lambda_\Pi \theta^2
-\lambda_{\Pi\pi} \sigma_{\alpha\beta}\sigma^{\alpha\beta} +
T\sum\limits_{ab}\zeta^{ab}_\Pi\,
\nabla^\sigma\alpha_a\nabla_\sigma\alpha_b,
\eea
where we defined the coefficients
\bea\label{eq:lambda_Pi}
\lambda_\Pi &=& \beta^2\!\int\! d^4x_1d^4x_2
\Big(\hat{p}^*(x),\hat{p}^*(x_1),\hat{p}^*(x_2)\Big),\\
\label{eq:lambda_1}
\lambda_{\Pi\pi} &=& -\frac{\beta^2 }{5}\!\int\! d^4x_1d^4x_2
\Big(\hat{p}^*(x),\hat{\pi}_{\gamma\delta}(x_1),
\hat{\pi}^{\gamma\delta}(x_2)\Big),\\
\label{eq:zeta1_ab}
\zeta^{ab}_\Pi &=& \frac{\beta}{3}\!\int\! d^4x_1d^4x_2
\Big(\hat{p}^*(x), \hat{\mathscr J}_{a\gamma}(x_1),
\hat{\mathscr J}^{\gamma}_b(x_2)\Big).
\eea

\subsubsection{Final result for the bulk-viscous pressure}
\label{sec:final_bulk}

Combining all contributions from Eqs.~\eqref{eq:bulk_av_21_3},
\eqref{eq:bulk_av_22_1}, and \eqref{eq:bulk_av_23_2} we obtain
according to Eq.~\eqref{eq:stat_average_2_new}
\bea\label{eq:bulk_av_2_final}
\langle \hat{p}^*(x)\rangle_2 &=&
\zeta\tau_\Pi D\theta -\zeta_\beta (\Pi\theta
-\pi^{\mu\nu}\sigma_{\mu\nu})
-\tilde{\zeta}_\beta \partial_\mu q^{\mu} +(\lambda_\Pi +\zeta^*)\theta^2 
\nonumber\\
&-&\lambda_{\Pi\pi} \sigma_{\alpha\beta}\sigma^{\alpha\beta}
+\sum\limits_a \zeta_{\alpha_a} \partial_\mu {\mathscr J}^{\mu}_a
+T\sum\limits_{ab}\zeta^{ab}_\Pi\,
\nabla^\sigma\alpha_a\nabla_\sigma\alpha_b
\nonumber\\
&+& q^\mu \Big[\zeta_\beta Du_\mu +\sum\limits_a \zeta_{\alpha_a} 
\nabla_\mu (n_ah^{-1})\Big].
\eea
A relaxation equation for the bulk-viscous pressure can be obtained by
approximating in the first term $\theta\simeq -\zeta^{-1}{\Pi}$, 
as we did in the case of the shear-stress tensor. 
We thus obtain ($\dot{\Pi}\equiv D\Pi$)
\bea\label{eq:bulk_relax}
\zeta\tau_\Pi D\theta &=& 
-\tau_\Pi \dot{\Pi} +\tau_\Pi{\Pi}\zeta^{-1}D\zeta
\no\\
&=&-\tau_\Pi \dot{\Pi}+\tau_\Pi \beta \zeta^{-1}
\bigg(\gamma\frac{\partial\zeta}{\partial\beta} 
-\sum\limits_a \delta_a\frac{\partial\zeta}
{\partial\alpha_a}\bigg)\theta \Pi,
\eea
where we used Eqs.~\eqref{eq:D_beta1} and \eqref{eq:D_alpha1}. 

Now combining Eqs.~\eqref{eq:bulk_sum}, \eqref{eq:bulk_av_2_final}, and
\eqref{eq:bulk_relax}, denoting $\dot{u}_\mu=Du_\mu$, and defining
\bea\label{eq:varsigma}
\varsigma &=& \lambda_\Pi +\zeta^*  +
\psi_{\epsilon\epsilon}\zeta_\epsilon^2 +
2\zeta_\epsilon \sum\limits_a\psi_{\epsilon a} \zeta_{a} +
\sum\limits_{ab}\psi_{ab} \zeta_{a} \zeta_{b},\\
\label{eq:tilde_lambda_Pi}
\tilde{\lambda}_\Pi &=& \tau_{\Pi}\beta \zeta^{-1}
\bigg(\gamma\frac{\partial\zeta}{\partial\beta} 
-\sum\limits_a \delta_a\frac{\partial\zeta}
{\partial\alpha_a}\bigg),
\eea
we obtain the final evolution equation for the bulk-viscous pressure,
\bea\label{eq:bulk_final_relax}
\tau_\Pi \dot{\Pi}+ \Pi &=&
-\zeta\theta 
+  \tilde{\lambda}_\Pi \theta \Pi 
+\zeta_\beta (\sigma_{\mu\nu}\pi^{\mu\nu} - \theta\Pi )
+\varsigma\theta^2  -\lambda_{\Pi\pi} \sigma_{\mu\nu}\sigma^{\mu\nu}
-\tilde{\zeta}_\beta \partial_\mu q^{\mu} \nonumber\\
 &+& 
\sum\limits_a \zeta_{\alpha_a} \partial_\mu {\mathscr J}^{\mu}_a
 +T\sum\limits_{ab}\zeta^{ab}_\Pi\,
\nabla^\mu \alpha_a\nabla_\mu\alpha_b 
+q^\mu \Big[\zeta_\beta \dot{u}_\mu +\sum\limits_a 
\zeta_{\alpha_a} \nabla_\mu (n_ah^{-1})\Big]. \qquad
\eea

\subsection{Second-order corrections to the diffusion currents}
\label{sec:2nd_diff}

\subsubsection{Non-local corrections from the two-point correlation function}
\label{sec:non_loc_diff}

Using Eqs.~\eqref{eq:charge_currents1}, \eqref{eq:op_C1}, and
\eqref{eq:stat_average_21} and again applying Curie's theorem we
obtain
\bea\label{eq:current_av_21}
\langle \hat{\mathscr J}_{c\mu}(x)\rangle_2^1 =
-\sum\limits_a\! \int\! d^4x_1 
\Big(\hat{\mathscr J}_{c\mu}(x),\hat{\mathscr J}_{a\sigma}(x_1)\Big)
\nabla_{x_1}^\sigma\alpha_a(x_1) -\sum\limits_a\chi_{ca}(x)\nabla_\mu \alpha_a(x).
\eea
Now we substitute the correlation function given by Eqs.~\eqref{eq:corr1_curr_2nd}
and \eqref{eq:projector_2nd_1} into Eq.~\eqref{eq:current_av_21} 
\bea\label{eq:current_av_21_1}
\langle \hat{\mathscr
  J}_{c\mu}(x)\rangle_2^1 &=& -\frac{1}{3} \sum\limits_a
\Delta_{\mu\beta}(x)\!
\int\! d^4x_1 \Big(\hat{\mathscr J}_{c}^{\lambda}(x),
\hat{\mathscr J}_{a\lambda}(x_1)\Big)\nabla_{x_1}^\beta\alpha_a(x_1) \no \\
& & -\sum\limits_a\chi_{ca}(x)\nabla_\mu \alpha_a(x).  
\eea
Next we use Eqs.~\eqref{eq:proj2_op} and \eqref{eq:diff_currents} to
express the operators $\mathscr{\hat{J}}_a^{\lambda}$ in terms of the
energy-momentum tensor and the charge currents, as we did in the
case of the bulk-viscous pressure,
\bea\label{eq:diff_currents1}
\mathscr{\hat{J}}_{a\lambda}(x_1)
=\Delta_{\lambda\mu}(x_1)\hat{N}^{\mu}_a(x_1)
-\frac{n_a(x_1)}{h(x_1)}\Delta_{\lambda\mu}(x_1)u_{\nu}(x_1)\hat{T}^{\mu\nu}(x_1).
\eea
Expanding all hydrodynamic quantities in Eq.~\eqref{eq:diff_currents1}
around their values at $x_1=x$, keeping only the first-order terms in
gradients, and using the decompositions \eqref{eq:T_munu_decomp} and
\eqref{eq:N_a_decomp} we obtain
\bea\label{eq:diff_currents_ex}
\mathscr{\hat{J}}_{a\lambda}(x_1)=
 \mathscr{\hat{J}}_{a\lambda}(x_1)_x +
(x_1-x)^\tau\partial_\tau\! \mathscr{\hat{J}}_{a\lambda}(x),
\eea
where $\mathscr{\hat{J}}_a^{\lambda}(x_1)_x$ is obtained from
$\mathscr{\hat{J}}_a^{\lambda}(x_1)$ via replacing the arguments $x_1$
of all hydrodynamic quantities with $x$, while the operators are taken
at $x_1$, and where
\bea\label{eq:deriv_currents}
\partial_\tau\! \mathscr{\hat{J}}_{a\lambda} &=&
-u_\lambda(\partial_\tau u_{\mu}) \hat{\mathscr J}^{\mu}_a-
\partial_\tau (n_ah^{-1}){\hat{q}}_{\lambda}
 -\hat{n}_a \partial_\tau u_{\lambda} \no\\ 
 & + & n_ah^{-1}\big(\hat{\epsilon}
 \partial_\tau u_{\lambda}+\hat{p}\partial_\tau u_{\lambda}
 -\hat{\pi}^\nu_\lambda \partial_\tau u_{\nu}\big).
\eea
Here, we have made use of Eq.~\eqref{eq:proj2_op} and the fact that to this order we are
allowed to approximate $u_\mu(x) \approx u_\mu(x_1)$.  As in 
Eq.~\eqref{eq:delta_p_star}, all hydrodynamic quantities are evaluated at $x$, while
all operators are evaluated at $x_1$, the difference being of higher order.
Substituting Eq.~\eqref{eq:diff_currents_ex} into
Eq.~\eqref{eq:current_av_21_1} and expanding also the thermodynamic
force $\nabla_{x_1}^\beta\alpha_a(x_1)$ around $x$ we obtain
\bea\label{eq:current_av_21_2}
\langle \hat{\mathscr J}_{c\mu}(x)\rangle_2^1 &=& 
-\frac{1}{3} \sum\limits_a \Delta_{\mu\beta}(x)\!
\int\! d^4x_1 \Big(\hat{\mathscr J}_{c}^{\lambda}(x), 
\mathscr{\hat{J}}_{a\lambda}(x_1)_x +
\partial_\tau\! \mathscr{\hat{J}}_{a\lambda}(x)
(x_1-x)^\tau\Big)\nonumber\\
&& \quad \times \left[\nabla^\beta\alpha_a(x)+
\partial_\tau (\nabla^\beta\alpha_a) (x_1-x)^\tau\right]
- \sum\limits_a\chi_{ca}(x)\nabla_\mu \alpha_a (x)\nonumber\\
&=&- 
\frac{1}{3} \sum\limits_a \Delta_{\mu\beta}(x)
\partial_\tau (\nabla^\beta\alpha_a)\!
\int\! d^4x_1 \Big (\hat{\mathscr J}_{c}^{\lambda}(x), 
\mathscr{\hat{J}}_{a\lambda}(x_1)_x \Big) (x_1-x)^\tau
\nonumber\\
&& -
\frac{1}{3} \sum\limits_a \nabla_\mu\alpha_a(x)\!
\int\! d^4x_1 \Big(\hat{\mathscr J}_{c}^{\lambda}(x), 
\partial_\tau\! \mathscr{\hat{J}}_{a\lambda}(x)\Big) (x_1-x)^\tau.
\eea
Here the first-order terms cancel each other in the same way
as in the case of bulk viscosity, see
Eqs.~\eqref{eq:bulk_av_21_1} and \eqref{eq:def_zeta_1st}.

Now substituting Eq.~\eqref{eq:deriv_currents} into Eq.~\eqref{eq:current_av_21_2}, 
taking into account Curie's theorem,
employing the orthogonality condition
$u^\lambda\!\! \hat{\mathscr J}_{c\lambda}=0$, and using
Eqs.~\eqref{eq:vector_I0_rest} -- \eqref{eq:vec_gen} in
\ref{app:Green_func} we obtain
\bea\label{eq:current_av_21_3}
\langle \hat{\mathscr J}_{c\mu}(x)\rangle_2^1 
&=&- 
\frac{1}{3} \sum\limits_a \Delta_{\mu\beta}(x)
\partial_\tau (\nabla^\beta\alpha_a)\!
\int\! d^4x_1 \Big(\hat{\mathscr J}_{c}^{\lambda}(x),
\mathscr{\hat{J}}_{a\lambda}(x_1)\Big) (x_1-x)^\tau
\nonumber\\
&& +
\frac{1}{3} \sum\limits_a \partial_\tau (n_ah^{-1})
\nabla_\mu\alpha_a(x)\!
\int\! d^4x_1 \Big (\hat{\mathscr J}_{c}^{\lambda}(x), 
{\hat{q}}_{\lambda}(x_1)\Big) (x_1-x)^\tau \nonumber\\
&=&
\sum\limits_a \tilde{\chi}_{ca}\Delta_{\mu\beta}(x)
D (\nabla^\beta\alpha_a) - \tilde{\chi}_{cq} 
\sum\limits_a D (n_ah^{-1})\nabla_\mu\alpha_a(x),
\eea
where 
\bea\label{eq:chi_ab_tilde}
\tilde{\chi}_{ac} &=& i\frac{d}{d\omega} 
{\chi}_{ac}(\omega)\bigg\vert_{\omega=0}=
\frac{T}{6} \frac{d^2}{d\omega^2} {\rm Re}
G^R_{\mathscr{\hat{J}}_{a}^{\lambda}
\mathscr{\hat{J}}_{c\lambda}}(\omega)\bigg\vert_{\omega=0},\\
\label{eq:chi_c_tilde}
\tilde{\chi}_{cq} &=& i\frac{d}{d\omega} 
{\chi}_{cq}(\omega)\bigg\vert_{\omega=0}=
\frac{T}{6} \frac{d^2}{d\omega^2} {\rm Re}
G^R_{\hat{\mathscr J}_{c}^{\lambda}, 
{\hat{q}}_{\lambda}}(\omega)\bigg\vert_{\omega=0},
\eea
and ${\chi}_{cq}$ is defined in the next subsection.  Because the
expression \eqref{eq:current_av_21_3} is already of second order,
we can use Eq.~\eqref{eq:ideal_hydro2} to replace
$D (n_ah^{-1})=-n_ah^{-2}Dp $. From Eqs.~\eqref{eq:ideal_hydro2} and
\eqref{eq:gamma_delta_a} we find
\bea
Dp=\gamma D\epsilon+\sum\limits_d \delta_d Dn_d
=-\Big(\gamma h+\sum\limits_d \delta_d n_d\Big)\theta.\no
\eea
Substituting these results into Eqs.~\eqref{eq:current_av_21_3} we obtain 
\bea\label{eq:current_av_21_4}
\langle \hat{\mathscr J}_{c\mu}\rangle_2^1 
= \sum\limits_a \tilde{\chi}_{ca}\Delta_{\mu\beta}
D (\nabla^\beta\alpha_a)-
\tilde{\chi}_{cq} h^{-2} \Big(\gamma h+\sum\limits_d
\delta_d n_d\Big)\theta \sum\limits_a n_a \nabla_\mu\alpha_a,
\eea
where we suppressed the $x$-dependence for the sake of brevity.

\subsubsection{Corrections from extended thermodynamic forces}
\label{sec:C2_diff}

Using Eqs.~\eqref{eq:op_C2} and \eqref{eq:stat_average_22} and
again taking into account Curie's theorem we obtain
\bea\label{eq:current_av_22}
\langle \hat{\mathscr J}_{c\mu}\rangle_2^2 
&=& \chi_{cq} \beta h^{-1}\big(-\nabla_\mu \Pi
+\Pi Du_\mu+ \Delta_{\mu\nu} D q^{\nu}
 + q^{\nu}\partial_\nu u_{\mu}+ q_{\mu}\theta 
 +\Delta_{\mu\sigma}\partial_\nu \pi^{\nu\sigma}\big), \qquad
\eea
where we used a relation analogous to the 
one given by Eq.~\eqref{eq:corr1_current}
\bea
\Big(\hat{\mathscr J}_{c\mu}(x),
\hat{q}^{\sigma}(x_1)\Big) =\frac{1}{3}\Delta_\mu^\sigma(x)
\Big(\hat{\mathscr J}_{c\alpha}(x),
\hat{q}^{\alpha}(x_1)\Big),\no
\eea
and defined new coefficients via
\bea\label{eq:kappa_c}
\chi_{cq} =-\frac{1}{3} \int\! d^4x_1
\Big (\hat{\mathscr J}_{c\alpha}(x),
\hat{q}^{\alpha}(x_1)\Big) =\frac{T}{3}\frac{d}{d\omega}
{\rm Im}G^R_{\hat{\mathscr{J}}_c^{\alpha}
  \hat{q}_{\alpha}}(\omega)\bigg\vert_{\omega=0}.
\eea
Denoting 
\bea\label{eq:dot_q_u}
\dot{u}_\mu = Du_\mu,\qquad
\dot{q}_\mu = \Delta_{\mu\nu} D q^{\nu},\no
\eea
Eq.~\eqref{eq:current_av_22} is written as
\bea\label{eq:current_av_22_1}
\langle \hat{\mathscr J}_{c\mu}\rangle_2^2 &=&
 \chi_{cq} \beta h^{-1}\big(-\nabla_\mu \Pi
 +\Pi \dot{u}_\mu + \dot{q}_{\mu}
 + q^{\nu}\partial_\nu u_{\mu}+ q_{\mu}\theta 
 +\Delta_{\mu\sigma}\partial_\nu \pi^{\nu\sigma}\big).
\eea

\subsubsection{Corrections from the three-point correlation function}
\label{sec:nonlin_diff}

Substituting Eq.~\eqref{eq:op_C1} into Eq.~\eqref{eq:stat_average_23}
we obtain
\bea\label{eq:current_av_23}
\langle \hat{\mathscr J}_{c\mu}(x)\rangle_2^3 =
\int\! d^4x_1 d^4x_2 \Big(\hat{\mathscr J}_{c\mu}(x),
\Big[-\beta \theta \hat{p}^*
+\beta\hat{\pi}_{\rho\sigma}\sigma^{\rho\sigma}
-\sum\limits_a\hat{\mathscr J}^{\sigma}_a
\nabla_\sigma\alpha_a \Big]_{x_1},\nonumber\\
\Big[-\beta \theta \hat{p}^*
 +\beta\hat{\pi}_{\alpha\beta}\sigma^{\alpha\beta}
-\sum\limits_b\hat{\mathscr J}^{\alpha}_b
\nabla_\alpha\alpha_b\Big]_{x_2}\Big).
\eea
The nonvanishing correlators in this case are 
\bea\label{eq:cor_current_mixed1}
\Big(\hat{\mathscr J}_{c\mu}(x),\hat{\mathscr J}_{a\sigma}(x_1),
\hat{p}^*(x_2)\Big) &=&
\frac{1}{3} \Delta_{\mu\sigma}(x)
\Big(\hat{\mathscr J}_{c\beta}(x),\hat{\mathscr J}^{\beta}_a(x_1),\hat{p}^*(x_2)\Big),\\
\label{eq:cor_current_mixed2}
\Big(\hat{\mathscr J}_{c\mu}(x),\hat{\mathscr J}_{a\sigma}(x_1),
\hat{\pi}_{\alpha\beta}(x_2)\Big)&=& \frac{1}{5}
\Delta_{\mu\sigma\alpha\beta}(x)
\Big(\hat{\mathscr J}^{\gamma}_c(x),\hat{\mathscr J}^{\delta}_a(x_1),
\hat{\pi}_{\gamma\delta}(x_2)\Big).
\eea
We now define the following coefficients
\bea\label{eq:zeta_3_ca}
\zeta_{\!{\mathscr J}}^{ca} &=& \frac{2\beta }{3}\!\int\!
 d^4x_1d^4x_2 \Big(\hat{\mathscr J}_{c\gamma}(x),
\hat{\mathscr J}_{a}^\gamma (x_1), \hat{p}^*(x_2)\Big),\\
\label{eq:lambda_3_ca}
\lambda_{\!{\mathscr J}}^{ca} &=& \frac{2\beta}{5}\!\int\!
 d^4x_1d^4x_2 \Big(\hat{\mathscr J}_{c}^{\gamma}(x),
\hat{\mathscr J}_{a}^{\delta}(x_1),
\hat{\pi}_{\gamma\delta}(x_2)\Big).
\eea
Then, from Eqs.~\eqref{eq:current_av_23} -- \eqref{eq:lambda_3_ca} and
from the symmetry property \eqref{eq:3_point_corr_sym} we obtain
\bea\label{eq:current_av_23_2}
\langle \hat{\mathscr J}_{c\mu}\rangle_2^3 =
\sum\limits_a\Big( \zeta_{\!{\mathscr J}}^{ca} 
\theta \nabla_\mu\alpha_a-\lambda_{\!{\mathscr J}}^{ca} 
\sigma_{\mu\nu} \nabla^\nu\alpha_a \Big).
\eea

\subsubsection{Final result for the diffusion currents}
\label{sec:final_diff}

Combining Eqs.~\eqref{eq:charge_currents1}, 
\eqref{eq:stat_average_2_new}, \eqref{eq:current_av_21_4},
\eqref{eq:current_av_22_1}, and \eqref{eq:current_av_23_2} 
we obtain the diffusion currents up to second order,
\bea\label{eq:current_total_final}
{\mathscr J}_{c\mu}
&=&\sum\limits_b\chi_{cb}\nabla_\mu \alpha_b +
\sum\limits_a \tilde{\chi}_{ca}\Delta_{\mu\beta}
D (\nabla^\beta\alpha_a)
-
\tilde{\chi}_{cq} h^{-2} \Big(\gamma h+\sum\limits_d 
\delta_d n_d\Big) \theta\sum\limits_b 
n_b \nabla_\mu\alpha_b\nonumber \\
&+&
\chi_{cq}\beta h^{-1}\big(-\nabla_\mu \Pi
+\Pi \dot{u}_\mu + \dot{q}_{\mu}
 + q^{\nu}\partial_\nu u_{\mu}+ q_{\mu}\theta 
 +\Delta_{\mu\sigma}\partial_\nu \pi^{\nu\sigma}\big)\nonumber \\
 &+&
\sum\limits_b\Big( \zeta_{\!{\mathscr J}}^{cb} 
\theta \nabla_\mu\alpha_b
-\lambda_{\!{\mathscr J}}^{cb} \sigma_{\mu\nu} 
\nabla^\nu\alpha_b \Big).
\eea
In order to obtain relaxation equations for the diffusion currents
 we invert Eq.~\eqref{eq:charge_currents1} as 
\bea
\nabla^\beta \alpha_a 
=\sum\limits_b(\chi^{-1})_{ab}\mathscr{J}_b^\beta,\no
\eea 
and employ it to modify the second term in
Eq.~\eqref{eq:current_total_final}.  Using Eqs.~\eqref{eq:D_beta1}, 
\eqref{eq:D_alpha1}, and \eqref{eq:dot_def} we obtain
\bea\label{eq:diff_relax1}
\sum\limits_a \tilde{\chi}_{ca}\Delta_{\mu\beta}
D (\nabla^\beta\alpha_a)
&=&-\sum\limits_b \tau_{\!\mathscr J}^{cb}
\mathscr{\dot{J}}_{b\mu}\no\\
&+&\beta\theta \sum\limits_{ab} \tilde{\chi}_{ca}
\left[\gamma\frac{\partial(\chi^{-1})_{ab}}{\partial \beta}
-\sum\limits_d \delta_d\frac{\partial(\chi^{-1})_{ab}}
{\partial \alpha_d}\right]\!\mathscr{J}_{ b\mu}, 
\eea 
where we defined the relaxation-time matrix 
\bea\label{eq:diff_ab_relax}
\tau_{\!\mathscr J}^{cb} = -
(\tilde{\chi}\chi^{-1})_{cb}=
-\sum\limits_a \tilde{\chi}_{ca}(\chi^{-1})_{ab}.
\eea
Introducing the coefficients
\bea\label{eq:lambda_tilde}
\tilde{\lambda}^{cb}_{\mathscr J} &=& \beta
\sum\limits_{a} \tilde{\chi}_{ca}
\left[\gamma\frac{\partial(\chi^{-1})_{ab}}{\partial \beta}
-\sum\limits_d \delta_d\frac{\partial(\chi^{-1})_{ab}}
{\partial \alpha_d}\right],\\
\label{eq:chi_star}
\chi^*_{cb} &=& 
\zeta_{{\mathscr J}}^{cb}-\tilde{\chi}_{cq}n_b h^{-2} 
\Big(\gamma h+\sum\limits_d \delta_d n_d\Big),
\eea
we obtain 
\bea\label{eq:current_final_relax}
\sum\limits_b \tau_{\!\mathscr J}^{ab}\!\mathscr{\dot{J}}_{b\mu}+ {\mathscr J}_{a\mu}
=\sum\limits_b\Big[\chi_{ab}\nabla_\mu \alpha_b 
+  \tilde{\lambda}^{ab}_{\!\mathscr J} 
\theta\!\mathscr{J}_{ b\mu} + \chi^*_{ab} 
\theta\nabla_\mu\alpha_b -
{\lambda}^{ab}_{\!\mathscr J} 
 \sigma_{\mu\nu} \nabla^\nu\alpha_b \Big]\nonumber\\
+
\chi_{aq}\beta h^{-1}\big(-\nabla_\mu \Pi
+\Pi \dot {u}_\mu + \dot{q}_{\mu}
 + q^{\nu}\partial_\nu u_{\mu}+ q_{\mu}\theta 
 +\Delta_{\mu\sigma}\partial_\nu \pi^{\nu\sigma}\big). 
\eea
If we have only one conserved charge species, then
Eq.~\eqref{eq:current_final_relax} simplifies to
\bea\label{eq:current_final_relax1}
\tau_{\!\mathscr J}\!\mathscr{\dot{J}}_{\mu}+ {\mathscr J}_{\mu}
&=&\chi \nabla_\mu \alpha 
+  \tilde{\lambda}_{\!\mathscr J} 
\theta\!\mathscr{J}_{\mu} + \chi^* 
\theta\nabla_\mu\alpha -
{\lambda}_{\!\mathscr J} 
 \sigma_{\mu\nu} \nabla^\nu\alpha \nonumber\\
&+&
\chi_{q}\beta h^{-1}\big(-\nabla_\mu \Pi
+\Pi \dot {u}_\mu + \dot{q}_{\mu}
 + q^{\nu}\partial_\nu u_{\mu}+ q_{\mu}\theta 
 +\Delta_{\mu\sigma}\partial_\nu \pi^{\nu\sigma}\big), 
\eea
where the current relaxation time is given by 
[see Eqs.~\eqref{eq:chi_ab_tilde} and \eqref{eq:diff_ab_relax}] 
\bea\label{eq:diff_current_relax}
\chi \tau_{\!\mathscr J} = 
 -i\frac{d}{d\omega}\chi(\omega)\bigg\vert_{\omega=0}
=-\frac{T}{6}\frac{d^2}{d\omega^2} {\rm Re}
G^R_{\hat{{\mathscr J}_{\mu}}
\hat{\mathscr J}^{\mu}}(\omega)\bigg\vert_{\omega=0}, 
\eea
and 
\bea\label{eq:lambda_tilde1}
\tilde{\lambda}_{\!\mathscr J} &=& \tau_{\!\mathscr J} 
\beta\chi^{-1}\left(\gamma\frac{\partial\chi}{\partial \beta}
-\delta\frac{\partial\chi}{\partial\alpha}\right),\\
\label{eq:chi_star1}
\chi^* &=& \zeta_{\!{\mathscr J}}-
\tilde{\chi}_{q}n h^{-2} \big(\gamma h+\delta n\big).
\eea
Recalling the relation
\bea\label{eq:kappa_chi}
\kappa=\left(\frac{h}{nT}\right)^2\chi
\eea
between the diffusion coefficient and the thermal conductivity and
Eq.~\eqref{eq:op_heat_flux}, we can write the relaxation time
$\tau_{\!\mathscr J}\equiv\tau_h$ also in the following form
\bea\label{eq:kappa_relax}
\kappa\tau_{h}=-i\frac{d}{d\omega}\kappa(\omega)\bigg\vert_{\omega=0}
=-\frac{\beta}{6}\frac{d^2}{d\omega^2} {\rm Re}
G^R_{\hat{h}_{\mu}\hat{h}^{\mu}}(\omega)\bigg\vert_{\omega=0}. 
\eea
The frequency-dependent coefficients $\chi$ and $\kappa$ in
Eqs.~\eqref{eq:diff_current_relax} and \eqref{eq:kappa_relax} are
defined according to the formula~\eqref{eq:I_XY} in
\ref{app:Green_func} with the pertaining choice of operators.

\section{Discussion}
\label{sec:hydro_discuss}

\subsection{General structure of the second-order dissipative
hydrodynamic equations}
\label{sec:discuss_eqs}

The complete set of evolution equations for the
dissipative currents obtained in the previous section reads [see
Eqs.~\eqref{eq:shear_final_relax}, \eqref{eq:bulk_final_relax}, and
\eqref{eq:current_final_relax}]
\bea\label{eq:shear_final}
\tau_\pi \dot{\pi}_{\mu\nu}+ {\pi}_{\mu\nu}
&=& 2\eta \sigma_{\mu\nu}
+\tilde{\lambda}_\pi\theta\pi_{\mu\nu}
+ \lambda \theta {\sigma}_{\mu\nu}\no\\
&&+ \lambda_\pi \sigma_{\rho<\mu}\sigma_{\nu>}^{\rho}
+\sum\limits_{ab}\lambda_{\pi{\!\mathscr
    J}}^{ab}\nabla_{<\mu}\alpha_a\nabla_{\nu>}
\alpha_b ,\\
\label{eq:bulk_final}
\tau_\Pi \dot{\Pi}+\Pi &=&
-\zeta\theta 
+  \tilde{\lambda}_\Pi \theta \Pi +\varsigma\theta^2
+\zeta_\beta (\sigma_{\mu\nu}\pi^{\mu\nu} - \theta\Pi )
-\lambda_{\Pi\pi} \sigma_{\mu\nu}\sigma^{\mu\nu}\no\\
 &&+ \sum\limits_a \zeta_{\alpha_a} 
\partial_\mu {\mathscr J}^{\mu}_a -\tilde{\zeta}_\beta \partial_\mu
q^{\mu}
 +q^\mu \Big[\zeta_\beta \dot{u}_\mu +\sum\limits_a \zeta_{\alpha_a} 
 \nabla_\mu (n_ah^{-1})\Big] \nonumber\\
 &&
 +T\sum\limits_{ab}\zeta^{ab}_\Pi
\nabla^\mu \alpha_a\nabla_\mu\alpha_b,\qquad\qquad\\
\label{eq:current_final}
\sum\limits_b\tau_{\!\mathscr J}^{ab}\!
\mathscr{\dot{J}}_{b\mu}+{\mathscr J}_{a\mu}
&=&\sum\limits_b\Big[\chi_{ab}\nabla_\mu \alpha_b 
+  \tilde{\lambda}^{ab}_{\!\mathscr J} 
\theta\! \mathscr{J}_{ b\mu} + \chi^*_{ab} 
\theta\nabla_\mu\alpha_b -{\lambda}^{ab}_{\!\mathscr J} 
 \sigma_{\mu\nu} \nabla^\nu\alpha_b \Big]\nonumber\\
&& +
\chi_{aq}\beta h^{-1}\big(-\nabla_\mu \Pi
+\Pi \dot {u}_\mu + \dot{q}_{\mu}
+ q^{\nu}\partial_\nu u_{\mu}+ q_{\mu}\theta 
 +\Delta_{\mu\sigma}\partial_\nu \pi^{\nu\sigma}\big),\qquad
\eea
where the dot denotes the comoving derivative
\bea\label{eq:corrent_dot}
&&\dot{\Pi}=D\Pi,\qquad
\dot{\pi}_{\mu\nu}=\Delta_{\mu\nu\rho\sigma}
D{\pi}^{\rho\sigma}, \qquad 
\dot{u}_\mu = Du_\mu,\\
&&\dot{q}_\mu = \Delta_{\mu\nu} D q^{\nu},\qquad
\mathscr{\dot{J}}_{a\mu}=\Delta_{\mu\nu} D\!\!\mathscr{J}^\nu_a. 
\eea

The first terms on the right-hand sides of
Eqs.~\eqref{eq:shear_final} -- \eqref{eq:current_final} represent the
corresponding Navier--Stokes contributions. The first-order
coefficients $\eta$, $\zeta$, and $\chi_{ab}$ are the shear viscosity,
the bulk viscosity, and the matrix of diffusion coefficients,
respectively; these coefficients are expressed in terms of retarded two-point 
correlation functions via the Kubo
formulas~\eqref{eq:shear_bulk_mod} -- \eqref{eq:green_func}.

The first terms on the left-hand sides of
Eqs.~\eqref{eq:shear_final} -- \eqref{eq:current_final} signify the
relaxation of the dissipative currents towards their leading-order (Navier--Stokes)
values, if they differ from those at the initial moment.  The
coefficients $\tau_\pi$, $\tau_\Pi$, and $\tau_{\! \mathscr J}^{ab}$
represent the characteristic time scales on which this relaxation
occurs. For example, in the case of $\sigma_{\mu\nu}=0$,
Eq.~\eqref{eq:shear_final} implies an exponential decay of the shear-stress 
tensor on the time scale given by $\tau_\pi$, \ie,
$\pi_{\mu\nu}\propto \exp(-t/\tau_\pi)$ (if we ignore the other
second-order terms).

The relaxation times $\tau_\pi$, $\tau_\Pi$, and
$\tau_{\! \mathscr J}^{ab}$ are related to the corresponding first-order
transport coefficients.  For example, the shear relaxation time is
given by a Kubo-type formula
\bea\label{eq:tau_pi}
\eta\tau_\pi = -i\frac{d}{d\omega}\eta(\omega)\bigg\vert_{\omega=0}
=\frac{1}{20}\frac{d^2}{d\omega^2} {\rm Re}G^R_{\hat{\pi}_{\mu\nu}
\hat{\pi}^{\mu\nu}}(\omega)\bigg\vert_{\omega=0},
\eea
where the retarded Green's function is defined in
Eq.~\eqref{eq:green_func}. Here $\eta(\omega)$ is the generalization of the
shear viscosity to nonvanishing frequencies and is defined via the
generalization of Eq.~\eqref{eq:shear_def} according to
Eq.~\eqref{eq:I_XY}. The positivity of $\tau_\pi$ can be anticipated
from Eqs.~\eqref{eq:tau_pi} and \eqref{eq:I_XY}, upon taking into
account that the $(\pi_{\mu\nu}, \pi^{\mu\nu})$ correlator, which determines
the shear viscosity $\eta$, should be positive.  Similar formulas as
Eq.~\eqref{eq:tau_pi}  hold also for the bulk and the
diffusion relaxation times [see Eqs.~\eqref{eq:zeta_relax},
\eqref{eq:chi_ab_tilde}, \eqref{eq:diff_ab_relax}, and
\eqref{eq:diff_current_relax}]. Our formulas for the shear 
and the bulk relaxation times are in general consistent with those 
obtained in Refs.~\cite{Baier2008JHEP,Romatschke2010CQGra,Moore2011PhRvL}, 
but may differ in particular details. A direct comparison 
is not straightforward because of the difference in methods and
approaches applied in those works.

The physical meaning of the formula~\eqref{eq:tau_pi} for $\tau_\pi$
is easy to understand.  As we showed in the previous section, the
relaxation terms originate from the non-local (memory) effects encoded
in the non-equilibrium statistical operator. In the case where these
memory effects are neglected, \ie, in the first-order theory, the
proportionality between $\pi_{\mu\nu}$ and $\sigma_{\mu\nu}$ is given
by the zero-frequency (static) limit of the shear viscosity, as seen
from Eqs.~\eqref{eq:shear_bulk_1} and \eqref{eq:shear_bulk_mod}. The
effects of finite memory in the dissipation of shear stresses imply 
actually a dispersion (\ie, frequency dependence) in the shear
viscosity, which at the leading order should be accounted for by the
first frequency derivative of $\eta(\omega)$, as seen from
Eq.~\eqref{eq:tau_pi}. Therefore, we conclude that the memory effects
naturally generate finite relaxation time scales in the transport
equations, as stressed earlier in
Refs.~\cite{Zubarev1972Phy,Koide2007PhRvC,Koide2008PhRvE}.

The second terms on the right-hand sides of
Eqs.~\eqref{eq:shear_final} -- \eqref{eq:current_final} arise as a
consequence of inhomogeneities in the first-order coefficients,
because these are functions of the temperature and the chemical
potentials, which vary in space and time.  The coefficients
$\tilde{\lambda}_\pi$, $\tilde{\lambda}_\Pi$, and
$\tilde{\lambda}^{ab}_{\!\mathscr J}$ involve derivatives of the
corresponding first-order transport coefficients with respect to 
temperature and chemical potentials, see
Eqs.~\eqref{eq:tilde_lambda_pi}, \eqref{eq:tilde_lambda_Pi}, and
\eqref{eq:lambda_tilde}.

In Eqs.~\eqref{eq:shear_final} -- \eqref{eq:current_final} we identify
three classes of second-order terms which are not of the
relaxation-type: (i) terms which contain products of the thermodynamic
forces with dissipative currents [\eg, the term $\propto\theta \Pi$ in
Eq.~\eqref{eq:bulk_final}]; (ii) terms which contain space-like
derivatives of the dissipative currents (\eg,
$\partial_\mu {\mathscr J}^{\mu}_a$); and (iii) terms which 
include a product of two thermodynamic forces (\eg,
$\sigma_{\mu\nu}\sigma^{\mu\nu}$).  The terms of the type (i)
originate either from the non-local corrections [second terms on the
right-hand sides of
Eqs.~\eqref{eq:shear_final} -- \eqref{eq:current_final}], or from the
second-order corrections to the operator $\hat{C}$ [see
Eq.~\eqref{eq:op_C2}].  The corrections of the type (ii) arise purely
from the operator $\hat{C}_2$.  As we discussed in
Sec.~\ref{sec:C_decomp2}, the operator $\hat{C}_2$ originates from the
dissipative terms in the hydrodynamic 
equations~\eqref{eq:hydro1} -- \eqref{eq:hydro3} and
can be viewed as an {\it extended thermodynamic force}. Thus, the
coefficients $\zeta_\beta$, $\tilde{\zeta}_\beta$, and
${\zeta}_{\alpha_a}$ in Eq.~\eqref{eq:bulk_final} and $\chi_{aq}$ in
Eq.~\eqref{eq:current_final} represent the mixing of the transport
equations with the conservation laws.  We also note that such mixed
terms are absent in the relaxation equation~\eqref{eq:shear_final} for
the shear-stress tensor.  The transport coefficients in the terms of
type (i) and (ii) are related to two-point correlation functions [see
Eqs.~\eqref{eq:zeta_beta}, \eqref{eq:zeta_alpha_a},
\eqref{eq:tilde_zeta_beta}, and \eqref{eq:kappa_c}].

 The corrections of the type (iii) contain all possible combinations
 which are quadratic in the thermodynamic forces $\sigma_{\mu\nu}$,
 $\theta$, and $\nabla_\mu\alpha_a$. For example, the relevant
 corrections for the shear-stress tensor are given by three terms
 which are allowed by the symmetries: $\theta {\sigma}_{\mu\nu}$,
 $\sigma_{\rho<\mu}\sigma_{\nu>}^{\rho}$, and
 $\nabla_{<\mu}\alpha_a\nabla_{\nu>}\alpha_b$.  The transport
 coefficients coupled to these terms involve three-point correlation
 functions, which account for nonlinear couplings between different
 dissipative processes.

Note that because of the terms of the type (iii) the transport
equations become parabolic, and, therefore, acausal and
unstable~\cite{Denicol2012PhRvD,Finazzo2015JHEP}. This problem
can be circumvented by modifying some of the nonlinear terms using the
Navier--Stokes equations, \eg, $\lambda_\pi \sigma_{\rho<\mu}\sigma_{\nu>}^{\rho}$
can be replaced by
$(\lambda_\pi/2\eta)
\pi_{\rho<\mu}\sigma_{\nu>}^{\rho}$~\cite{Denicol2012PhRvD,Finazzo2015JHEP}.
In this case, we recover most of the second-order terms 
derived in Ref.~\cite{Betz2011EPJWC}.

\subsection{Comparison with other studies}
\label{sec:shear_compare}

In this subsection, we discuss in more detail the second-order
expression for the shear-stress tensor and compare it with the results
of other studies.

For the sake of simplicity, we will consider a fluid without conserved
charges. Equation~\eqref{eq:gamma_delta_a} then implies 
$\gamma \equiv c_s^2$, with $c_s$ being the speed of sound.  It is
more suitable to use here the expression for the shear-stress tensor
given by Eq.~\eqref{eq:shear_total_final}
\bea\label{eq:shear_discuss}
{\pi}_{\mu\nu}
=  2\eta \sigma_{\mu\nu}
-2\eta\tau_\pi (\dot{\sigma}_{\mu\nu}+ c_s^2 \theta \sigma_{\mu\nu})
+ 2\lambda_{\pi\Pi}\theta \sigma_{\mu\nu}
+ \lambda_\pi \sigma_{\alpha<\mu}\sigma_{\nu>}^{\alpha},
\eea
where
$\dot{\sigma}_{\mu\nu}\equiv\Delta_{\mu\nu\rho\sigma}D
\sigma^{\rho\sigma}$.  We recall that the terms in the parentheses are
those which arise from the non-locality of the thermodynamic
forces in the statistical operator (see Sec.~\ref{sec:2nd_shear}).
The next two terms in Eq.~\eqref{eq:shear_discuss} arise from the
quadratic terms of the expansion of the statistical operator in
thermodynamic forces.  Thus, the second-order corrections to the shear-stress 
tensor in the absence of conserved charges contain three new
coefficients: $\tau_\pi$, $\lambda_\pi$, and $\lambda_{\pi\Pi}$. 

The second-order terms in the parentheses in Eq.~\eqref{eq:shear_discuss}
have a simple physical interpretation. The effects of non-locality
generate two distinct terms in the shear-stress tensor. The first term
in the parentheses involves the comoving derivative of the
thermodynamic force $\sigma_{\mu\nu}$ and incorporates the effect of
the acceleration of the fluid on account of the shear stresses. In other words,
this term contains information about the velocity stresses
$\sigma_{\mu\nu}$ from the previous moments in time.  The relaxation
time $\tau_\pi$ measures how long this information remains in the
``memory" of the shear-stress tensor $\pi_{\mu\nu}$.  Thus, the first
term in parentheses in Eq.~\eqref{eq:shear_discuss} can be associated 
with the non-locality of the statistical operator {\it in time}, \ie, 
it is related to {\it memory effects}.

We argue that the second term in parentheses in
Eq.~\eqref{eq:shear_discuss} accounts for {\it spatially} non-local
effects. Indeed, this term involves the product of the thermodynamic
force $\sigma_{\mu\nu}$ with the fluid expansion rate
$\theta=\partial_\mu u^\mu$, which can be regarded as a relevant
(scalar) measure of how strong the spatial ``non-locality" in the
fluid velocity field is. This term describes how the shear-stress
tensor is distorted by uniform expansion or contraction of the fluid.

Next, we discuss the last two terms in Eq.~\eqref{eq:shear_discuss}.
These terms are quadratic in the thermodynamic forces
$\sigma_{\mu\nu}$ and $\theta$. The relevant second-order transport
coefficients are expressed via three-point correlation functions by
the formulas \eqref{eq:lambda_pi} and \eqref{eq:lambda_2}.
The coefficient $\lambda_\pi$ describes the nonlinear effects of two
velocity stresses on $\pi_{\mu\nu}$.  By analogy with the relevant
linear transport coefficient $\eta$, which measures the correlation
between two shear stresses, the second-order coefficient $\lambda_\pi$
measures the correlation between three shear stresses. The coefficient
$\lambda_{\pi\Pi}$ describes the nonlinear coupling between the shear-
and the bulk-viscous processes. Similarly, this coefficient is given
by a three-point correlation function between two shear stresses and
the bulk-viscous pressure.

We remark that $\lambda_{\pi\Pi}$ term in Eq.~\eqref{eq:shear_discuss}
has the same gradient structure $\theta\sigma_{\mu\nu}$ as the second
term in the parentheses. However, despite this formal similarity, these
two terms have different origins and, therefore, different physical
interpretations. As explained above, the relevant term
$\propto \tau_\pi$ originates from {\it non-local} effects in the
statistical distribution, whereas the term $\propto \lambda_{\pi\Pi}$
stands purely for {\it nonlinear} coupling between the bulk- and the
shear-viscous effects.  In this sense, it is natural to regard as
nonlinear only the term $\propto \lambda_{\pi\Pi}$, but not the term
$\propto\tau_\pi$. A similar classification of the second-order terms
was suggested also in Ref.~\cite{Moore2012JHEP}.

It is instructive to compare our expression for the shear-stress
tensor~\eqref{eq:shear_discuss} to the one in
Ref.~\cite{Baier2008JHEP} for conformal fluids.  The most general
second-order expression for the shear-stress tensor of a conformal
fluid in flat space-time reads \footnote{Note that
  Ref.~\cite{Baier2008JHEP} uses a metric convention which differs by
  an overall sign from ours,
  and their definition of the shear viscosity differs from ours by a
  factor of 2.}
\bea\label{eq:shear_Baier2008}
  \pi_{\mu\nu}^{c} = 2\eta \sigma_{\mu\nu}
  - 2\eta\tau_\pi \left( \dot{\sigma}_{\mu\nu}
    + \frac{1}{3}\theta\sigma_{\mu\nu} \right)
  + \lambda_1 {\sigma}_{\alpha <\mu} \sigma_{\nu>}^{\alpha}
  + \lambda_2 {\sigma}_{\alpha<\mu} \omega_{\nu>}^{\alpha}
  + \lambda_3 {\omega}_{\alpha<\mu} \omega_{\nu>}^{\alpha},
\eea 
where
$\omega_{\alpha\beta}=(\nabla_{\alpha}u_{\beta}-
\nabla_{\beta}u_{\alpha})/2$ is the vorticity tensor. Note that
  we neglected the vorticity tensor from the outset assuming that
  the fluid is irrotational. As a consequence, our transport equations
  \eqref{eq:shear_final} -- \eqref{eq:current_final} do not contain
  terms involving vorticity. These, however, can be restored if the
  energy-momentum tensor in the local-equilibrium distribution is
  extended to include non-zero vorticity~\cite{Buzzegoli2017,Buzzegoli2018}.  

In the case of a conformal fluid we have $ c_s^2=1/3$.  Furthermore,
because the conformal invariance implies vanishing bulk-viscous
pressure, it is natural to expect that the correlations involving the
relevant operator $\hat{p}^*$ [see Eq.~\eqref{eq:p_star}] vanish as
well, \ie, $\lambda_{\pi\Pi}=0$ in this case. We then recover from
Eq.~\eqref{eq:shear_discuss} the term involving $\tau_\pi$ in
Eq.~\eqref{eq:shear_Baier2008}. Hence we conclude that the terms
$\propto\tau_\pi$ in Eq.~\eqref{eq:shear_Baier2008} given in
Ref.~\cite{Baier2008JHEP} have a non-local origin.  For the rest of
the terms we identify $\lambda_1=\lambda_{\pi}$,
$\lambda_2=\lambda_3=0$.

In the case of non-conformal fluids Eq.~\eqref{eq:shear_Baier2008}
has two additional terms (in flat
space-time)~\cite{Romatschke2010CQGra}. One of these terms shares the
same structure with the term $-2\eta\tau_\pi\theta\sigma_{\mu\nu}/3$
and can be written, after converting to our notations, as
$-2\eta\tau_\pi^*\theta\sigma_{\mu\nu}/3$. Comparing with our
expression~\eqref{eq:shear_discuss}, we identify
$\tau^*_\pi=\tau_\pi(3c_s^2-1)-3\lambda_{\pi\Pi}/\eta$. Our formula
for $\tau^*_\pi$ contains an additional term $\propto\lambda_{\pi\Pi}$
compared to the formula given in Ref.~\cite{Romatschke2010CQGra} for a
special class of strongly coupled fluids.

\section{Concluding remarks}
\label{sec:hydro_conclusions}

In this work, we provided a novel derivation of relativistic second-order
dissipative hydrodynamics for strongly correlated systems. We adopted
Zubarev's non-equilibrium statistical-operator formalism and extended
the existing studies of relativistic dissipative hydrodynamics within
this formalism up to second order in thermodynamic gradients. We
considered a multicomponent quantum system in the hydrodynamic regime,
where it is described utilizing the energy-momentum tensor and the
currents of conserved charges.

Our starting point is Zubarev's method of deriving the full
non-equilibrium statistical operator from the quantum Liouville
equation.  Starting from the exact solution of the Liouville equation
in the form of a non-local functional of thermodynamic parameters and
their space-time derivatives we performed an expansion of the
statistical operator up to second order with respect to the
thermodynamic gradients to obtain approximate solutions which capture
the low-frequency and the long-wavelength dynamics of the system.  In
this manner wex derived second-order evolution equations for the
shear-stress tensor, the bulk-viscous pressure, and the
flavor-diffusion currents under the assumption that the fluid is
irrotational. In particular, we obtained new non-local terms which do
not appear in the first-order treatments of
Refs.~\cite{Zubarev1979TMP,Hosoya1984AnPhy,Huang2011AnPhy,Hayata2015}.
We classified the second-order terms by observing that they arise from
two different sources: (i) the quadratic terms in the Taylor expansion
of the statistical operator; (ii) the linear terms of this expansion
with proper inclusion of effects of memory and non-locality. The terms
of the type (i) generate corrections which are quadratic (nonlinear)
in thermodynamic forces. The terms of the type (ii) generate
relaxation terms for the dissipative currents, which are required for
reasons of causality. Apart from these non-local terms, the
corrections from class (ii) include also additional second-order terms
which account for the mixing of the transport equations with the
conservation laws. These last types of corrections contribute only to
the bulk-viscous pressure and to the diffusion currents.

We obtained also formal expressions for all second-order transport
coefficients in terms of certain two- and three-point equilibrium
correlation functions, the computation of which can be performed by
applying standard thermal field-theory
methods~\cite{le_bellac_1996}. In particular, we derived Kubo-type
formulas for the relaxation times, which are given via the
frequency derivatives of the relevant first-order transport
coefficients, taken in the relevant zero-frequency limit.

It was demonstrated that in the absence of diffusion currents the
second-order expression for the shear-stress tensor contains in
general three second-order transport coefficients: the
shear relaxation time $\tau_\pi$ and two other coefficients, which are
responsible for nonlinear couplings between shear and bulk viscous
effects.

It would be interesting to compute the second-order transport
coefficients derived in this work for strongly interacting matter, for
example, in the framework of the Nambu--Jona-Lasinio model of
QCD~\cite{Lang2015EPJA,Harutyunyan2017PhRvD_a,Harutyunyan2017PhRvD_b}.
Of special importance are the shear and the bulk relaxation times,
which are necessary for hydrodynamic simulations of heavy-ion
collisions as well as binary neutron-star mergers.

\section{Acknowledgments}

This research was funded by the collaborative research grant No. 97029
of Volkswagen Foundation (Hannover, Germany). The work of A.S.\ is  supported by
the Deutsche Forschungsgemeinschaft (DFG, German Research Foundation)
Grant No. SE 1836/5-1 and the  European COST Action “PHAROS” (CA16214).
The work of D.H.R.\ is supported by the Deutsche Forschungsgemeinschaft 
(DFG, German Research Foundation) through the Collaborative Research Center 
CRC-TR 211 ``Strong-interaction matter under extreme conditions'' -- 
project number 315477589 - TRR 211.

\appendix

\section{Frames}
\label{app:frames}

In ideal hydrodynamics, the energy-momentum and the charge currents are always
parallel, and the fluid rest frame is defined as the frame where all
these currents vanish.  The simultaneous presence of energy- and
charge-diffusion currents in the case of dissipative fluids makes the
definition of the fluid velocity and the fluid rest frame
ambiguous. There are two natural ways to define the fluid rest frame,
which we will discuss in this appendix.

{\it Landau frame.}  One of the natural choices of the fluid rest
frame is the frame where the 3-momentum current is zero (Landau frame or
L-frame)~\cite{Landau1987}. In this case, $u^\mu$ is chosen to be the
time-like eigenvector of $T^{\mu\nu}$,
\bea\label{eq:vel_Landau}
u_L^\mu =\frac{u_{L\nu} T^{\mu\nu}}{\sqrt{u_{L\nu} 
T^{\mu\nu} u^{\lambda}_L T_{\mu\lambda}}}, 
\eea
which together with Eqs.~\eqref{eq:T_munu_decomp}, \eqref{eq:orthogonality},
and \eqref{eq:proj1_op} implies
\bea\label{eq:eps_Landau}
\epsilon_L =\sqrt{u_{L\nu} T^{\mu\nu} u^\lambda_L T_{\mu\lambda}},\qquad
u_{L\nu} T^{\mu\nu}=\epsilon_L u_{L}^\mu,
\qquad q^\mu_L =0,
\eea
where the index $L$ labels the quantities evaluated according to the
Landau definition of $u^\mu$.  Thus, with this choice of the velocity
field the energy-diffusion current vanishes, and heat-transport
phenomena are expressed via the charge-diffusion currents
$j_{La}^\mu$.  The 4-momentum current  
\bea\label{eq:momentum_dens_diss}
P^\mu \equiv u_\nu T^{\mu\nu}+pu^\mu  =h u^\mu +q^\mu
\eea
in the L-frame is parallel to the fluid velocity (because $q_L^\mu = 0$), which allows 
one to write Eq.~\eqref{eq:vel_Landau} in an alternative way
\bea\label{eq:vel_Landau1}
u_L^\mu =\frac{P^{\mu}_L}{\sqrt{P^{\mu}_L P_{L\mu}}}.
\eea

To find the relation between $u^\mu_L$ and a generic velocity
$u^\mu$ we note that in a generic fluid rest frame the current
\eqref{eq:momentum_dens_diss} reads $P^{\mu}=(h,q^i)$, therefore the
boost velocity from an arbitrarily defined rest frame ($u^i=0$) to the
Landau rest frame ($u^i_L=0$) is $v^i_L=q^i/h={\cal O}_1$ (following
Refs.~\cite{Israel1976AnPhy,Israel1979AnPhy}, here we introduced
the symbol ${\cal O}_n$ to denote the quantities of $n$th order
in deviations from equilibrium). The transformation of the charge currents
into the L-frame then reads
\bea\label{eq:N_i_trans}
N^i_{La} = N^i_a-v^i_LN^0_a+{\cal O}_2.\no
\eea
Substituting here $N^i_a=j^i_{a}$, $N^i_{La}=j^i_{La}$, and
$N^0_a =n_a$, we obtain the charge-diffusion currents measured in the
Landau rest frame,
\bea\label{eq:diff_curr_Li}
{j}^{i}_{La}=
{j}^{i}_{a}-\frac{n_a}{h}q^i +{\cal O}_2.
\eea
Note that the transformed current ${j}^{i}_{La}$ is evaluated at
the transformed coordinate $x'$, but the difference
${j}^{i}_{La}(x')-{j}^{i}_{La}(x)\simeq (x'-x)^\tau\partial_\tau
{j}^{i}_{La}\propto v_L  |\nabla {j}^{i}_{La}|$ 
is already of order ${\cal O}_3$ 
and can be ignored.  Taking into account that in
the fluid rest frame ${j}^{0}_{La}=j^0_a=q^0=0$, we can cast
Eq.~\eqref{eq:diff_curr_Li} into a covariant form
\bea\label{eq:diff_curr_L}
{j}^{\mu}_{La}
={j}^{\mu}_{a}-\frac{n_a}{h}q^\mu +{\cal O}_2,
\eea
which is valid in an arbitrary frame, 
\ie, not only in the fluid rest frame. The 4-currents
\bea\label{eq:diff_curr1}
\mathscr{J}_a^\mu={j}^{\mu}_{a}-\frac{n_a}{h}q^\mu
= {N}^{\mu}_{a}-\frac{n_a}{h}P^\mu
\eea
are the charge-diffusion currents with respect to the 4-momentum flow, \ie,
the charge currents in the absence of energy-diffusion currents. 
Although the energy-diffusion current $q^\mu$ and the charge-diffusion 
currents $j^\mu_a$ depend on the choice of the velocity field,
the combination \eqref{eq:diff_curr1} remains invariant under
first-order changes in $u^\mu$. 

The remaining thermodynamic variables entering
Eqs.~\eqref{eq:T_munu_decomp} and \eqref{eq:N_a_decomp} change only
in second order in thermodynamic gradients under the change of
$u^\mu$~\cite{Israel1976AnPhy,Israel1979AnPhy}.  We then obtain
the following relation between $u$ and $u_L$
\bea\label{eq:u_Landau}
u^\mu_L =u^\mu +\frac{q^\mu}{h}+{\cal O}_2,
\eea
which follows immediately from a comparison of
Eqs.~\eqref{eq:momentum_dens_diss} and \eqref{eq:vel_Landau1}.

{\it Eckart frame.}  According to the Eckart definition the velocity
field is chosen to be parallel to one of the conserved currents
$N^\mu_a$ (E-frame).  In the case where there is only one species of
conserved charge (for example, the net particle number), we have
$N^\mu=n u^\mu+j^\mu$, and the 4-velocity is defined
as~\cite{Eckart1940PhRv}
\bea\label{eq:vel_Eckart}
u_E^\mu=\frac{N^\mu}{\sqrt{N^\mu N_\mu}}, 
\eea
which together with Eqs.~\eqref{eq:T_munu_decomp}, 
\eqref{eq:orthogonality}, and \eqref{eq:proj1_op} implies
\bea\label{eq:n_Eckart}
n_E=\sqrt{N^\mu N_\mu},\qquad
N^\mu =n_E u_E^\mu,\qquad
j^\mu_E =0,
\eea
\ie, the particle-diffusion flux is absent in this case.  The index
$E$ in Eqs.~\eqref{eq:vel_Eckart} and \eqref{eq:n_Eckart} labels the
E-frame.

The boost velocity from a generic rest frame to the Eckart rest frame
is $v^i_E=j^i/n$, and the velocities $u^\mu$ and $u^\mu_E$ are related
via 
\bea\label{eq:u_Eckart}
u^\mu_E =u^\mu +\frac{j^\mu}{n}+{\cal O}_2.
\eea
Transforming the 4-momentum current into the Eckart rest frame we obtain
\bea\label{eq:P_i_trans}
P_E^{i}= P^{i}- \frac{j^i}{n}P^0+{\cal O}_2,
\eea
therefore the energy-diffusion current in the E-frame reads
\bea\label{eq:heat_curr_E}
{q}^{\mu}_{E}=
q^\mu-\frac{h}{n}{j}^{\mu}+{\cal O}_2.
\eea
The quantity
\bea\label{eq:heat_flux}
h^{\mu}=q^\mu-\frac{h}{n}{j}^{\mu}
=-\frac{h}{n}\mathscr{J}^\mu
\eea
is the energy flow with respect to the particle flow, and, therefore,
it is natural to call it heat flux. The relation \eqref{eq:heat_flux}
shows that {\it heat conduction and particle diffusion are the same
  phenomena observed from different reference frames}, in the case
where only first-order deviations from equilibrium are taken into
account.  From Eqs.~\eqref{eq:u_Landau} and \eqref{eq:u_Eckart} we
find the relation between the L-frame and the E-frame to order
${\cal O}_1$
\bea\label{eq:u_EL}
u^\mu_L -u^\mu_E =
\frac{h^\mu}{h}=-\frac{\mathscr{J}^\mu}{n}.
\eea

The generalization to the case of multiple conserved charges is
straightforward. We can connect a reference frame to each of these
species via the definition
\bea\label{eq:vel_Eckart_a}
u_{a}^\mu=\frac{N^\mu_a}{\sqrt{N^\mu_a N_{a\mu}}}, 
\eea
which implies that the corresponding diffusion current vanishes, \ie,
$j^\mu_a=0$.

\section{The entropy-production rate (H-theorem)}
\label{app:H-theorem}

It is interesting to compute also the entropy-production 
rate using the formalism of the non-equilibrium statistical 
operator. Differentiating Eq.~\eqref{eq:entropy_op} with 
respect to time, we obtain
\bea\label{eq:entropy_op_deriv_t}
\frac{d}{dt}\hat{S}(t) = -\frac{d}{dt}\Omega_l(t)+
\int\! d^3x \frac{d}{dt}\Big[\beta^\nu(x)\hat{T}_{0\nu}(x)
-\sum\limits_a \alpha_a(x) \hat{N}^0_a(x)\Big].
\eea
The first term can be computed from Eq.~\eqref{eq:Om_eq},
\bea\label{eq:Om_deriv_t} 
\frac{d}{dt}\Omega_l(t) = \int\! d^3x\,
\Big\langle\frac{d}{dt}\Big[\beta^\nu(x)\hat{T}_{0\nu}(x)
-\sum\limits_a \alpha_a(x) \hat{N}^0_a(x)\Big]\Big\rangle_l, 
\eea
where we used Eqs.~\eqref{eq:stat_op_eq} and
\eqref{eq:stat_av_l}.  The time derivative of the integrand in
Eqs.~\eqref{eq:entropy_op_deriv_t} and \eqref{eq:Om_deriv_t} was
already computed in Eq.~\eqref{eq:part_int}, where the surface term
can be dropped. Recalling the definition of the thermodynamic force
given by Eq.~\eqref{eq:C_op} we obtain for
Eq.~\eqref{eq:entropy_op_deriv_t}
\bea\label{eq:entropy_op_deriv_t1}
\frac{d}{dt}\hat{S}(t) = \int\! d^3x 
\left[\hat{C}(x)-\big\langle \hat{C}(x)
\big\rangle_l\right].
\eea
Averaging Eq.~\eqref{eq:entropy_op_deriv_t1} over the full
non-equilibrium statistical operator~\eqref{eq:stat_full_2nd_order}
and using Eq.~\eqref{eq:stat_average} we obtain the entropy-production rate
\bea\label{eq:entropy_op_deriv_t2}
\frac{d}{dt}{S}(t)&=&
\int\! d^3x \,d^4x_1\Big(\hat{C}(x),\hat{C}(x_1)\Big)\no\\
&+&
\int\! d^3x\, d^4x_1d^4x_2\Big(\hat{C}(x),\hat{C}(x_1),
\hat{C}(x_2)\Big)+\dots,
\eea
where the ellipsis stands for higher-order terms.

Thus, the entropy-production rate can be computed in principle at {\it
  any order} using {\it equilibrium} correlators between several
operators $\hat{C}(x)$. We see from Eq.~\eqref{eq:entropy_op_deriv_t1}
that the source of the irreversible entropy production is given by the
deviation of the operator $\hat{C}$ from its local-equilibrium
value. The physical meaning of this result is intuitively clear:
entropy can be produced only if the system deviates from local
thermodynamic equilibrium.

In first-order approximation we can substitute the expression
\eqref{eq:C_decompose} into Eq.~\eqref{eq:entropy_op_deriv_t1},
\bea\label{eq:entropy_op_deriv_t3}
\frac{d}{dt}{S}(t) &=& 
\int\! d^3x \Big( - \beta \theta
{\Pi}+\beta{\pi}_{\rho\sigma}\sigma^{\rho\sigma}
-\sum\limits_a\mathscr{J}_a^\sigma\nabla_\sigma\alpha_a \Big).
\eea
Substituting Eqs.~\eqref{eq:shear_bulk_1} and \eqref{eq:charge_currents1} 
into Eq.~\eqref{eq:entropy_op_deriv_t3} we obtain
\bea\label{eq:entropy_op_deriv_t4}
\frac{d}{dt}{S}(t) =
\int\! d^3x \Big( \zeta\beta \theta^2 
+2\eta\beta{\sigma}_{\mu\nu}\sigma^{\mu\nu}
-\sum\limits_{ab}\chi_{ab}\nabla_\mu\alpha_a \nabla^\mu\alpha_b\Big).
\eea
The integrand of Eq.~\eqref{eq:entropy_op_deriv_t4} is always
positive if $\eta$, $\zeta$ are positive and $\chi_{ab}$ is 
positive semidefinite, which guarantees an increase in entropy
due to non-equilibrium processes.

\section{Derivation of the Kubo formulas}
\label{app:Green_func}

In this Appendix, we derive the relations between the first-order
transport coefficients defined in Sec.~\ref{sec:trans} and retarded
Green's functions, closely following similar derivations in
Refs.~\cite{Hosoya1984AnPhy,Huang2011AnPhy}.  We recall that
in the evaluation of the transport coefficients any non-uniformities
in the thermodynamic parameters can be neglected, \ie, the 
{\it local-equilibrium} distribution can be replaced by a {\it global-equilibrium} 
distribution with some average temperature
$T=\beta^{-1}$ and chemical potentials $\mu_a$.
 
Consider a generic two-point correlator given by
Eq.~\eqref{eq:2_point_corr}.  In equilibrium and in the fluid rest
frame we have $\hat{A} =\beta\hat{K}$,
$\hat{K}=\hat{H}-\sum\limits_a\mu_a\hat{\cal N}_a$ [see
Eq.~\eqref{eq:Gibbs_dist}], therefore from Eqs.~\eqref{eq:B_tau} 
and \eqref{eq:2_point_corr} we obtain
\bea\label{eq:2_point_corr1}
\Big(\hat{X}(\bm x, t),\hat{Y}(\bm x_1, t_1)\Big)=
\int_0^1\!\! d\tau \Big\langle\hat{X}(\bm x, t)
\left[e^{-\beta\tau \hat{K}}\hat{Y}(\bm x_1,t_1)
e^{\beta\tau \hat{K}} - \big\langle \hat{Y}(\bm
x_1,t_1)\big\rangle_l\right]\Big\rangle_l.
\eea
The time evolution of any operator in the Heisenberg picture is
governed by the equation
\bea\label{eq:op_heisenberg} 
\hat{Y}(\bm x,t)=e^{i\hat{K}t}\hat{Y}(\bm x,0)
e^{-i\hat{K}t},
\eea
therefore we have $\hat{Y}(\bm x,t+\delta t)=e^{i\hat{K}(t+\delta t)}\hat{Y}(\bm x,0)
e^{-i\hat{K}(t+\delta t)}=e^{i\hat{K}\delta t}\hat{Y}(\bm
x,t)e^{-i\hat{K}\delta t}$.
Performing an analytic continuation 
$\delta t\to i\tau'$ we obtain
\bea\label{eq:op_heisenberg1}
\hat{Y}(\bm x,t+i\tau')=e^{-\hat{K}\tau'}
\hat{Y}(\bm x,t)e^{\hat{K}\tau'},
\eea
from which we obtain the relations
\bea\label{eq:Y_l}
\big\langle \hat{Y}(\bm x,t+i\tau')\big\rangle_l &=& 
\big\langle \hat{Y}(\bm x,t)\big\rangle_l,\\
\label{eq:KMS}
\big\langle \hat{X}(\bm  x, t) \hat{Y}
(\bm  x_1,t'+i\beta)\big\rangle_l &=& 
\big\langle\hat{Y}(\bm  x_1,t') 
\hat{X}(\bm  x, t) \big\rangle_l.
\eea 
The relation \eqref{eq:KMS} is known as 
Kubo--Martin--Schwinger relation.

Performing a variable change $\beta\tau =\tau'$ in
Eq.~\eqref{eq:2_point_corr1} and employing
Eqs.~\eqref{eq:op_heisenberg1} and \eqref{eq:Y_l} we obtain
\bea\label{eq:2_point_corr2}
\Big(\hat{X}(\bm x, t),\hat{Y}(\bm x_1, t_1)\Big)
= \frac{1}{\beta} \int_0^\beta\!\! d\tau' 
\Big\langle\hat{X}(\bm x, t)\left[\hat{Y}(\bm x_1,t_1+i\tau')
- \big\langle \hat{Y}(\bm x_1,t_1+i\tau')
\big\rangle_l\right]\Big\rangle_l.
\eea
Assuming that the correlations vanish in the limit
$t_1\to -\infty$~\cite{Hosoya1984AnPhy,Huang2011AnPhy}, \ie,
\bea\label{eq:corr_vanish2}
\lim_{t_1\to -\infty}\left(\big\langle\hat{X}(\bm x, t)
\hat{Y}(\bm x_1,t_1+i\tau')\big\rangle_l -
\big\langle\hat{X}(\bm x, t) \big\rangle_l 
\big\langle\hat{Y}(\bm x_1,t_1+i\tau')\big\rangle_l\right)=0,
\eea
we can modify the integrand in Eq.~\eqref{eq:2_point_corr2} as follows 
\bea\label{eq:itgr_mod2}
&&\hspace{2cm}\big\langle\hat{X}(\bm x, t)
\hat{Y}(\bm x_1,t_1+i\tau')\big\rangle_l-
\big\langle\hat{X}(\bm x, t)\big\rangle_l
\big\langle \hat{Y}(\bm x_1,t_1+i\tau')\big\rangle_l \nonumber\\
&&=\Big\langle\hat{X}(\bm x, t) \int_{-\infty}^{t_1}\! dt' 
\frac{d}{dt'}\hat{Y}(\bm  x_1,t'+i\tau')\Big\rangle_l -
\big\langle\hat{X}(\bm x, t)\big\rangle_l
\int_{-\infty}^{t_1}\! dt' \frac{d}{dt'}
\big\langle \hat{Y}(\bm  x_1,t'+i\tau')\big\rangle_l \nonumber\\
&&= -i\int_{-\infty}^{t_1}\! dt' 
\big\langle\hat{X}(\bm x, t)\frac{d}{d\tau'}
\hat{Y}(\bm  x_1,t'+i\tau')\big\rangle_l +
i\int_{-\infty}^{t_1}\! dt' 
\big\langle\hat{X}(\bm x, t)\big\rangle_l\frac{d}{d\tau'}
\big\langle \hat{Y}(\bm  x_1,t'+i\tau')\big\rangle_l.\nonumber
\eea 
Substituting this back into Eq.~\eqref{eq:2_point_corr2} and using the
relations~\eqref{eq:Y_l} and \eqref{eq:KMS} we obtain
\bea\label{eq:2_point_corr3}
\Big(\hat{X}(\bm x, t),\hat{Y}(\bm x_1, t_1)\Big)
=\frac{i}{\beta} \int_{-\infty}^{t_1}\! dt' 
\big\langle\big[\hat{X}(\bm x, t),
\hat{Y}(\bm  x_1,t')\big]\big\rangle_l,
\eea
where the square brackets denote the commutator.
Taking into account that $t'\leq t_1 \leq t$, we 
can write for Eq.~\eqref{eq:2_point_corr3}
\bea\label{eq:2_point_corr5}
\Big(\hat{X}(\bm x, t),\hat{Y}(\bm x_1, t_1)\Big)
= -\frac{1}{\beta} \int_{-\infty}^{t_1}\! dt'\, 
G^R_{\hat{X}\hat{Y}}(\bm x-\bm x_1, t-t'),
\eea
where 
\bea\label{eq:retarded_corr}
G^R_{\hat{X}\hat{Y}}(\bm x-\bm x', t-t')=
-i\theta(t-t')\big\langle\big[\hat{X}(\bm x, t),
\hat{Y}(\bm  x',t')\big]\big\rangle_l
\eea
is the retarded two-point Green's function for a uniform medium.

Now consider a generic transport coefficient given by the integral
\bea\label{eq:I_XY}
I[{\hat{X},\hat{Y}}](\omega') = \beta\!\int\! d^3x_1\!
\int_{-\infty}^t\!\! dt_1 e^{i\omega'(t-t_1)}
e^{\varepsilon(t_1-t)}\Big(\hat{X}(\bm x, t),\hat{Y}
(\bm x_1, t_1)\Big),
\eea
where we also introduced a nonzero frequency $\omega'> 0$ for the sake of
convenience; we will take the limit $\omega'\to 0$ at
the end of the calculations.  According to Eq.~\eqref{eq:2_point_corr5} we can write
Eq.~\eqref{eq:I_XY} as
\bea\label{eq:I_XY1}
I[{\hat{X},\hat{Y}}](\omega')
=-\int_{-\infty}^0\! dt' e^{(\varepsilon-i\omega') t'}\!\!
\int_{-\infty}^{t'}\! dt\! \int\! d^3x\,
G^R_{\hat{X}\hat{Y}}(-\bm x, -t).
\eea
Considering the Fourier transformation
\bea\label{eq:green_fourier1}
G^R_{\hat{X}\hat{Y}}(\bm x, t) =
\int\! \frac{d^3k}{(2\pi)^3}\int_{-\infty}^{\infty}
\frac{d\omega}{2\pi}e^{-i(\omega t-\bm k\cdot \bm x)}
G^R_{\hat{X}\hat{Y}}(\bm k,\omega),\no
\eea
we obtain
\bea
\int\! d^3x\, G^R_{\hat{X}\hat{Y}}(-\bm x, -t)=
\int_{-\infty}^{\infty} \frac{d\omega}{2\pi}
e^{i\omega t}G^R_{\hat{X}\hat{Y}}(\omega),\no
\eea
where
$G^R_{\hat{X}\hat{Y}}(\omega)\equiv\lim_{\bm k\to 0}
G^R_{\hat{X}\hat{Y}}(\bm k,\omega)$.  
In Eq.~\eqref{eq:I_XY1} we now encounter the integral
$\int_{-\infty}^{t'} dt\, e^{i\omega t}$,
which we compute by a shift $\omega \rightarrow \omega - i \delta$, $\delta >0$, taking
the limit $\delta \rightarrow 0^+$ at the end,
\bea
\int_{-\infty}^{t'} dt\, e^{i\omega t}
= \lim_{\delta \to 0^+} \int_{-\infty}^{t'} dt\, e^{(i\omega + \delta)t}
= \lim_{\delta \to 0^+} \frac{e^{(i\omega + \delta) t'}}{i \omega + \delta} .
\eea
Then we have from Eq.~\eqref{eq:I_XY1}
\bea\label{eq:I_XY2}
I[{\hat{X},\hat{Y}}] (\omega')
&=&-\lim_{\delta \to 0^+} \int_{-\infty}^{\infty}\frac{d\omega}{2\pi}\,
G^R_{\hat{X}\hat{Y}}(\omega)\!\int_{-\infty}^0\! dt' 
e^{(\varepsilon-i\omega') t'}
\frac{e^{(i\omega+ \delta) t'}}{i\omega+ \delta}\nonumber\\
&=&\lim_{\delta \to 0^+}\frac{i}{\omega'+i\varepsilon}
\oint\frac{d\omega}{2\pi i}
\left(\frac{1}{\omega -\omega'-i(\varepsilon+ \delta)}
  -\frac{1}{\omega-i \delta}\right)G^R_{\hat{X}\hat{Y}}(\omega),\no
\eea
where the integral is closed in the upper half-plane, where the
retarded Green's function is analytic.  
Note that the contribution from the upper half-circle at 
infinity vanishes if the retarded Green's function goes 
to zero sufficiently rapidly, namely, not slower than 
$\omega^{-1}$, which we assume to be the case here.
Applying Cauchy's integral formula and performing the 
limits $\delta \to 0^+,\, \varepsilon\to 0^+$ we obtain
\bea\label{eq:I_XY3}
I[{\hat{X},\hat{Y}}] (\omega')
= \frac{i}{\omega'}\left[G^R_{\hat{X}\hat{Y}}
(\omega')-G^R_{\hat{X}\hat{Y}}(0)\right].
\eea
 Going to the zero-frequency limit $\omega'\to 0$
 we obtain the final formula
\bea\label{eq:I_XY4}
I[{\hat{X},\hat{Y}}](0)=i\frac{d}{d\omega}
G^R_{\hat{X}\hat{Y}}(\omega)\bigg\vert_{\omega=0},
\eea
with
\bea\label{eq:green_fourier2}
G^R_{\hat{X}\hat{Y}}(\omega) 
= -i\int_{0}^{\infty}\!\! dt e^{i\omega t}\!\!
\int\! d^3x\, \big\langle\big[\hat{X}(\bm x, t),
\hat{Y}(\bm 0,0)\big]\big\rangle_l.
\eea
From Eqs.~\eqref{eq:I_XY3} and \eqref{eq:green_fourier2}
we find that
\bea\label{eq:parity_green}
\big\{G^R_{\hat{X}\hat{Y}}(\omega)\big\}^* =
G^R_{\hat{X}\hat{Y}}(-\omega),\qquad
\big\{I[\hat{X},\hat{Y}](\omega)\big\}^* =
I[\hat{X},\hat{Y}](-\omega).
\eea
Indeed, since $\hat{X}(\bm x, t)$ and 
$\hat{Y}(\bm x, t)$
are hermitian operators, we have the property
\bea
\big\langle\big[\hat{X}(\bm x, t),
\hat{Y}(\bm  x',t')\big]\big\rangle_l^*=
-\big\langle\big[\hat{X}(\bm x, t),
\hat{Y}(\bm  x',t')\big]\big\rangle_l,
\eea
therefore the retarded Green's function given by
Eq.~\eqref{eq:retarded_corr} is real, which is used to obtain the
first relation in Eq.~\eqref{eq:parity_green}. From
Eq.~\eqref{eq:parity_green} we have also
\bea\label{eq:parity_green1}
{\rm Re}G^R_{\hat{X}\hat{Y}}(-\omega)=
{\rm Re}G^R_{\hat{X}\hat{Y}}(\omega),\qquad
{\rm Im}G^R_{\hat{X}\hat{Y}}(-\omega)=
-{\rm Im}G^R_{\hat{X}\hat{Y}}(\omega),
\eea
therefore from Eqs.~\eqref{eq:I_XY} and \eqref{eq:I_XY4} we obtain in
the zero-frequency limit
\bea\label{eq:I_XY5}
I[{\hat{X},\hat{Y}}](0)= \beta\!\int\! d^4x_1
\Big(\hat{X}(x),\hat{Y}(x_1)\Big) =-\frac{d}{d\omega}
{\rm Im}G^R_{\hat{X}\hat{Y}}(\omega)\bigg\vert_{\omega=0},
\eea
where we used the short-hand notation 
defined in Eq.~\eqref{eq:int_short}.

Now let us show that the Green's function \eqref{eq:green_fourier2} is
symmetric in its arguments if the operators $\hat{X}$ and $\hat{Y}$
have the same parity under time reversal. We have
\bea\label{eq:Green_trans}
G^R_{\hat{Y}\hat{X}}(\omega) 
&=& i\!\int_{0}^{\infty}\!\! dt e^{i\omega t}\!\! 
\int\! d^3x\,\big\langle\big[\hat{X}(\bm 0,0),
\hat{Y}(\bm x,t)\big]\big\rangle_l\nonumber\\
&=& i\!\int_{0}^{\infty}\!\! dt e^{i\omega t}\!\!
\int\! d^3x\,\big\langle\big[\hat{X}(-\bm x,-t),
\hat{Y}(\bm 0,0)\big]\big\rangle_l \nonumber\\
&=& i\!\int_{0}^{\infty}\!\! dt e^{i\omega t}\!\!
\int d^3x\,\big\langle\big[\hat{X}(\bm x,-t),
\hat{Y}(\bm 0,0)\big]\big\rangle_l,
\eea
where we used the uniformity of the medium.
For hermitian operators, we have the following
transformation rule under time reversal
\bea\label{eq:time_reversal}
\hat{X}_T(\bm x, t)=\eta_X\hat{X}(\bm x, -t),
\qquad\hat{Y}_T(\bm x, t)=\eta_Y\hat{Y}(\bm x, -t),\no
\eea 
with $\eta_{X,Y}=\pm 1$ for even/odd parity under time reversal. 
For Eq.~\eqref{eq:Green_trans} we then have
\bea
G^R_{\hat{Y}\hat{X}}(\omega) &=&
i\eta_X\eta_Y\! \int_{0}^{\infty}\! dt e^{i\omega t}\!
\int\! d^3x\,\big\langle\big[\hat{X}_T(\bm x,t),
\hat{Y}_T(\bm 0,0)\big]\big\rangle_l\nonumber\\
 &=& i\eta_X\eta_Y\! \int_{0}^{\infty}\! dt e^{i\omega t}\!
\int\! d^3x\, \big\langle\big[\hat{X}(\bm x,t),
\hat{Y}(\bm 0,0)\big]\big\rangle_{l,T}.\no
\eea
Finally, taking into account that the statistical average of a
commutator of hermitian operators is purely imaginary and the operator
of time reversal is antiunitary (\ie, transforms a number to its
complex conjugate), we obtain
\bea\label{eq:green_symmetry}
G^R_{\hat{Y}\hat{X}}(\omega) = -i\eta_X\eta_Y\! \int_{0}^{\infty}\!
dt e^{i\omega t}\!\int\! d^3x\,\big\langle\big[\hat{X}(\bm x,t),
\hat{Y}(\bm 0,0)\big]\big\rangle_l = 
\eta_X\eta_Y G^R_{\hat{X}\hat{Y}}(\omega).
\eea
Thus, if $\eta_X=\eta_Y$, we obtain
$G^R_{\hat{Y}\hat{X}}(\omega) =G^R_{\hat{X}\hat{Y}}(\omega)$, and,
therefore, $I[\hat{Y},\hat{X}](\omega) =I[\hat{X},\hat{Y}](\omega)$,
which is Onsager's symmetry principle for transport
coefficients. Using now Eq.~\eqref{eq:I_XY5} and the definitions of
the transport coefficients given by Eqs.~\eqref{eq:shear_def},
\eqref{eq:bulk_def}, \eqref{eq:chi_ab_def} and \eqref{eq:kappa_def},
we obtain the formulas \eqref{eq:shear_bulk_mod} and
\eqref{eq:kappa_mod} of the main text.

In the derivation of the second-order equations of motion for
the dissipative currents we encounter integrals of the type 
\bea\label{eq:I_XY_tau}
I^\tau[{\hat{X},\hat{Y}}](\omega)
= \beta\! \int\! d^4x_1 e^{i\omega(t-t_1)}
\Big(\hat{X}(x),\hat{Y}(x_1)\Big)(x_1-x)^\tau,
\eea
where we used again the short-hand notation~\eqref{eq:int_short}.  The
correlator $\big(\hat{X}(x),\hat{Y}(x_1)\big)$ evaluated in the local
rest frame depends on the spatial coordinates only via the difference
$|\bm x-\bm x_1|$, \ie, it is an even function of $\bm x-\bm
x_1$. Then Eq.~\eqref{eq:I_XY_tau} implies that the spatial components
of the vector $I^\tau$ vanish in that frame, and for the temporal
component we have
\bea\label{eq:vector_I0_rest}
I^0[{\hat{X},\hat{Y}}](\omega) &=& \beta\! 
\int\! d^4x_1 e^{i\omega(t-t_1)} \Big(\hat{X}(x),
\hat{Y}(x_1)\Big)(t_1-t)\nonumber\\
&=& i\beta\frac{d}{d\omega}\!\int\! d^4x_1
e^{i\omega(t-t_1)}\Big(\hat{X}(x),\hat{Y}(x_1)\Big)
= i\frac{d}{d\omega}I[{\hat{X},\hat{Y}}](\omega),
\eea
where we used Eq.~\eqref{eq:I_XY}.  From Eqs.~\eqref{eq:I_XY3} and
\eqref{eq:vector_I0_rest} we obtain in the limit $\omega\to 0$
\bea \label{eq:vec_I}
I^0[{\hat{X},\hat{Y}}](0) =K[{\hat{X},\hat{Y}}],
\eea
where we defined
\bea\label{eq:K_XY}
K[\hat{X},\hat{Y}] 
 \equiv -\frac{1}{2}\frac{d^2}{d\omega^2} G^R_{\hat{X}
\hat{Y}}(\omega)\bigg\vert_{\omega=0}
 =-\frac{1}{2}\frac{d^2}{d\omega^2} {\rm Re}G^R_{\hat{X}
\hat{Y}}(\omega)\bigg\vert_{\omega=0}.
\eea
Note that in Eqs.~\eqref{eq:I_XY4} and \eqref{eq:K_XY} the Green's
function should be evaluated in the fluid rest frame.  The relation
\eqref{eq:vec_I} can also be cast into a covariant form
\bea\label{eq:vec_gen}
\beta\!\int\! d^4x_1 \Big(\hat{X}(x),\hat{Y}
(x_1)\Big)(x_1-x)^\tau =K[{\hat{X},\hat{Y}}]u^\tau.
\eea


\section{Properties of the non-local projectors}
\label{app:projectors}

Using Eqs.~\eqref{eq:prop_proj} and \eqref{eq:projector_2nd_1} we find
the following properties for the second-rank projector
$\Delta_{\mu\nu}(x,x_1)$
\bea\label{eq:prop_projectors_2}
\Delta_{\mu\nu}(x,x_1) = \Delta_{\nu\mu}(x_1,x),&&
u^\mu(x)\Delta_{\mu\nu}(x,x_1) =
\Delta_{\mu\nu}(x,x_1)u^\nu(x_1) = 0,\nonumber\\
\Delta_{\mu\nu}(x,x) = \Delta_{\mu\nu}(x),&&
\Delta^{\mu}_\alpha(x)\Delta_{\mu\nu}(x,x_1)
=\Delta_{\alpha\nu}(x,x_1).
\eea
From Eqs.~\eqref{eq:prop_projector4_1}, \eqref{eq:prop_projector4_2},
and \eqref{eq:projector_2nd_2} we find for the fourth-rank projector 
$\Delta_{\mu\nu\rho\sigma}(x,x_1)$
\bea
\Delta_{\mu\nu\rho\sigma}(x,x_1) = 
\Delta_{\nu\mu\rho\sigma}(x,x_1)=
\Delta_{\nu\mu\sigma\rho}(x,x_1)=
\Delta_{\rho\sigma\mu\nu}(x_1,x),\hspace{2cm}\nonumber\\
\label{eq:prop_projectors_4}
 u^\mu(x)\Delta_{\mu\nu\rho\sigma}(x,x_1) =0,\quad
\Delta_{\mu\nu\rho\sigma}(x,x_1)u^\rho(x_1) = 0,\hspace{3cm}\\
\Delta_{\mu\nu\rho\sigma}(x,x)=
\Delta_{\mu\nu\rho\sigma}(x),\quad
\Delta^\mu_{~\mu\rho\sigma}(x,x_1)=0,\quad
\Delta_{\mu\nu\rho\sigma}(x,x_1)
\Delta^{\rho\sigma}_{\alpha\beta}(x_1)=
\Delta_{\mu\nu\alpha\beta}(x,x_1).\nonumber
\eea

For our calculations, it is sufficient to expand the non-local
projectors around the point $x_1=x$ keeping only the linear terms in
the difference $x_1-x$. We thus approximate
\bea
u_\mu(x_1)\simeq u_\mu(x)+(x_1-x)^\alpha\partial_\alpha u_\mu(x),\no
\eea
which due to the identity $u^\mu\partial_\alpha u_\mu =0$ gives
$u^\mu(x)u_\mu(x_1)\simeq 1$.
In this approximation, we find for the projector~\eqref{eq:projector_2nd_1} 
up to terms of second order in $x_1 - x$
\bea\label{eq:projector_1_approx}
\Delta_{\mu\nu}(x,x_1) =
\Delta_{\mu\nu}(x)-\left[ u_\mu(x_1)- u_\mu (x) \right] u_\nu(x_1)
=\Delta_{\mu\nu}(x) -u_\nu(x)(x_1-x)^\alpha\partial_\alpha u_\mu(x),
\eea
and from Eqs.~\eqref{eq:prop_projectors_2}
and \eqref{eq:projector_1_approx} we obtain
\bea
\Delta_{\mu}^{\mu}(x,x_1) = 3,\qquad
\Delta^\alpha_{\nu}(x)\Delta_{\mu\alpha}(x,x_1)
=\Delta_{\mu\nu}(x),\hspace{2.3cm}\nonumber\\
\label{eq:prop_projectors_approx1}
\Delta_{\mu\nu}(x,x_1)\Delta^{\mu\nu}(x,x_1)=
\Delta_{\mu\lambda}(x)\Delta^{\mu\nu}(x,x_1)
\Delta^\lambda_{\nu}(x_1) =
\Delta_{\mu\lambda}(x)\Delta^{\mu\lambda}(x,x_1)=3,\\
\Delta_{\mu\nu}(x,x_1)\Delta^{\nu\mu}(x,x_1)=
\Delta_{\mu\lambda}(x)\Delta^{\nu\mu}(x,x_1)
\Delta^\lambda_{\nu}(x_1) =
\Delta^{\nu\lambda}(x,x_1)\Delta_{\nu\lambda}(x_1)=3. 
\nonumber
\eea
In the same approximation the fourth-rank projector 
$\Delta_{\mu\nu\rho\sigma}(x,x_1)$ can be written as
\bea\label{eq:projector_2_approx}
\Delta_{\mu\nu\rho\sigma}(x,x_1)
=
\frac{1}{2}\big[\Delta_{\mu\rho}(x,x_1)
\Delta_{\nu\sigma}(x,x_1)
+\Delta_{\mu\sigma}(x,x_1)\Delta_{\nu\rho}(x,x_1)\big]-
\frac{1}{3}\Delta_{\mu\nu}(x)\Delta_{\rho\sigma}(x_1),
\eea
which together with the properties~\eqref{eq:prop_projectors_approx1} gives 
\bea\label{eq:prop_projectors_approx2}
\Delta_{\mu\nu}^{~~~\mu\nu}(x,x_1)=5.
\eea
Equation \eqref{eq:prop_projectors_approx2} together with the first
relation in Eq.~\eqref{eq:prop_projectors_approx1} was used in
Eqs.~\eqref{eq:corr1_curr_2nd} and \eqref{eq:corr1_stress_2nd} for the
normalization of the corresponding correlation functions.

Using the relation $\partial_\alpha \Delta_{\gamma\delta} =
-u_\gamma\partial_\alpha u_\delta -u_\delta\partial_\alpha u_\gamma$, 
from Eq.~\eqref{eq:projector_delta4} we obtain
\bea\label{eq:delta_deriv1}
\partial_\alpha\Delta_{\gamma\delta\rho\sigma}
&=&
-\frac{1}{2}\Big[
\Delta_{\gamma\rho}(u_\sigma\partial_\alpha u_\delta
+u_\delta\partial_\alpha u_\sigma)+
\Delta_{\delta\sigma}(u_\gamma\partial_\alpha u_\rho
+u_\rho\partial_\alpha u_\gamma)+(\rho\leftrightarrow\sigma)\Big]\nonumber\\
&& +\frac{1}{3}\Big[
\Delta_{\gamma\delta}(u_\rho\partial_\alpha u_\sigma
+u_\sigma\partial_\alpha u_\rho)+
\Delta_{\rho\sigma}(u_\gamma\partial_\alpha u_\delta
+u_\delta\partial_\alpha u_\gamma)\Big].\no
\eea
Multiplying this by $\Delta_{\mu\nu}^{\gamma\delta}(x)$ 
and using the properties \eqref{eq:prop_projector4_1} 
and \eqref{eq:prop_projector4_2} we obtain
\bea\label{eq:delta_deriv2_app}
\frac{\partial}{\partial x_1^\alpha}
\Delta_{\mu\nu\rho\sigma}(x,x_1)\bigg\vert_{x_1=x}
=
\Delta_{\mu\nu}^{\gamma\delta}\partial_{\alpha}
\Delta_{\gamma\delta\rho\sigma}=
-(\Delta_{\mu\nu\rho\beta}u_{\sigma}+
\Delta_{\mu\nu\sigma\beta}u_{\rho})
\partial_\alpha u^\beta,
\eea
where we recalled Eq.~\eqref{eq:projector_2nd_2}.

\newpage

\providecommand{\href}[2]{#2}\begingroup\raggedright\endgroup

\end{document}